\documentclass[reprint,twocolumn,superscriptaddress]{revtex4-2}
\usepackage{graphicx,amsmath,amssymb,amsthm,amsfonts,epsfig,bm,xcolor,longtable,booktabs,soul}
\usepackage{float}
\usepackage[colorlinks=true, urlcolor=blue, citecolor=blue, linkcolor=blue]{hyperref}
\usepackage[version=4]{mhchem}
\usepackage[utf8]{inputenc}
\def\oden{Oden Institute for Computational Engineering and Sciences, The University of Texas at Austin, 201 E. 24$^{th}$ Street, Austin, TX 78712, USA}
\def\physics{Department of Physics, The University of Texas at Austin, Austin, TX 78712, USA}
\newcommand{\figref}[1]{Fig.~\ref{#1}}
\newcommand{\tabref}[1]{Table~\ref{#1}}
\newcommand{\mob}{cm$^2$/Vs}
\newcommand{\aibte}{\textit{ai}BTE}
\def\bk{\mathbf k}
\def\bq{\mathbf q}
\def\bG{\mathbf G}
\def\kt{k_{\rm B}T}
\newcommand{\ratenm}{\Gamma_{n\bk\rightarrow m\bk+\bq}}
\newcommand{\ratemn}{\Gamma_{m\bk+\bq\rightarrow n\bk}}

\begin{document}

\title{High-throughput screening of 2D materials identifies \\[3pt] \textit{p}-type monolayer WS$_2$ as potential ultra-high mobility semiconductor}

\author{Viet-Anh Ha}
\affiliation{\oden}
\affiliation{\physics}
\author{Feliciano Giustino}%
\affiliation{\oden}
\affiliation{\physics}
\affiliation{Corresponding author E-mail: fgiustino@oden.utexas.edu}

\date{\today}

\begin{abstract}
2D semiconductors offer a promising pathway to replace silicon in next-generation electronics. Among their many advantages, 2D materials possess atomically-sharp surfaces and enable scaling the channel thickness down to the monolayer limit. However, these materials exhibit comparatively lower charge carrier mobility and higher contact resistance than 3D semiconductors, making it challenging to realize high-performance devices at scale. In this work, we search for high-mobility 2D materials by combining a high-throughput screening strategy with state-of-the-art calculations based on the \textit{ab initio} Boltzmann transport equation. Our analysis singles out a known transition metal dichalcogenide, monolayer WS$_2$, as the most promising 2D semiconductor, with the potential to reach ultra-high room-temperature hole mobilities in excess of 1300~\mob\ should Ohmic contacts and low defect densities be achieved. Our work also highlights the importance of performing full-blown \textit{ab initio} transport calculations to achieve predictive accuracy, including spin-orbital couplings, quasiparticle corrections, dipole and quadrupole long-range electron-phonon interactions, as well as scattering by point defects and extended defects.
\end{abstract}

\maketitle


%
%
\section{Introduction}

Ever since the early days of integrated circuits, the number of transistors per chip has nearly doubled every two years, a trend that is known as Moore's law~\cite{Mack2011}. For this trend to continue beyond the current 5\,nm technology node, it will be necessary to develop GAAFETs (gate-all-around field-effect transistors) and ribbonFETs with sub-nanometer channel thickness~\cite{OBrien2023}. In this ultra-scaled regime, the carrier mobility of silicon is severely degraded by quantum confinement, surface roughness, and dangling bonds, falling below 50~\mob\ at room temperature~\cite{Uchida2002, Tsutsui2006}. To mitigate this effect, van der Waals 2D materials are being investigated as potential transistor channels. The main appeal of 2D materials is that they possess atomically-sharp surfaces and can be scaled down to the monolayer limit~\cite{Liu2021,Su2021,Zhang2023}. Additionally, the use of 2D materials enables the reduction of the gate length as compared to silicon finFETs, leading to lower switching capacitance and reduced power consumption~\cite{OBrien2021}. 

Despite their unique promise, the use of 2D materials in transistor channels faces significant challenges. For one, reduced dimensionality leads to intrinsically high density of electronic states at the band edges, which leads to high carrier scattering rates and low phonon-limited mobilities~\cite{Cheng2018, Li2019, Cheng2019, Cheng2020}. Furthermore, the very lack of surface dangling bonds makes it harder to form Ohmic contacts, leading to high contact resistance~\cite{OBrien2023}. On top of these difficulties, native defects in 2D materials and the lack of suitable dopants make it challenging to avoid Fermi level pinning. As a result, most 2D materials exhibit low field-effect mobilities, typically in the range 50-100~\mob\ ~\cite{Liu2021, Su2021}.

Advanced \textit{ab initio} computational methods can play an important role in this area by shedding light on the microscopic mechanisms that hinder carrier transport in 2D materials, and by helping to identify high-mobility compounds via computational screening. Several groups have embarked in the computational search for high-mobility 2D materials~\cite{Sohier2018, Cheng2018, Cheng2019, Cheng2020, Song2023, Zhang2023b}. For example, in Ref.~\citenum{Song2023}, the authors have performed mobility calculations for two dozen putative binary borides, nitrides, and oxides in the monolayer hexagonal lattice, as well as III-V and II-VI semiconductors in the bilayer hexagonal lattice. They identified several candidates with very high mobility, in excess of 1000~\mob. However, these calculations are based on the relaxation time approximation and do not include spin-orbit coupling except for BSb. In Ref.~\citenum{Sohier2018}, the authors considered electron-doped MoS$_2$, WS$_2$, WSe$_2$, phosphorene, and arsenene, as well as hole-doped phosphorene, but spin-orbit coupling was not accounted for. They identified phosphorene as the most promising $n$-type channel, however this compound tends to be unstable toward oxidation under ambient conditions. In Ref.~\citenum{Cheng2019}, the authors investigated elemental pnictogen compounds in their hexagonal monolayer form (Sb, As, P, Bi), and predicted a very high hole mobility for antimonene, in excess of 1300~\mob\ at room temperature. In this case, calculations were performed within the relaxation time approximation, which may not be as accurate as full Boltzmann transport calculations~\cite{Ponce2021}.

Large-scale high-throughput (HT) searches for high-mobility 2D materials have recently been made possible by the development of 2D materials repositories such as the Materials Cloud 2D Database (MC2D)~\cite{Mounet2018, Campi2023} and the Computational 2D materials database (C2DB)~\cite{Haastrup2018, Gjerding2021}. The MC2D database was generated by considering all 3D bulk inorganic compounds experimentally synthesized, and by identifying the subset of layered compounds that are amenable to exfoliation. The C2DB database was generated by considering known 2D structure polytypes, and decorating these lattices combinatorially with every element of the periodic table. These libraries provides us with thousands of promising 2D materials to screen in search for high-mobility compounds. For example, the authors of Ref.~\citenum{Zhang2023b} recently analyzed over 4000 putative compounds from the C2DB dataset in search for high-mobility semiconductors; they identified several candidates with room-temperature mobilities in excess of 1400~\mob, including BSb monolayer which was also identified in Ref.~\citenum{Song2023}.

In this broad context, the aim of the present work is threefold: (i) To establish a systematic HT screening approach for identifying high-mobility 2D materials; (ii) To report reproducible \textit{ab initio} data on the carrier mobilities of top materials candidates; (iii) To explain why real 2D materials exhibit carrier mobilities that are far below their theoretical intrinsic limit. In order to complement existing studies, we here focus on the MC2D database whose practical advantage is that all 3D parent compounds are experimentally available, therefore the likelihood of realizing the corresponding 2D monolayers is high. This database was also screened in the previous work~\cite{Sohier2020} but using different criteria. In particular, in Ref.~\citenum{Sohier2020} SOC was not considered in the band structure screening and in the calculation of transport properties (except for \ce{WSe2}). As we show in this manuscript, SOC plays an important role in mobility calculations and in the determination of the hole effective masses in most cases.

The distinctive feature of the present study is that, for the most promising compounds, we do not limit ourselves to the relaxation time approximation, but we perform full-blown \textit{ab initio} Boltzmann transport calculations (including spin-orbit couplings and a recent development for Fr\"ohlich electron-phonon interactions in 2D~\cite{Sio2022}) in order to provide best theoretical estimates for the carrier mobility. For the case of WS$_2$, which emerges from the present unbiased search as the most promising $p$-type compound, we also perform GW quasiparticle calculations and include quadrupole corrections as well as scattering by ionized impurities and extended defects in order to obtain the most accurate estimate of the carrier mobility reported thus far.
%
%
%
\section{Results}

\begin{figure}
  \begin{center}
  \includegraphics[width=0.9\linewidth]{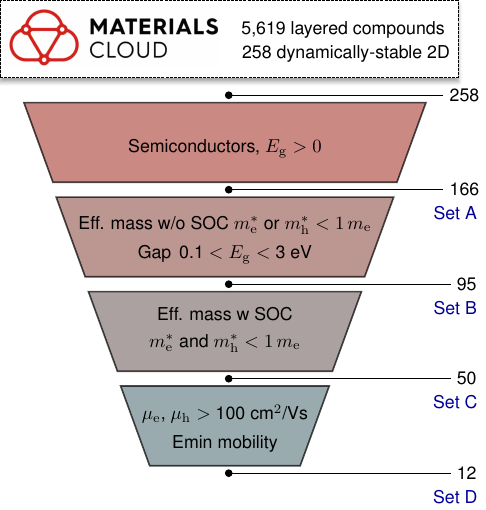}\vspace{-5pt}
  \caption{\textbf{High-throughput screening workflow}. Schematic of the down-selection procedure employed to identify high-mobility 2D materials within the MC2D database~\cite{Mounet2018}. Starting from this database, we perform structure optimization for 258 compounds; we compute band structures and identify a subset of 166 compounds with finite band gap (Set~A); we calculate conductivity effective masses without SOC and down-select 95 compounds (Set~B); for this set, we re-calculate effective masses with SOC and down-select to 50 compounds (Set~C); we estimate mobilities via Emin's formula and down-select 12 compounds for high-accuracy \textit{ab initio} Boltzmann transport calculations (Set~D).\vspace{-10pt}}
  \label{HT_tier}
  \end{center}
\end{figure}

We present our findings in three steps. First, we outline our procedure for pre-selecting high-mobility candidate materials from the MC2D database (see \figref{HT_tier}). Second, we report high-precision calculations of carrier mobility for sixteen top candidates based on the \textit{ab initio} Boltzmann transport equation (\aibte). Third, we proceed to an in-depth analysis of WS$_2$, which emerges from the present search as the highest-mobility 2D compound within the MC2D database. For this latter case, we report on high-level many-body calculations and investigate the role of carrier concentration, spin-orbital couplings (SOC), quadrupole corrections, impurity scattering, and scattering by extended defects. Full details on the methodology and calculation parameters are reported in the Methods section. 

%
%
\vspace{20pt}
\noindent\textbf{High-throughput screening}
\vspace{5pt}
Figure~\ref{HT_tier} shows a schematic of the HT screening workflow employed in this work. The workflow starts from the MC2D database, which reports 2D materials that can be obtained from exfoliation of 3D compounds already synthesized~\cite{Mounet2018}. Parent 3D compounds were sourced from the Inorganic Crystal Structure Database (ICSD)~\cite{ICSD_url} and the Crystallographic Open Database (COD)~\cite{Graulis2012}. The database contains 5,619 layered materials, and a subset of 1,036 compounds are labeled as ``easily exfoliable'' based on their computed exfoliation energy ($< 30$~meV/\AA$^2$). From this subset, phonon dispersion relations were computed, and 258 compounds were found to exhibit no soft phonons, i.e., to be dynamically stable. 166 compounds of this set are non-metallic at the DFT level. We name this collection of 166 non-metallic compounds ``Set~A'' (Table~\ref{tableI}).

\begin{figure}
  \begin{center}
  \includegraphics[width=0.9\columnwidth]{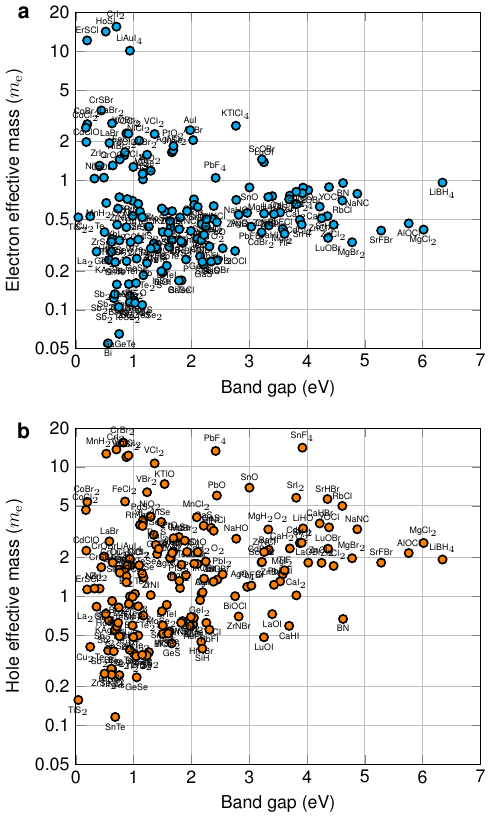}\vspace{-5pt}
  \caption{\textbf{Conductivity effective masses without SOC}. 
  Computed conductivity effective masses for the 165 compounds of Set~A. (a) Electron masses, (b) Hole masses. Effective masses are calculated at 300~K and for a carrier concentration of 10$^{10}$~cm$^{-2}$. In the case of anisotropic effective mass tensors, we report the isotropic average. These calculations do not include SOC.\vspace{-10pt}}
  \label{fig.masses.nosoc}
  \end{center}
\end{figure}

We perform DFT calculations for all 2D materials in Set A; in particular, we compute ground-state structure, band structures, and conductivity effective masses. Ref.~\citenum{Mounet2018} reports magnetic order for 25 compounds of this set. In these cases, we tested both ferro- and antiferromagnetic order; these compounds are indicated in Table~\ref{tableI}. 
The conductivity effective masses of electrons and holes computed for compounds in Set A are shown in Fig.~\ref{fig.masses.nosoc}, as a function of the density-functional theory (DFT) gap. Masses are evaluated for a carrier concentration of 10$^{10}$~cm$^{-2}$ and a temperature of 300~K. We have checked that these masses do not change significantly when considering a higher carrier concentration of 10$^{12}$~cm$^{-2}$. At this stage we do not include SOC in order to contain computational cost. Calculated masses span a very broad range, from 0.05\,$m_{\rm e}$ to $\sim$300\,$m_{\rm e}$, where $m_{\rm e}$ is the free electron mass. For clarity, only compounds with masses smaller than 20\,$m_{\rm e}$ are shown in the figure, and all data are reported in Table~\ref{tableI}. As in the case of 3D bulk semiconductors \cite{Hautier2013}, hole masses are typically heavier than electron masses: the average electron mass in Set~A is 2.0\,$m_{\rm e}$, while the average hole mass is 4.1\,$m_{\rm e}$. Conversely, unlike in simple $\bf k \cdot \bf p$ models, which predict a monotonic dependence of the effective mass on the band gap, Fig.~\ref{fig.masses.nosoc} shows that the correlation between masses and band gap is relatively weak for the compounds investigated here.

In the next step of the down-selection process, we retain compounds which exhibit low effective masses and narrow or intermediate band gaps. Specifically, we retain compounds with either the electron or the hole mass $< 1\,m_{\rm e}$ based on typical values for conventional 3D semiconductors, and band gaps in the range $0.1 < E_{\rm g} < 3$~eV. Given the band gap underestimation in DFT, this lower bound ensures that we only retain compounds with (expected) experimental band gap $\gtrsim$0.5~eV. These criteria narrow down the collection to 94 compounds; we call this reduced subset ``Set~B'' (Table~\ref{tableS2}).

\begin{figure}
  \begin{center}
  \includegraphics[width=0.9\columnwidth]{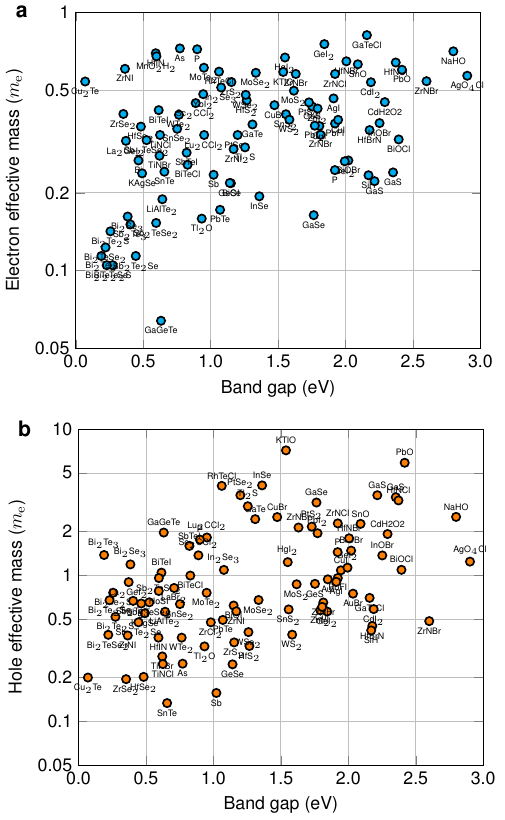}\vspace{-5pt}
  \caption{\textbf{Conductivity effective masses including SOC}. Computed conductivity effective masses for the 95 compounds of Set~B. (a) Electron masses, (b) Hole masses. Effective masses are calculated at 300~K and for a carrier concentration of 10$^{10}$~cm$^{-2}$. In the case of anisotropic effective mass tensors, we report the isotropic average. SOC is included in these calculations. Table~\ref{tableS2} reports directional masses for each compound.\vspace{-10pt}}
  \label{fig.masses.soc}
  \end{center}
\end{figure}

\begin{figure}
  \begin{center}
  \includegraphics[width=0.9\columnwidth]{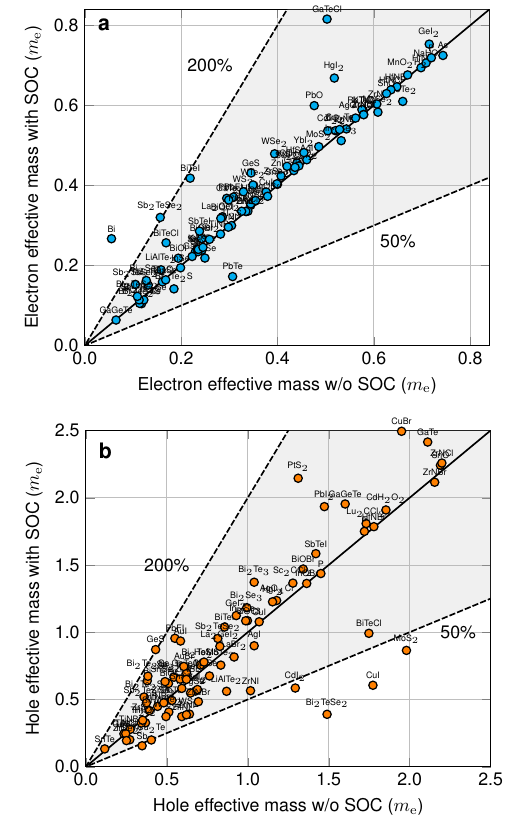}\vspace{-5pt}
  \caption{ 
  \textbf{Effect of SOC on the conductivity effective masses.}
  Comparison between the conductivity effective masses computed for the 95 compounds of Set~B, without SOC (horizontal axis) and with SOC (vertical axis). (a) Electron masses; (b) Hole masses. If SOC and non-SOC masses were identical, all data points would fall on the diagonal line at $45^\circ$. For compounds with anisotropic mass tensors, we report the isotropic average. The shaded areas bounded by the dashed lines indicate the region where the ratio of non-SOC and SOC effective masses is between 50\% and 200\%. All effective masses are evaluated at 300~K for a carrier concentration of 10$^{10}$~cm$^{-2}$. The complete data set is reported in Table~\ref{tableS2}.\vspace{-10pt}}
  \label{fig.masses.comparison}
  \end{center}
\end{figure}

Starting from Set~B, we re-calculate the conductivity effective masses of 95 compounds, this time by taking into account SOC, which is considerably more demanding computationally. The computed effective masses are shown in Fig.~\ref{fig.masses.soc}, and reported in full in Table~\ref{tableS2}. As expected, SOC modifies the effective masses significantly in most materials containing heavy elements, as shown in Fig.~\ref{fig.masses.comparison}. In this figure we see that the modification of the effective masses induced by SOC ranges from a reduction of 50\% to an enhancement of up to 200\%. In average, this effect is most pronounced in the case of the hole effective masses, for which the mean absolute relative error (MARE) between non-SOC and SOC calculations over Set~B is 27.6\%; in comparison, the MARE is 10.5\% for electrons. Furthermore, several compounds such as \ce{CdI2}, BiTeCl, CuI, \ce{MoS2}, and \ce{Bi2TeSe2} show anomalously large reduction of the hole mass upon including SOC, in the range $2\times$ to $4\times$ (Table~\ref{tableS2}). From this dataset, we retain compounds exhibiting both electron and hole effective masses including SOC below $1\,m_{\rm e}$. Upon application of this filter, we obtain 50 compounds which we denote as ``Set~C'' (compounds marked in bold font in Table~\ref{tableS2}). We note that this dataset does not contain any magnetic compounds.

\begin{figure}
\begin{center}
\includegraphics[width=0.9\columnwidth]{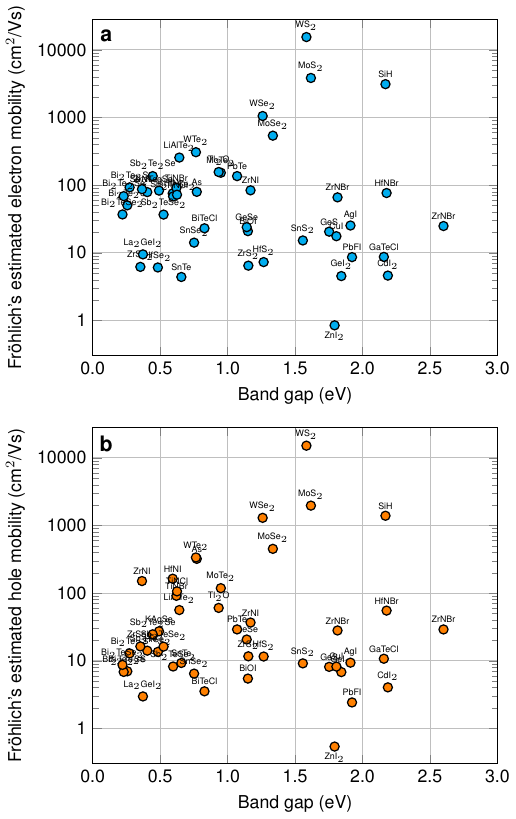}
\caption{\textbf{Estimates of carrier mobilities using the 2D Fr\"ohlich model}. (a) Electron mobilities of 48 compounds in Set~C, estimated using the polar phonon scattering model given by Eqs.~\eqref{eq.frohlichmodel}-\eqref{eq.taufr} of the Methods, at room temperature. The dataset contains 50 compounds, but antimonene and bismuthene are excluded as they are non-polar. (b) Estimated hole mobilities for the same dataset. The 2D Fr\"ohlich model yields unrealistically high mobilities. The parameters used to generate these data are given in Table~\ref{tableS3}.\vspace{-10pt}}
\label{fig.frohlich}
  \end{center}
\end{figure}

For compounds in Set~C, we estimate the carrier mobilities using simple analytical formulas. We start from a 2D Fr\"ohlich model, which assumes scattering of carriers by polar phonons in 2D, and corresponds to Eqs.~\eqref{eq.frohlichmodel}-\eqref{eq.taufr} of the Methods. The parameters necessary for this model are the highest phonon frequency, the high-frequency and static dielectric constants of the 2D materials, and the layer thickness. These quantities are computed by performing $\Gamma$-point phonon calculations, and by extracting the thickness and dielectric constants using the same procedure as in Ref.~\citenum{Sohier2016}. All calculated parameters are reported in full in Table~\ref{tableS3}. Figure~\ref{fig.frohlich} shows that the room-temperature mobilities estimated with this model extend up to unrealistically high values ($>$10,000~\mob) for well-known transition metal dichalcogenides (TMDs). Therefore this simple model is insufficient to describe carrier lifetimes and mobilities in these materials.

\begin{figure}
\begin{center}
\includegraphics[width=0.9\columnwidth]{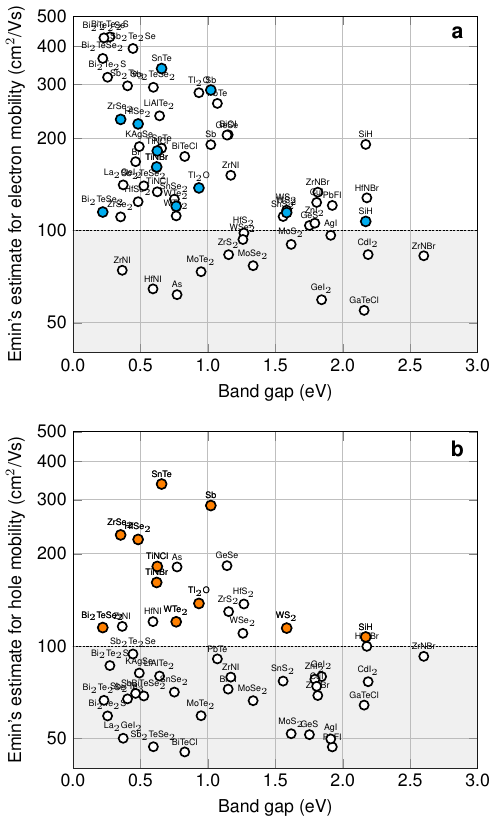}
\caption{\textbf{Estimates of carrier mobilities using Emin's formula}. (a) Electron mobilities of 50 compounds in Set~C, stimated using Emin's formula [Eq.~\eqref{eq.emin} of the Methods], at room temperature. (b) Estimated hole mobilities for the same dataset. These estimates fall within the expected range for TMDs, lending support to this simple model. The shaded regions mark the mobility threshold for identifying potential high-mobility candidates. 2D materials with both electron and hole mobility $> 100$\mob\ are highlighted in color, and constitute Set~D. The effective masses used to generate these data are given in Table~\ref{tableS3}.\vspace{-10pt}}
\label{fig.emin}
  \end{center}
\end{figure}

\begin{figure*}
\begin{center}
\includegraphics[width=0.8\linewidth]{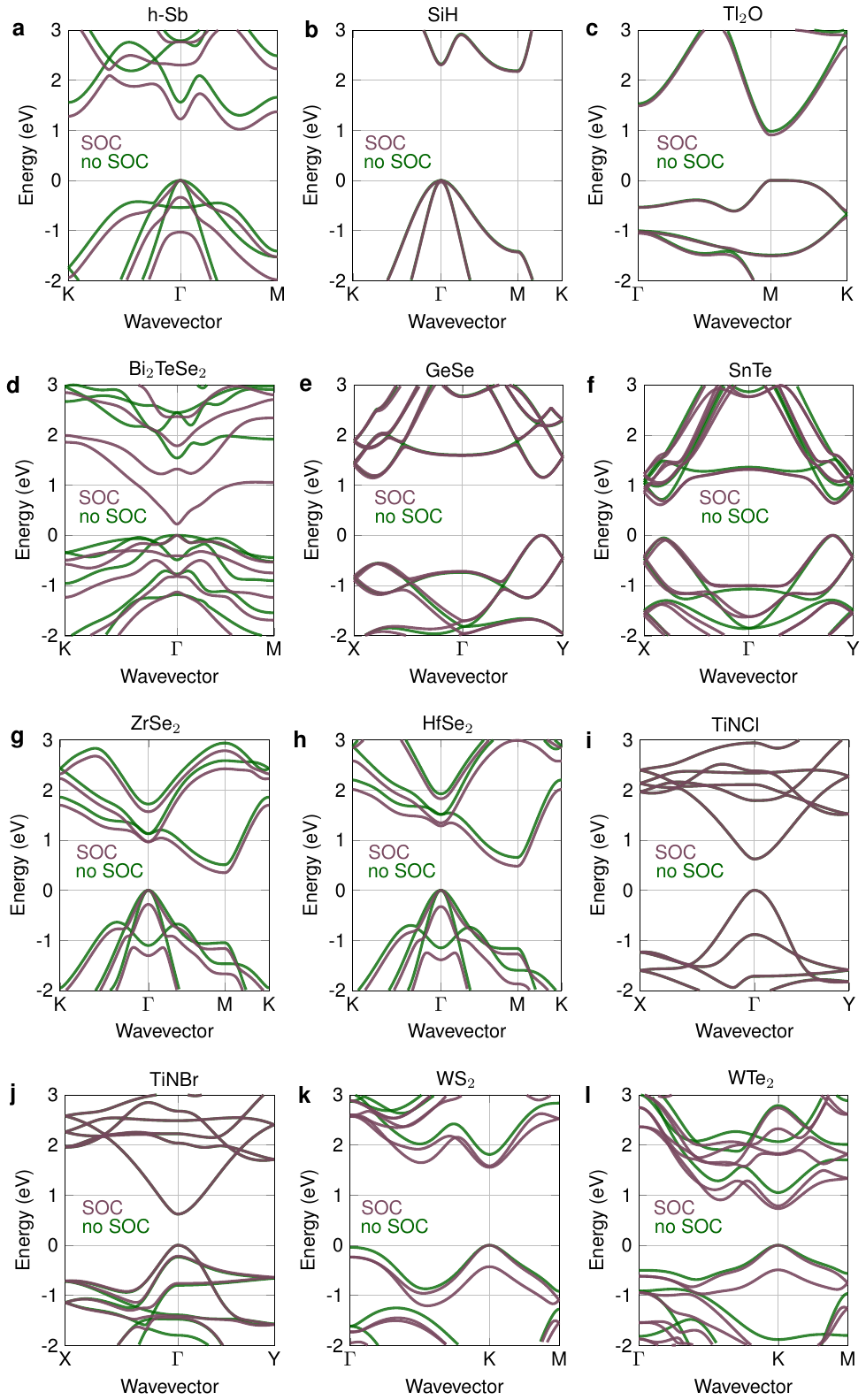}
\end{center}
\vspace{-10pt}
\caption{\textbf{Band structures of 2D materials in Set~D}. We compare DFT calculations performed without SOC (green) and with SOC (pink). We confirmed the fine structure of the bands of h-Sb and \ce{Bi2TeSe2} near the band edges via separate calculations using the HSE hybrid functional~\cite{Heyd2003, Heyd2006}, see Figure~\ref{hse.bs.Sb-Bi-BTS}(a) and (b).
}
\label{bandstr_wo_wSOC}
\end{figure*}

As an alternative model to estimate mobilities, we employ the simple Emin formula shown in Eq.~\eqref{eq.emin} of the Methods. Estimates based on this model are repored in Table~\ref{tableS3}. Figure~\ref{fig.emin} shows that room-temperature mobilities estimated with the Emin model for TMDs fall in the expected range~\cite{Liu2021, Su2021}. Electron mobilities are generally higher than hole mobilities, reflecting the corresponding effective masses shown in Fig.~\ref{fig.masses.soc}. We also note that, for this subset of compounds, higher mobilities are found for materials with narrower band gaps, in line with expectations. Based on these observations, we use Fig.~\ref{fig.emin} as a zero-th order approximation to the mobility of Set~C. We can now down-select compounds with \textit{both} the electron and the hole mobility above the threshold $\mu_{\rm th}=100$~\mob. This threshold is chosen so that 2D materials will be competitive with ultrathin-body Si MOSFETs scaled down to a thickness of 2.5~nm, which exhibit mobilities near $\mu_{\rm th}$~\cite{Uchida2002}. Compounds that fulfill this criterion are highlighted in color in Fig.~\ref{fig.emin}. After this selection, we are left with 12 candidates 2D materials with potential for balanced ambipolar transport, namely: Sb, GeSe, SnTe, \ce{ZrSe2}, \ce{HfSe2}, \ce{WS2}, \ce{WTe2}, SiH, \ce{Tl2O}, \ce{Bi2TeSe2}, TiNCl, and TiNBr. We note that skippenite \ce{Bi2TeSe2} is the only candidate identified in this work that overlaps with the selection of Ref.~\citenum{Sohier2020}. The difference between our lists originates from our choice of including SOC in the screening of band effective masses. These compounds constitute Set~D. The band structures of these compounds, both with and without SOC, are reported in Figure~\ref{bandstr_wo_wSOC}. For these materials candidates, we perform full-blown \textit{ai}BTE calculations. For comparative purposes, we also perform \textit{ai}BTE calculations for monolayer MoS$_2$, Bi, GeS, and WSe$_2$, as discussed below.
%

%
%
\vspace{20pt}
\noindent\textbf{Carrier mobilities via the \textit{ab initio} Boltzmann transport equation}
\vspace{5pt}

In this section, we first report on calculations for MoS$_2$, for which there are many experimental data available. This comparison serves as a validation test and allows us to gauge the predictive accuracy of our our methodology. Then, we discuss in turn our \aibte\ calculations for all materials in Set~D. SOC is included in all calculations.

\begin{figure*}
\begin{center}
\includegraphics[width=0.95\textwidth]{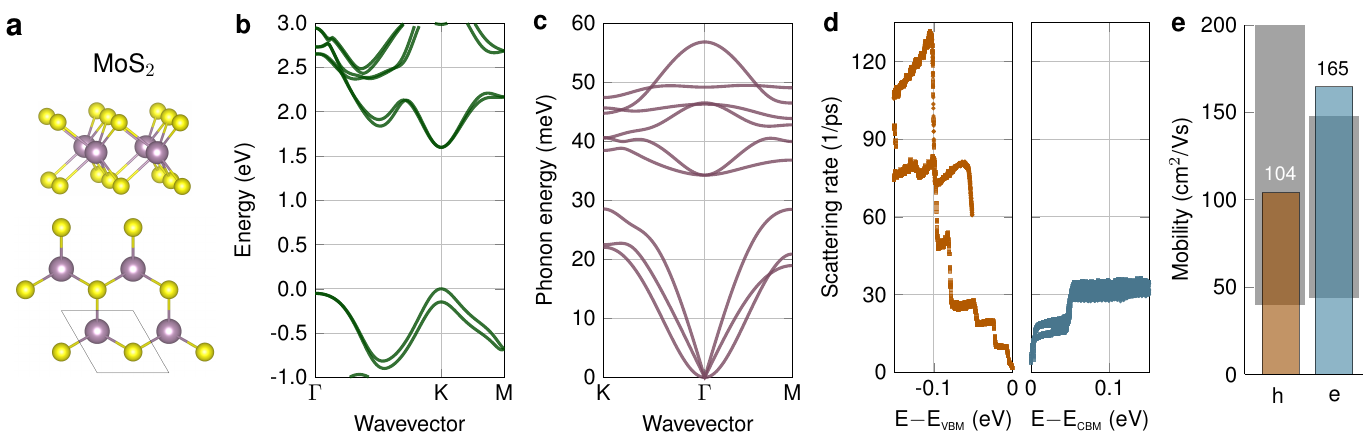}
\caption{\label{MoS2_mobility}\textbf{Carrier mobility of monolayer \ce{MoS2}}. (a) Side view and top view of a ball-stick model of \ce{MoS2}. Mo is in pink, S is in yellow. (b) DFT electronic band structure. (c) Phonon dispersion relations. (d) Carrier scattering rates from electron-phonon coupling. The left panel shows the scattering rates of holes (orange), the right panel is for electrons (blue). (e) BTE phonon-limited hole mobility (orange) and electron mobility (blue) of \ce{MoS2}. Scattering rates and mobilities are evaluated at 300~K for a carrier concentration of 10$^{10}$~cm$^{-2}$. The gray bars in (e) indicate the range of experimental values reported for monolayer \ce{MoS2}, from Refs.~\onlinecite{Zhang_2012, Lembke_2015, Yu_2015, Cui_2015, Chuang2016, Huo_2018}.\vspace{-10pt}}
\end{center}
\end{figure*}

\begin{figure*}
\begin{center}
\includegraphics[width=\linewidth]{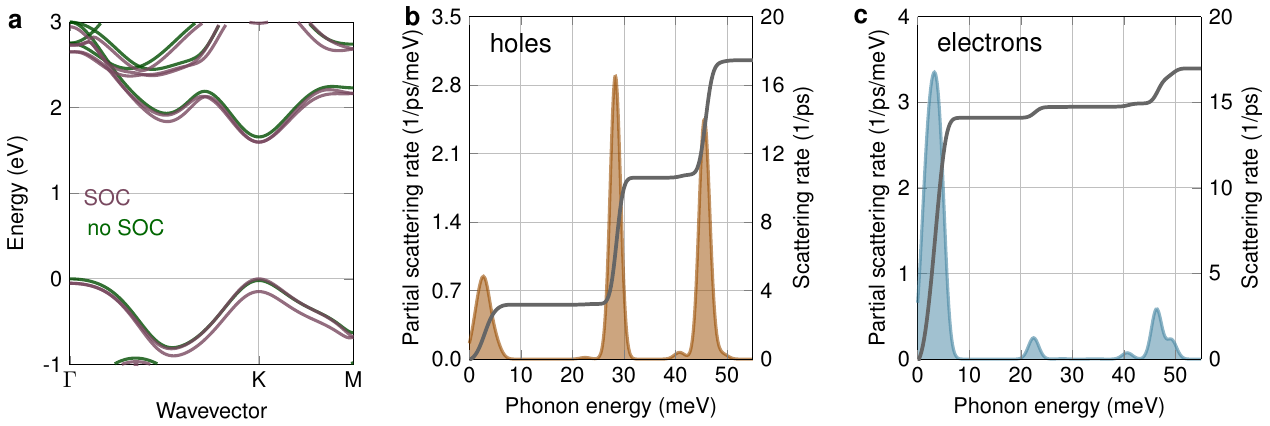}
\end{center}
\caption{\textbf{SOC effect in \ce{MoS2} and analysis of scattering rates}. (a) Comparison of the electron band structures of \ce{MoS2} without (green) and with (pink) SOC.
(b) Partial scattering rates of holes by phonons in \ce{MoS2}, resolved by phonon mode energy (orange line and filled area). The gray line is the cumulative scattering rate, i.e. the running integral of the blue curve, as reported on the right axis. (c) Same as (b), but for electrons in \ce{MoS2}. Rates are calculated as averages over carriers with energy $k_{\rm B} T$ from the band edges, with $T=300$\,K, and a carrier concentration of $10^{10}$\,cm$^{-2}$.}
\label{MoS2_bandstr}
\end{figure*}

\begin{figure*}
\begin{center}
\includegraphics[width=0.8\linewidth]{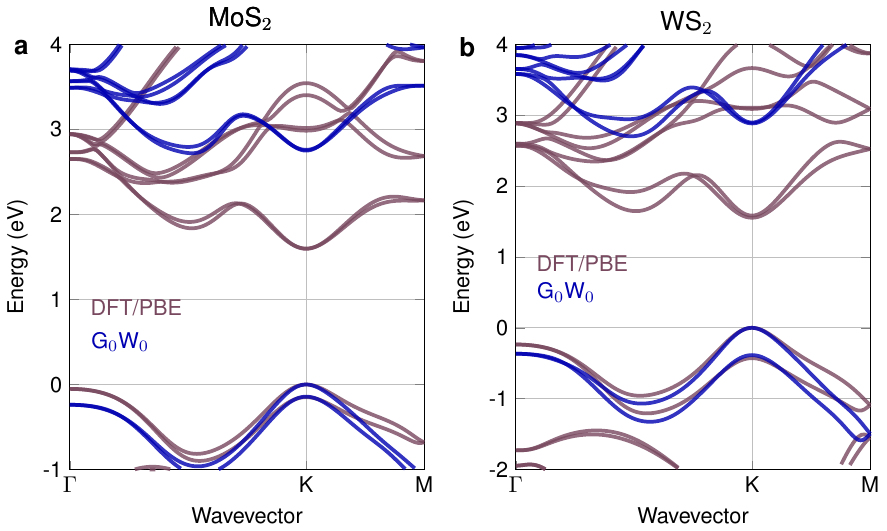}
\end{center}
\caption{\textbf{GW quasiparticle band structures}. (a) Comparison between the band structures of \ce{MoS2} calculated using DFT/PBE (red) and the G$_0$W$_0$ method (blue). In the GW band structure, the $\Gamma$-valley is lowered by 240~meV, leading to a suppression of $\Gamma$-K inter-valley scattering. \textit{ai}BTE calculations including GW corrections yield a hole mobility of 323~\mob\ (at room temperature and 10$^{13}$\,cm$^{-2}$ carriers); this value constitute a $3\times$ enhancement over the DFT mobility reported in Fig.~7 of the main text, and is slightly higher than the measured record value of 200\,\mob~\cite{Chuang2016}. We expect that by including defect scattering our calculated mobility will match closely the experimental value. (b) Comparison between DFT/PBE and G$_0$W$_0$ band structures of \ce{WS2}. SOC is included in all calculations.}
\label{gw.bs.MoS2.WS2}
\end{figure*}

%
%
\textbf{Molybdenum disulfide}. Figure~\ref{MoS2_mobility} shows the structure, electronic band structure, phonon dispersions, scattering rate, and carrier mobilities of monolayer \ce{MoS2}. For a very low carrier concentration of 10$^{10}$~cm$^{-2}$, we obtain room-temperature hole and electron mobilities of 104~\mob\ and 165~\mob, respectively. These values are consistent with hole mobilities around 70~\mob\ and electron mobilities in the range 130-180~\mob\ reported in previous theoretical works~\cite{Cheng2020, Ponce2023b, Zhang2023b}. In addition, we note that our calculated electron mobility is significantly lower than the value of 410~\mob\ obtained in the first \textit{ab initio} calculation of carrier mobility in 2D \ce{MoS2}~\cite{Kaasbjerg2012}. This difference originates from the fact that in Ref.~\citenum{Kaasbjerg2012} SOC was not taken into account and various approximations for the electron-phonon matrix elements and the solution of the BTE were employed. Our results fall in the middle of the range of reported experimental mobilities for holes, from 40~\mob\ to 200~\mob~\cite{Zhang_2012,Lembke_2015,Chuang2016}; and slightly above the range of reported mobilities for electrons, from 44~\mob\ to 148~\mob~\cite{Zhang_2012, Lembke_2015, Yu_2015, Cui_2015, Huo_2018}. The comparatively lower mobility of holes with respect to electrons is primarily due to the heavier masses (0.87 $m_e$ for holes vs. 0.50 $m_e$ for electrons, Table~\ref{tableS2}), while the scattering rates are comparable up to 50~meV from the band edges [Fig.~\ref{MoS2_mobility}(d)]. SOC plays an important role in the hole mobility, since it splits the $K$-valleys by 148~meV and lowers the $\Gamma$ valley by 51~meV [Figure~\ref{MoS2_bandstr} (a)]; these values are comparable to the highest phonon energy [Fig.~\ref{MoS2_mobility}(c)], therefore intervalley scattering is partly suppressed by SOC. A spectral analysis of the scattering rates shows that acoustic phonons are primarily responsible electron scattering (90\%) while optical optical phonons dominate hole scattering (82\%), Figure~\ref{MoS2_bandstr} 2(b) and (c). Overall, our calculations are in very good agreement with existing experimental and theoretical values, thus showing the reliability of the present methodology.
Further improvements on our calculations can be obtained by using GW band structures, as discussed in Figure~\ref{gw.bs.MoS2.WS2}.

\begin{figure*}
\begin{center}
\includegraphics[width=0.95\textwidth]{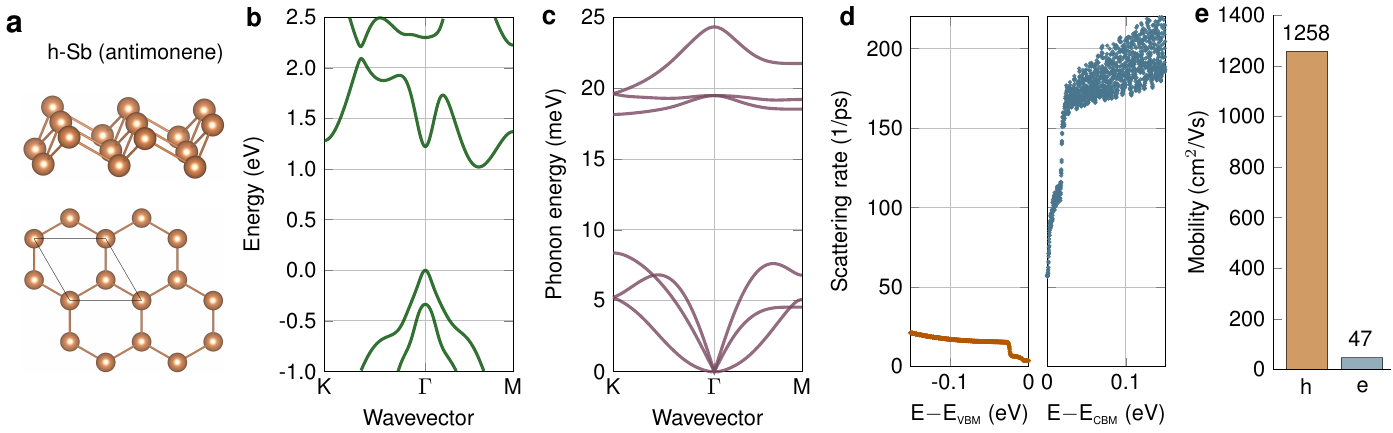}\vspace{-5pt}
\caption{\label{Sb_mobility}\textbf{Carrier mobility of antimonene}. (a) Side and top views of antimonene. (b) DFT electronic band structure. (c) Phonon dispersion relations. (d) Carrier-phonon scattering rates of holes (orange) and electrons (blue). (e) Phonon-limited carrier mobility of antimonene from the \textit{ai}BTE solution: holes (orange) and electrons (blue). Scattering rates and mobilities are evaluated at 300~K and a carrier density of 10$^{10}$~cm$^{-2}$. The confirmation of electronic band structure near band edges using the HSE hybrid functional~\cite{Heyd2003, Heyd2006} is provided in Figure~\ref{hse.bs.Sb-Bi-BTS}(a).\vspace{-10pt}}
\end{center}
\end{figure*}

\begin{figure*}
\begin{center}
\includegraphics[width=0.8\linewidth]{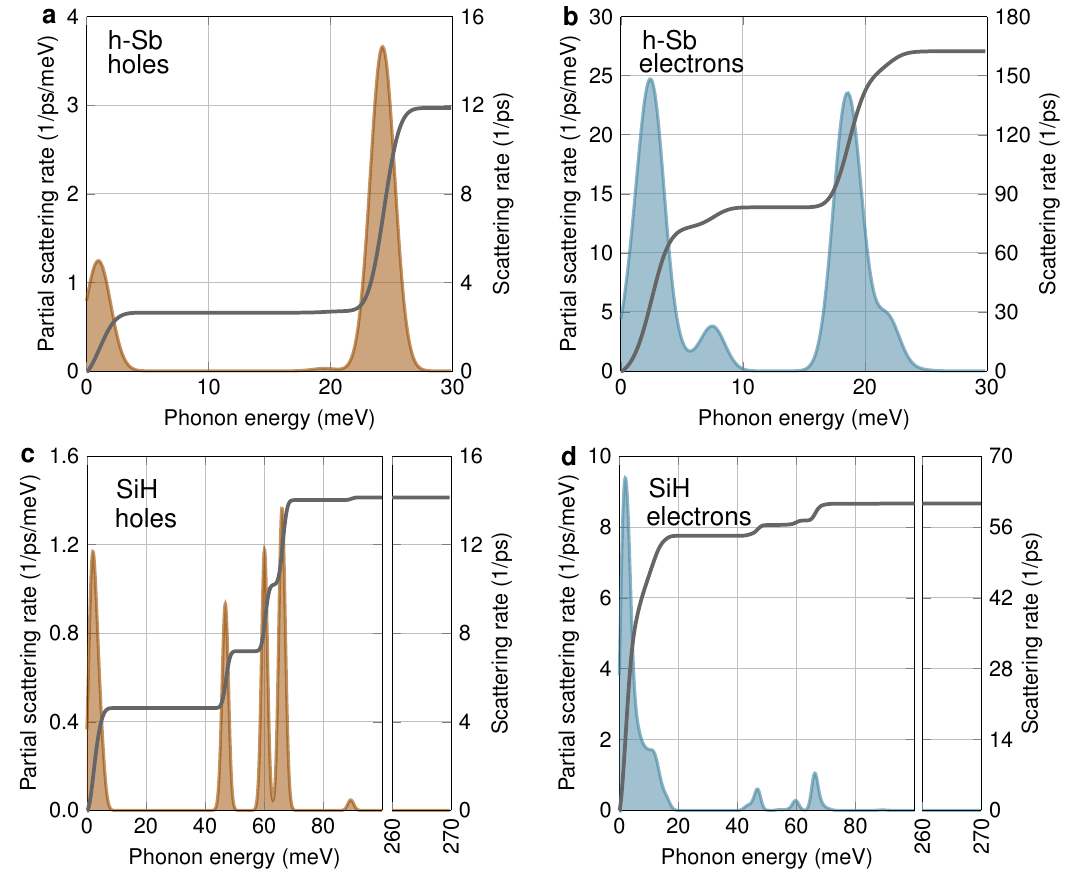}
\end{center}
\caption{\textbf{Analysis of scattering rates for h-Sb and SiH}. (a) Partial scattering rates of holes by phonons in h-Sb, resolved by phonon mode energy (orange line and filled area). The gray line is the cumulative scattering rate, i.e. the running integral of the blue curve, as reported on the right axis. (b) Partial scattering rates of electrons by phonons in h-Sb (blue line and filled area). The gray line is the cumulative rate (right axis). (c) and (d): Same format as for (a) and (b), but for holes and electrons in SiH, respectively. All rates are calculated as averages over carriers with energy $k_{\rm B} T$ from the band edges, with $T=300$\,K, for a carrier concentration of $10^{10}$\,cm$^{-2}$.}
\label{spectral_decomp_SbSiH}
\end{figure*}

%
%
\textbf{Antimonene}. Antimonene, or monolayer h-Sb, was predicted to exist in both a tetragonal $\alpha$ phase and an hexagonal $\beta$ phase~\cite{Wang2015}. Both phases have successfully been synthesized~\cite{Xue2021}. The structure of the hexagonal phase, which is identified by our screening as a potential high-mobility material, is shown in \figref{Sb_mobility}(a). This phase was successfully synthesized in few-layer as well as monolayer form~\cite{Ji2016, FortinDeschnes2017, MartnezPerin2018, Xiao2020, Ares2021, Gupta2021}, and has been investigated for diverse applications ranging from optoelectronics to energy storage~\cite{Wang2018, Zhong2022, Carrasco2023}. To the best of our knowledge, experimental mobility data on this compound have not been reported yet. Our calculated room-temperature phonon-limited mobilities are 1258~\mob\ and 47~\mob\ for holes and electrons, respectively [Fig.~\ref{Sb_mobility}(e)]. Both hole and electron effective masses are very low (0.16\,$m_e$ and 0.23\,$m_e$, respectively, Table~\ref{tableS2}); the hole scattering rates are unusually low, in the range of 10\,ps$^{-1}$ [Fig.~\ref{Sb_mobility}(d)], contributing to an exceptionally high hole mobility. The low hole-phonon scattering rates can be attributed to the SOC splitting the valence band maximum at $\Gamma$ by more than 300~meV, causing a suppression of intravalley interband scattering. In contrast, intervalley scattering between the six degenerate conduction band minima located half-way along the $\Gamma M$ line is allowed, leading to poor electron mobility. Spectral analysis of the scattering rates [Figure~\ref{spectral_decomp_SbSiH} (a) and (b)] reveals that holes are primarily scattered by the highest-energy optical phonons [78\%, Fig.~\ref{Sb_mobility}(c)], while both acoustic and optical phonons contribute equally to electron scattering. We note that h-Sb was already recognized as a potential high-mobility 2D materials in a previous computational study~\cite{Cheng2019, Zhang2023b}, which reported a hole mobility of 1100-1330~\mob, in good agreement with our value.

\begin{figure*}
\begin{center}
\includegraphics[width=0.95\linewidth]{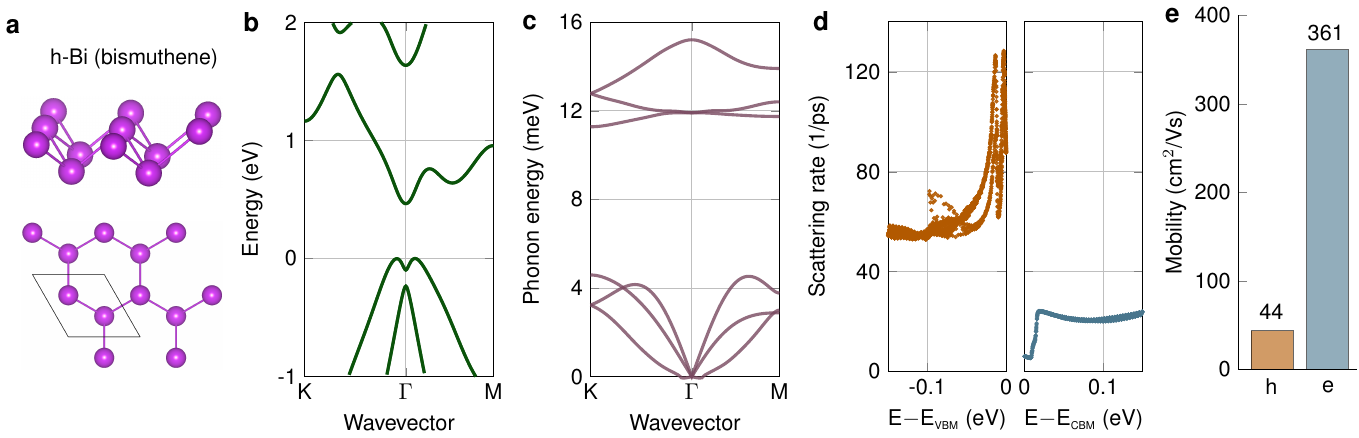}
\end{center}
\vspace{-10pt}
\caption{\label{Bi_mobility}\textbf{Carrier mobility of bismuthene}. (a) Side and top views of bismuthene. (b) DFT electronic band structure. (c) Phonon dispersion relations. (d) Carrier-phonon scattering rates of holes (orange) and electrons (blue). (e) Phonon-limited carrier mobility of bismuthene from the \textit{ai}BTE solution: holes (orange) and electrons (blue). Scattering rates and mobilities are evaluated at 300~K and a carrier density of 10$^{10}$~cm$^{-2}$. The shape of the top of the valence bands is reminiscent of the Rashba-Dresselhaus effect and is due to SOC, however there is no spin-splitting since the system is inversion-symmetric. A comparison between band structure with SOC and without SOC is shown in Figure~\ref{bandstr_wo_wSOC_3}(a). The ring-like structure of the valence band causes a 1D-like density of states, which is reflected in the scattering rates [panel (d), left] and significantly enhances intravalley scattering. Conversely, the conduction bands exhibits a standard 2D-like density of states, which is reflected in the step-like scattering rates [panel (d), right]. Our calculated mobilities are consistent with prior theoretical work~\cite{Cheng2019} reporting mobilities of 47~\mob\ for holes and 348~\mob\ for electrons. Bismuthene has been synthesized~\cite{Yang2020, Liu2020d, Zhao2022}, and few-layer films have been used as transistor channels, with reported mobilities between 220~\mob\ and 322~\mob~\cite{Yang2019, Zhong2020, Sun2021}, in line with our calculations.}
\end{figure*}

\begin{figure*}
\begin{center}
\includegraphics[width=0.9\linewidth]{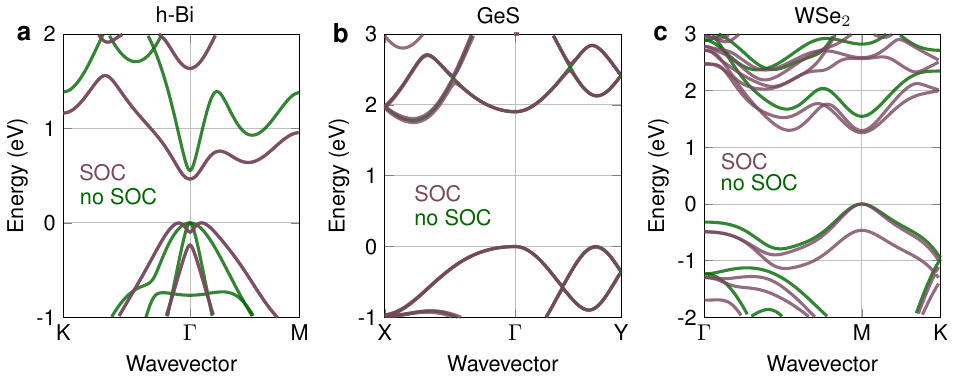}
\end{center}
\vspace{-10pt}
\caption{\textbf{Band structures of additional 2D materials not contained in Set~D}. DFT band structures of three additional 2D materials that are closely related to compounds in  Set~D: h-Bi, \ce{GeSe}, and \ce{WSe2}. We compare calculations performed without SOC (green) and with SOC (pink). For h-Bi, we performed additional calculations using the HSE hybrid functional~\cite{Heyd2003, Heyd2006} and confirmed the fine structure near the band edges, as seen in Figure~\ref{hse.bs.Sb-Bi-BTS}(c).
}
\label{bandstr_wo_wSOC_3}
\end{figure*}

A related compound, bismuthene h-Bi, was filtered out by Emin's estimate for the hole mobility [\figref{fig.emin}(b)], and is discussed for completeness in Figures~\ref{Bi_mobility},~\ref{bandstr_wo_wSOC_3}, and~\ref{spectral_decomp_WS2.WTe2}.

\begin{figure*}
\begin{center}
\includegraphics[width=0.95\textwidth]{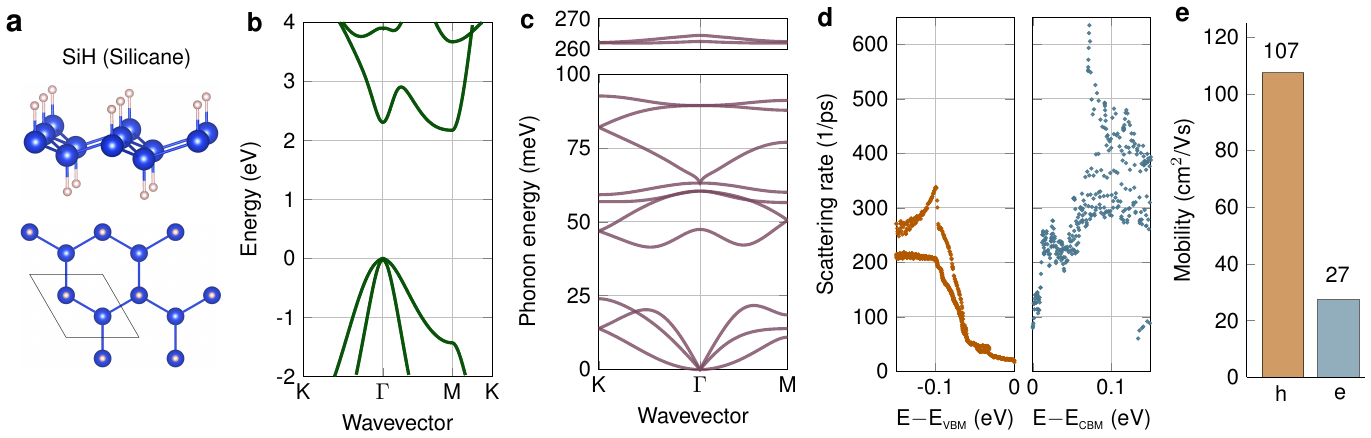}\vspace{-5pt}
\caption{\label{SiH_mobility}\textbf{Carrier mobility of silicane}. (a) Side and top views of silicane. Si is blue, H is white. (b) DFT electronic band structure. (c) Phonon dispersion relations. (d) Carrier-phonon scattering rates of holes (orange) and electrons (blue). (e) Phonon-limited carrier mobility of silicane from the \textit{ai}BTE: holes (orange) and electrons (blue), evaluated at 300~K and a carrier density of 10$^{10}$~cm$^{-2}$.\vspace{-10pt}}
\end{center}
\end{figure*}

%
%
\textbf{Silicane}. Silicane (SiH) is a hydrated monolayer of silicon, as shown in \figref{SiH_mobility}(a). This compound has been investigated extensively from the computational standpoint~\cite{LewYanVoon2010, Houssa2011, Khatami2019, Wani2022, Nguyen2022}, but only partially-hydrogenated layers~\cite{Qiu2015_1, Qiu2015_2, Wang2016, Wei2017} and \ce{Ge_{0.5}Si_{0.5}H} alloys~\cite{Roy2023} have been synthesized. In silicane, Si atoms maintain a tetrahedral coordination as in bulk Si, but two out of four neighbors are H atoms. As all bonds are saturated, this compound exhibits a band gap~\cite{Zhao2016}, as seen in \figref{SiH_mobility}(b). The gap is indirect, between the VBM at $\Gamma$ and the CBM at $M$. Figure~\ref{SiH_mobility}(c) shows the phonon dispersion relations, which consist of three sets of modes: Si-Si vibrations in the range up to 60~meV, H wagging modes between 60-100~meV, and Si-H stretching modes near 260~meV with dominant H displacements, as expected from the mass difference. These latter modes do not contribute to the scattering of carriers [Figure~\ref{spectral_decomp_SbSiH}(c) and (d)]. The predominant scattering channel is from acoustic phonons for electrons (89\%) and from optical phonons for holes (67\%). Our calculated room-temperature mobilities (for $10^{10}$~carriers/cm$^{-2}$) are 107~\mob\ for holes, and 27~\mob\ for electrons [\figref{SiH_mobility}(d)]. These values are in good agreement with previous calculations which reported 109~\mob\ and 53~\mob\ for holes and electrons, respectively~\cite{Khatami2019}. Given the relatively low phonon-limited theoretical mobilities, silicane does not constitute a promising 2D semiconductor.

\begin{figure*}
\begin{center}
\includegraphics[width=0.95\textwidth]{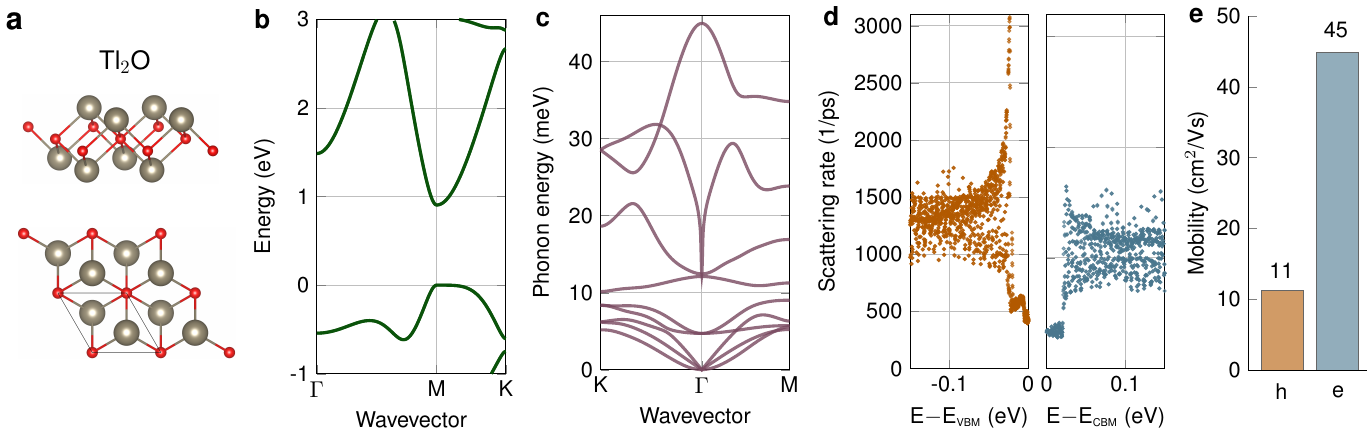}\vspace{-5pt}
\caption{\label{Tl2O_mobility}\textbf{Carrier mobility of monolayer thallium oxide}. (a) Side and top views of monolayer \ce{Tl2O}. Tl is in gold, O is in red. (b) DFT electronic band structure. (c) Phonon dispersion relations. (d) Carrier-phonon scattering rates of holes (orange) and electrons (blue). (e) Phonon-limited carrier mobility of monolayer \ce{Tl2O} from the \textit{ai}BTE: holes (orange) and electrons (blue), evaluated at 300~K and a carrier density of 10$^{10}$~cm$^{-2}$.\vspace{-10pt}}
\end{center}
\end{figure*}

\begin{figure*}
\begin{center}
\includegraphics[width=0.8\linewidth]{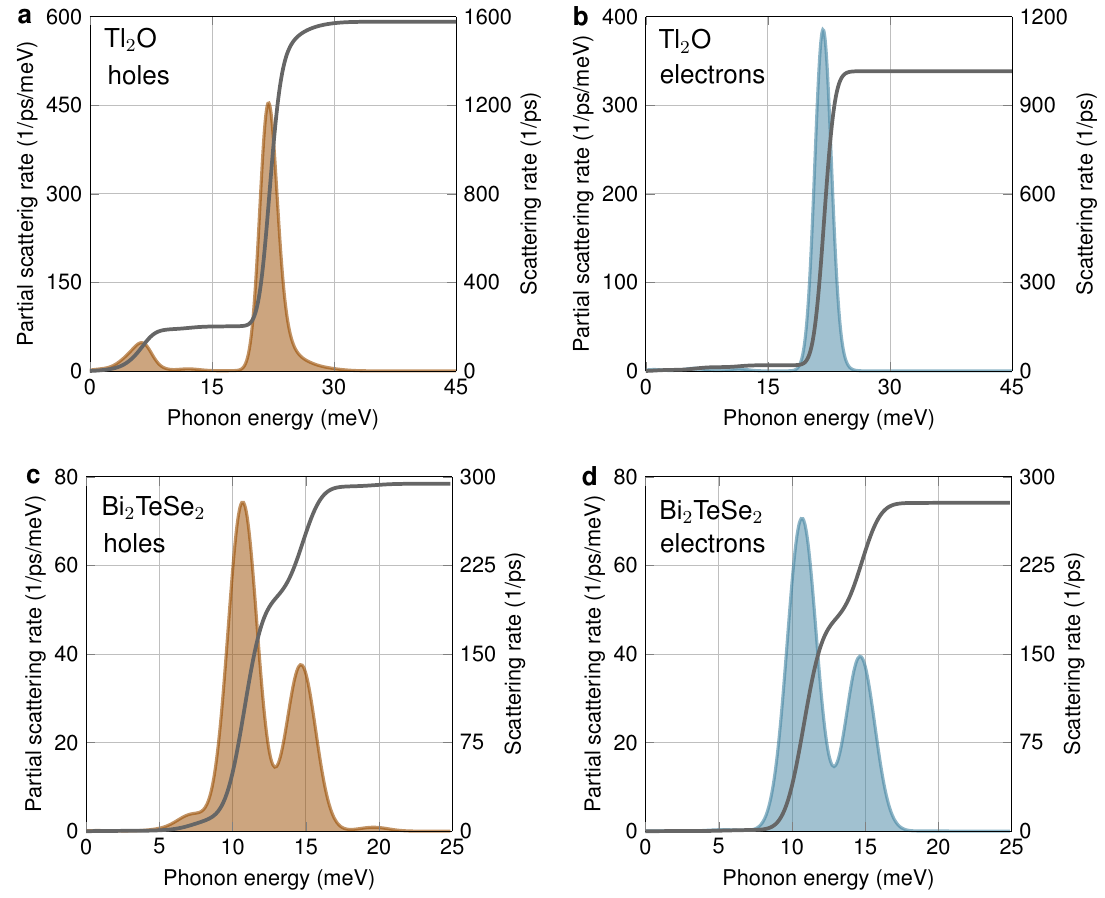}
\end{center}
\caption{\textbf{Analysis of scattering rates for \ce{Tl2O} and \ce{Bi2TeSe2}}. (a) Partial scattering rates of holes by phonons in \ce{Tl2O}, resolved by phonon mode energy (orange line and filled area). The gray line is the cumulative scattering rate, i.e. the running integral of the blue curve, as reported on the right axis. (b) Partial scattering rates of electrons by phonons in \ce{Tl2O} (blue line and filled area). The gray line is the cumulative rate (right axis). (c) and (d): Same format as for (a) and (b), but for holes and electrons in \ce{Bi2TeSe2}, respectively. All rates are calculated as averages over carriers with energy $k_{\rm B} T$ from the band edges, with $T=300$\,K, for a carrier concentration of $10^{10}$\,cm$^{-2}$.}
\label{spectral_decomp_TOBTS}
\end{figure*}

%
%
\textbf{Thallium monoxide}. \ce{Tl2O} represents the ﬁrst 2D semiconductor that was predicted to be stable with metal cations located on the outer layers (hence dubbed ``metal-shrouded'')~\cite{Ma2017}. The structure is the same as that of 1T-\ce{MoS2}, with O and Tl replacing Mo and S, respectively [\figref{Tl2O_mobility}(a)]. It has been investigated intensively from the theoretical standpoint, to assess its potential for applications in thermoelectrics and spintronics~\cite{Ma2018, Huang2019, Sajjad2019, Xu2019, Gao2019, Wang2019, Pan2020, Huang2020, Li2020, Yan2021}. However, we are unaware of any synthesis attempts, which may be due to the toxicity of thallium. Figure~\ref{Tl2O_mobility}(b) shows that \ce{Tl2O} exhibits a direct band gap at $M$, and that electrons are lighter than holes (Table~\ref{tableS2}). The main source of carrier scattering are the optical phonons near 21~meV [Figure~\ref{spectral_decomp_TOBTS}(a) and (b)], which correspond to in-plane longitudinal-optical (LO) vibrations of the O sublattice. Our calculated mobilities at room temperature are 11~\mob\ for holes and 45~\mob\ for electrons. Based on these \textit{ai}BTE data, our assessment is that \ce{Tl2O} does not constitute a promising 2D semiconductor. We also note that prior work predicted very high carrier mobilities for this compound, in the order of 4,000~\mob\ at room temperature~\cite{Ma2017}; this prior work was based on the acoustic deformation potential model, which misses the dominant contribution from optical phonons to the scattering rates and leads to a significant overestimation of the mobility. 

\begin{figure*}
\begin{center}
\includegraphics[width=0.95\textwidth]{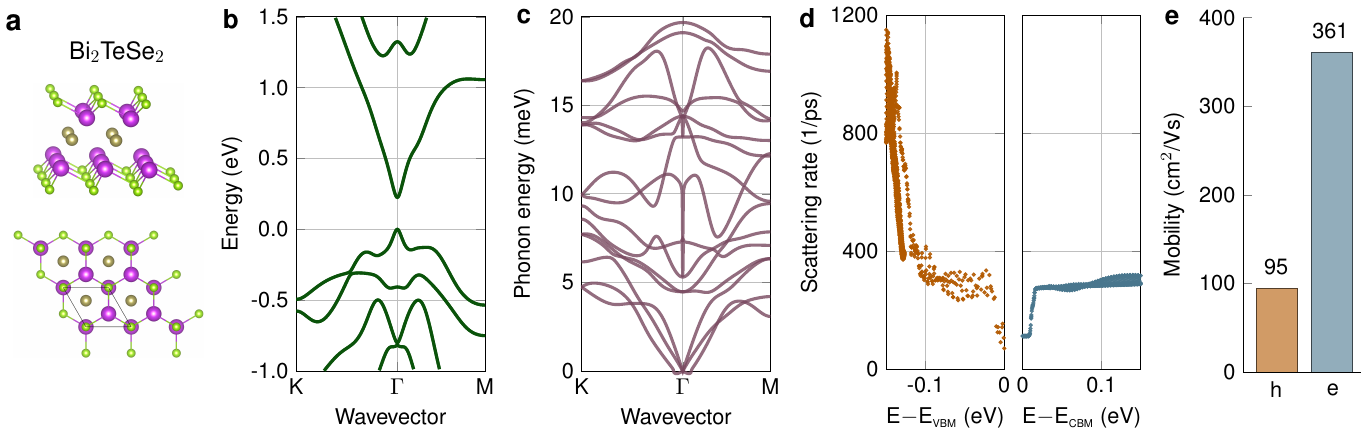}\vspace{-5pt}
\caption{\label{Bi2TeSe2_mobility}\textbf{Carrier mobility of monolayer skippenite}. (a) Side and top views of monolayer \ce{Bi2TeSe2}. Bi is pink, Se is green, and Te brown. (b) DFT electronic band structure. (c) Phonon dispersion relations. (d) Carrier-phonon scattering rates of holes (orange) and electrons (blue). (e) Phonon-limited carrier mobility of monolayer \ce{Bi2TeSe2} from the \textit{ai}BTE: holes (orange) and electrons (blue), evaluated at 300~K and for a carrier concentration of 10$^{10}$~cm$^{-2}$. The confirmation of electronic band structure near band edges using the HSE hybrid functional~\cite{Heyd2003, Heyd2006} is provided in Figure~\ref{hse.bs.Sb-Bi-BTS}(b).\vspace{-10pt}}
\end{center}
\end{figure*}

%
%
\textbf{Skippenite}. The monolayer form of skippenite, \ce{Bi2TeSe2}, consists of a single layer of Te sandwiched between two hexagonal BiSe bilayers, as shown in \figref{Bi2TeSe2_mobility}(a). This compound attracted interest for its predicted transport and thermoelectric properties~\cite{Lu2019, Liu2020b, Wang2022, Peng2024}, but only multi-layer films have been synthesized thus far~\cite{Yuan2015_mat, Wang2016_scirep}. Our calculations indicate that SOC is especially important in this system, as expected from the presence of Bi. For example, upon including SOC the band gap decreases from 1.5~eV to 0.2~eV, and the hole mass decreases concomitantly from 1.49\,$m_e$ to 0.39\,$m_e$ [Figure~\ref{bandstr_wo_wSOC}(d)].  
Both holes and electrons are primarily scattered by two optical phonons near 10~meV and 15~meV, which correspond to the in-plane LO vibrations of the Te and Se layers, respectively [Figure~\ref{spectral_decomp_TOBTS}(c) and (d)]. Our predicted hole and electron mobilities are 95~\mob\ and 361~\mob\, respectively. These values are significantly lower than previously reported theoretical mobilities in the range 754-1329~\mob\ for holes and 1983-2510~\mob\ for electrons~\cite{Lu2019, Liu2020b, Peng2024}. We ascribe this discrepancy to the fact that previous work employed the acoustic deformation potential model, which neglects the key role of optical phonons in this compound. Based on our calculations, $n$-type \ce{Bi2TeSe2} should be a promising 2D semiconductor. 

\begin{figure*}
\begin{center}
\includegraphics[width=0.95\textwidth]{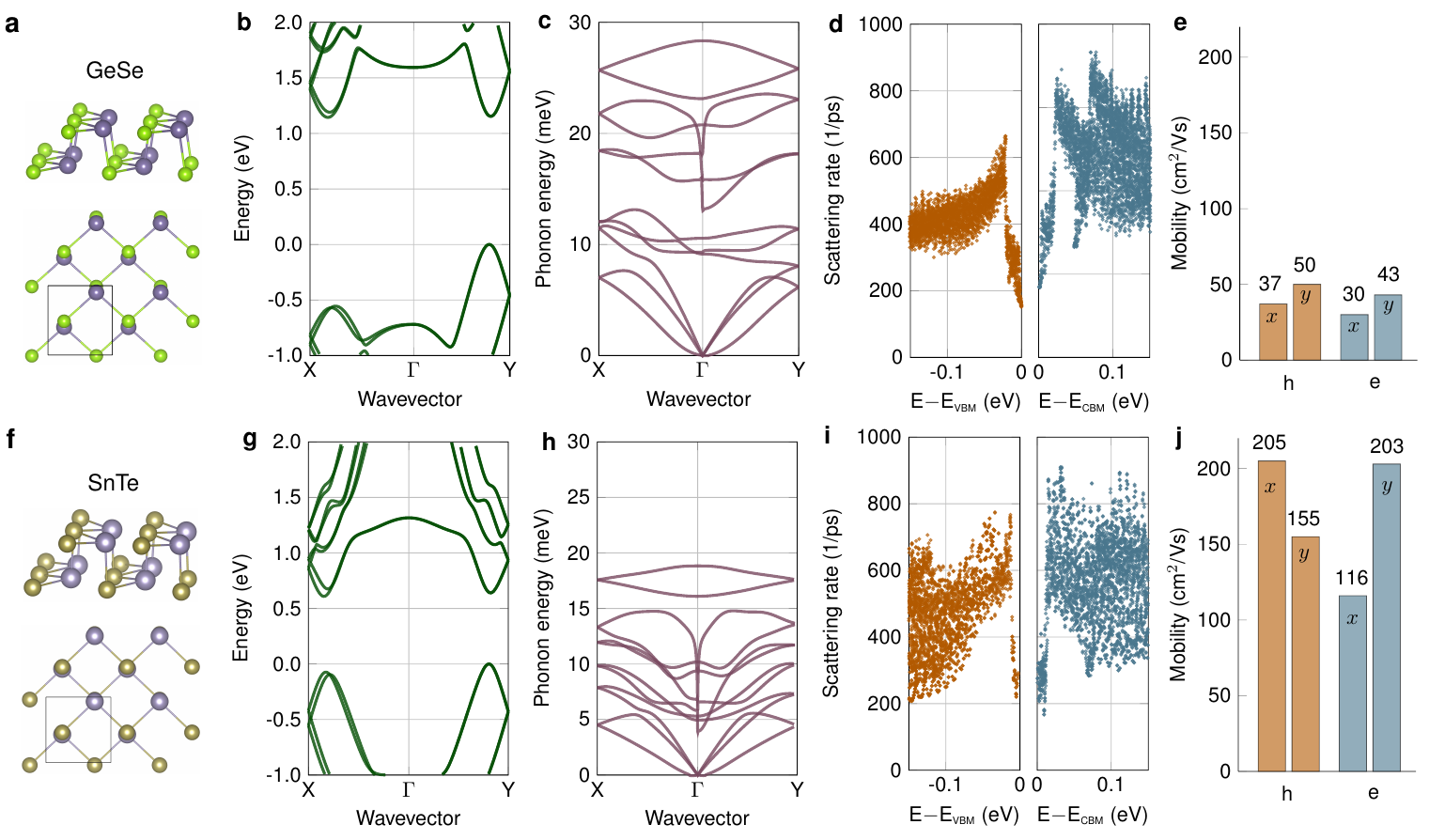}\vspace{-5pt}
\vspace{-5pt}
\caption{\label{GeSe_SnTe_mobility}\textbf{Carrier mobilities of GeSe and SnTe monolayers}. 
(a) Side and top views of monolayer GeSe. Ge is purple, Se is green. (b) DFT band structure. (c) Phonon dispersion relations. (d) Carrier-phonon scattering rates of holes (orange) and electrons (blue). (e) Phonon-limited carrier mobility of monolayer GeSe from the \textit{ai}BTE: holes (orange) and electrons (blue), evaluated at 300~K and for a carrier concentration of 10$^{10}$~cm$^{-2}$. $x$ and $y$ denote the armchair and zigzag directions, respectively. (f) Side and top views of monolayer SnTe. Sn is gray, Te is dark green. (g)-(j) Same as (b)-(e) but for SnTe.\vspace{-10pt}} 
\end{center}
\end{figure*}

\begin{figure*}
\begin{center}
\includegraphics[width=0.95\linewidth]{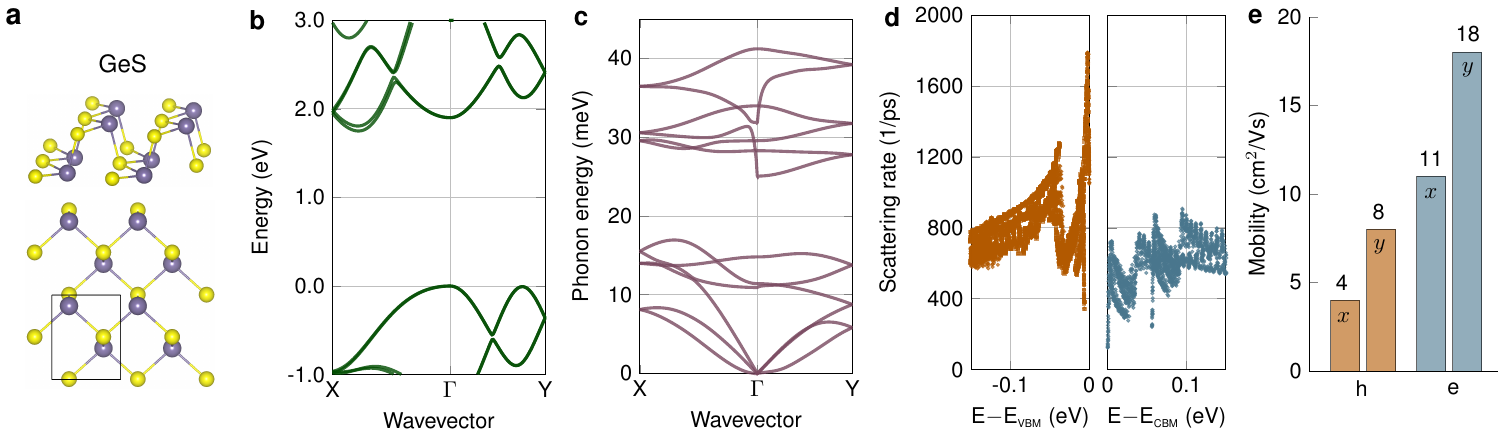}
\end{center}
\vspace{-10pt}
\caption{\label{GeS_mobility}\textbf{Carrier mobilities of GeS monolayer}. 
(a) Side and top views of monolayer GeS. Ge is purple, S is yellow. (b) DFT electronic band structure. (c) Phonon dispersion relations. (d) Carrier-phonon scattering rates of holes (orange) and electrons (blue). (e) Phonon-limited carrier mobility of monolayer GeS from the \textit{ai}BTE: holes (orange) and electrons (blue), evaluated at 300~K and for a carrier concentration of 10$^{10}$~cm$^{-2}$. The scattering rates are very high as a result of strong electron-phonon couplings and significant degree of valley degeneracy. Prior theoretical work predicted very high mobilities~\cite{Li2016, Cheng2021}, but neglected optical phonon scattering. When including optical phonons, values closer to our present data have been obtained~\cite{Yang2020}. Low mobility values, of the order of 0.001\,\mob, were measured for few-layer films~\cite{Ulaganathan2016}.}
\end{figure*}

%
%
\textbf{Germanium selenide and tin telluride}. These materials belong to the broader family of IV-VI 2D semiconductors. Material Set~C (Table~\ref{tableS3}) contains three candidates from this family: GeS, GeSe, and SnTe. Of these candidates, GeS is filtered out based on Emin's estimate for the hole mobility [\figref{fig.emin}(b)]; for completeness, we discuss its properties in Figures~\ref{GeS_mobility},~\ref{bandstr_wo_wSOC_3}, and~\ref{spectral_decomp_WS2.WTe2}. 3D parent compounds of these materials have been investigated extensively owing to potential applications as photodetectors and ferroelectrics~\cite{Hu2019}. In their bulk form, GeSe and SnTe crystallize in layered orthorhombic structures with Cmcm or Pnma space groups~\cite{Hu2019}. The 2D versions are obtained by exfoliation of the Pnma phase, and the resulting structures are shown in \figref{GeSe_SnTe_mobility}(a) and (f), respectively. Synthesis of few-layer and thin-film GeSe is well established~\cite{Mukherjee2013, Ramasamy2016, Wang2017, Ye2017, Zhou2018, Hu2019_b, Ma2019, Muhammad2022, Tan2022, Wang2023}, while monolayer crystals are far less common~\cite{Zhao2017}. Conversely, several studies reported the synthesis of monolayer SnTe~\cite{Chang2016, Chang2018, Chang2019, Chang2019b, Liu2020c, BarrazaLopez2021, Su2023}, which attracted considerable interest in the field of 2D ferroelectrics.

\begin{figure*}
\begin{center}
\includegraphics[width=0.8\linewidth]{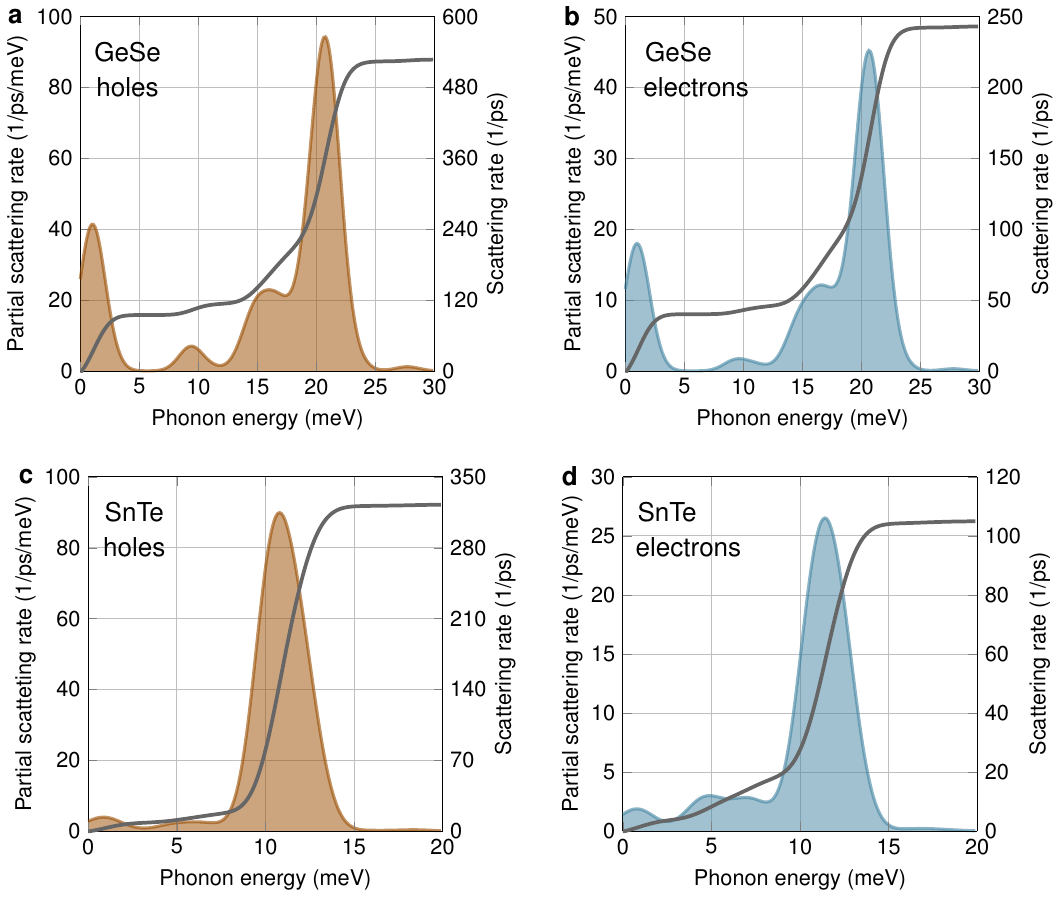}
\end{center}
\caption{\textbf{Analysis of scattering rates for GeSe and SnTe}. (a) Partial scattering rates of holes by phonons in GeSe, resolved by phonon mode energy (orange line and filled area). The gray line is the cumulative scattering rate, i.e. the running integral of the blue curve, as reported on the right axis. (b) Partial scattering rates of electrons by phonons in GeSe (blue line and filled area). The gray line is the cumulative rate (right axis). (c) and (d): Same format as for (a) and (b), but for holes and electrons in SnTe, respectively. All rates are calculated as averages over carriers with energy $k_{\rm B} T$ from the band edges, with $T=300$\,K, for a carrier concentration of $10^{10}$\,cm$^{-2}$.}
\label{spectral_decomp_GeSeSnTe}
\end{figure*}

Electron-phonon coupling in GeSe is dominated by LO stretching vibrations near 20~meV, leading to unusually high scattering rates near the band edges [Figure~\ref{spectral_decomp_GeSeSnTe}(a) and (b)]. Thus, despite the relatively light carrier masses (Table~\ref{tableI}), this compound exhibits an unimpressive room-temperature phonon-limited hole mobility of 44~\mob\ (isotropic average value) and electron mobility of 37~\mob, as shown in \figref{GeSe_SnTe_mobility}(e). Our results agree with recent computational studies which reported 45~\mob\ and 18~\mob, respectively~\cite{Yang2020b}. To the best of our knowledge, measured mobilities have been reported only for few-layer GeSe, and the values are very low, in the range 0.9-5~\mob\ \cite{Zhou2018, Zhao2019, Hussain2020, Mao2023}; we are unaware of mobility measurements for monolayer GeSe.

In the case of SnTe, SOC induces significant band band splitting both in the valence and in the conduction bands (70-80~meV), as seen in \figref{GeSe_SnTe_mobility}(g); since the highest phonon energy is 18~meV [\figref{GeSe_SnTe_mobility}(h)], SOC splitting suppresses inter-valley scattering by phonons. LO stretching phonons around 11~meV couple most strongly with both electrons and holes, leading once again to unusually high scattering rates as for GeSe [Figure~\ref{spectral_decomp_GeSeSnTe}(c) and (d)]. We calculate directionally-averaged room-temperature mobilities of 180~\mob\ for holes, and 160~\mob\ for electrons, in reasonable agreement with a very recent computational study which reported 220~\mob\ and 120~\mob, respectively~\cite{Zhang2023b}. We are unaware of experimentally-measured mobilities for monolayer SnTe. Overall, among IV-VI monolayer semiconductors, SnTe offers the highest promise as a 2D channel material owing to its moderately high ambipolar carrier mobilities at room temperature.

\begin{figure*}
\begin{center}
\includegraphics[width=0.95\textwidth]{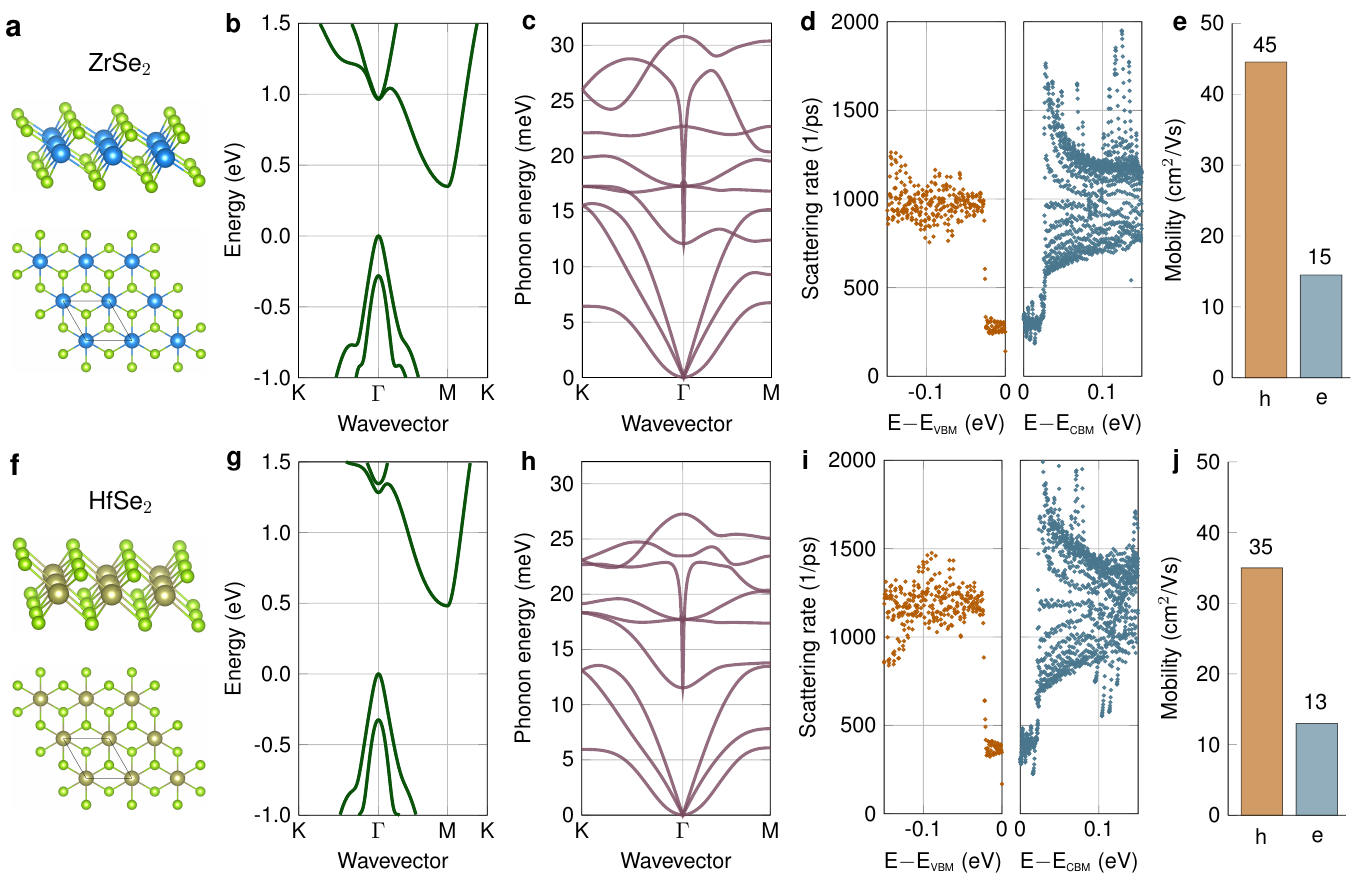}\vspace{-5pt}
\vspace{-5pt}
\caption{\label{ZrSe2_HfSe2_mobility}\textbf{Carrier mobilities of \ce{ZrSe2} and \ce{HfSe2} monolayers}. 
(a) Side and top views of monolayer \ce{ZrSe2}. Zr is blue, Se is green. (b) DFT band structure. (c) Phonon dispersion relations. (d) Carrier-phonon scattering rates of holes (orange) and electrons (blue). (e) Phonon-limited carrier mobility of monolayer \ce{ZrSe2} from the \textit{ai}BTE: holes (orange) and electrons (blue), evaluated at 300~K and for a carrier concentration of 10$^{10}$~cm$^{-2}$. (f) Ball-stick models of \ce{HfSe2}, with Hf and Se in brown and green, respectively. (g)-(j) Same as (b)-(e) but for \ce{HfSe2}.\vspace{-10pt}} 
\end{center}
\end{figure*}

\begin{figure*}
\begin{center}
\includegraphics[width=0.8\linewidth]{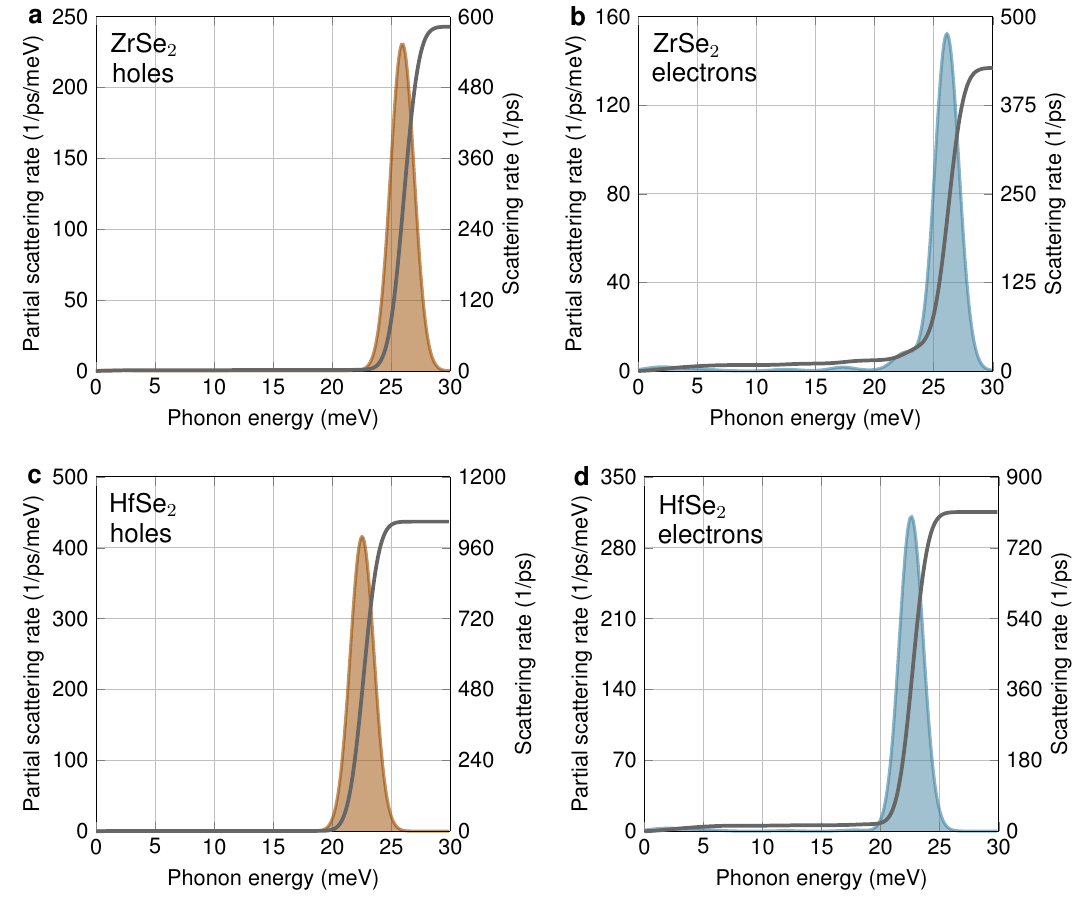}
\end{center}
\caption{\textbf{Analysis of scattering rates for \ce{ZrSe2} and \ce{HfSe2}}. (a) Partial scattering rates of holes by phonons in \ce{ZrSe2}, resolved by phonon mode energy (orange line and filled area). The gray line is the cumulative scattering rate, i.e. the running integral of the blue curve, as reported on the right axis. (b) Partial scattering rates of electrons by phonons in \ce{ZrSe2} (blue line and filled area). The gray line is the cumulative rate (right axis). (c) and (d): Same format as for (a) and (b), but for holes and electrons in \ce{HfSe2}, respectively. All rates are calculated as averages over carriers with energy $k_{\rm B} T$ from the band edges, with $T=300$\,K, for a carrier concentration of $10^{10}$\,cm$^{-2}$.}
\label{spectral_decomp_ZrSe2HfSe2}
\end{figure*}

%
%
\textbf{Zirconium and hafnium diselenides}. \ce{ZrSe2} and \ce{HfSe2} crystallize in the 1T-\ce{CdI2} structure, as seen in \figref{ZrSe2_HfSe2_mobility}(a) and (f). Both compounds have been synthesized as few-layer or nano-sheets (\ce{ZrSe2} \cite{MaasValero2016, Mleczko2017, Najafi2022,Tian2021}, \ce{HfSe2}~\cite{Yin2016, Mleczko2017, Tsai2018, Yin2020, Kang2015, Liu2021b}).
Despite having relatively light effective masses, as well as significant SOC splitting in the valence bands [\figref{ZrSe2_HfSe2_mobility}(b) and (g)], the calculated mobilities of both materials remain low due to a very strong electron-phonon coupling which causes exceptionally high scattering rates [\figref{ZrSe2_HfSe2_mobility}(d) and (i)]. In particular, the coupling is dominated by Fr\"ohlich interactions with modes at 26~meV and 22~meV for for \ce{ZrSe2} and \ce{HfSe}, respectively (Figure~\ref{spectral_decomp_ZrSe2HfSe2}). The same coupling mechanism has also been shown to lead to the formation of large polarons in these materials~\cite{Sio2023}. Our calculated room-temperature hole mobilities are 45~\mob\ and 35~\mob\ for \ce{ZrSe2} and \ce{HfSe2}, respectively, while the corresponding electron mobilities are 15~\mob\ and 13~\mob\ [\figref{ZrSe2_HfSe2_mobility}(e) and (j)]. We are not aware of experimental measurements or previous calculations for \ce{ZrSe2}. Conversely, in the case of \ce{HfSe2}, measured mobilities in the range 0.2-6~\mob\ have been reported for nanosheets and few-layer systems~\cite{Kang2015,Yin2016}. Furthermore, a previous theoretical study of \ce{HfSe2} obtained carrier mobilities below 2~\mob~\cite{Keshri2023}; this prior work employed the relaxation time approximation, which might explain the small discrepancy from our calculations.

\begin{figure*}
\begin{center}
\includegraphics[width=0.95\textwidth]{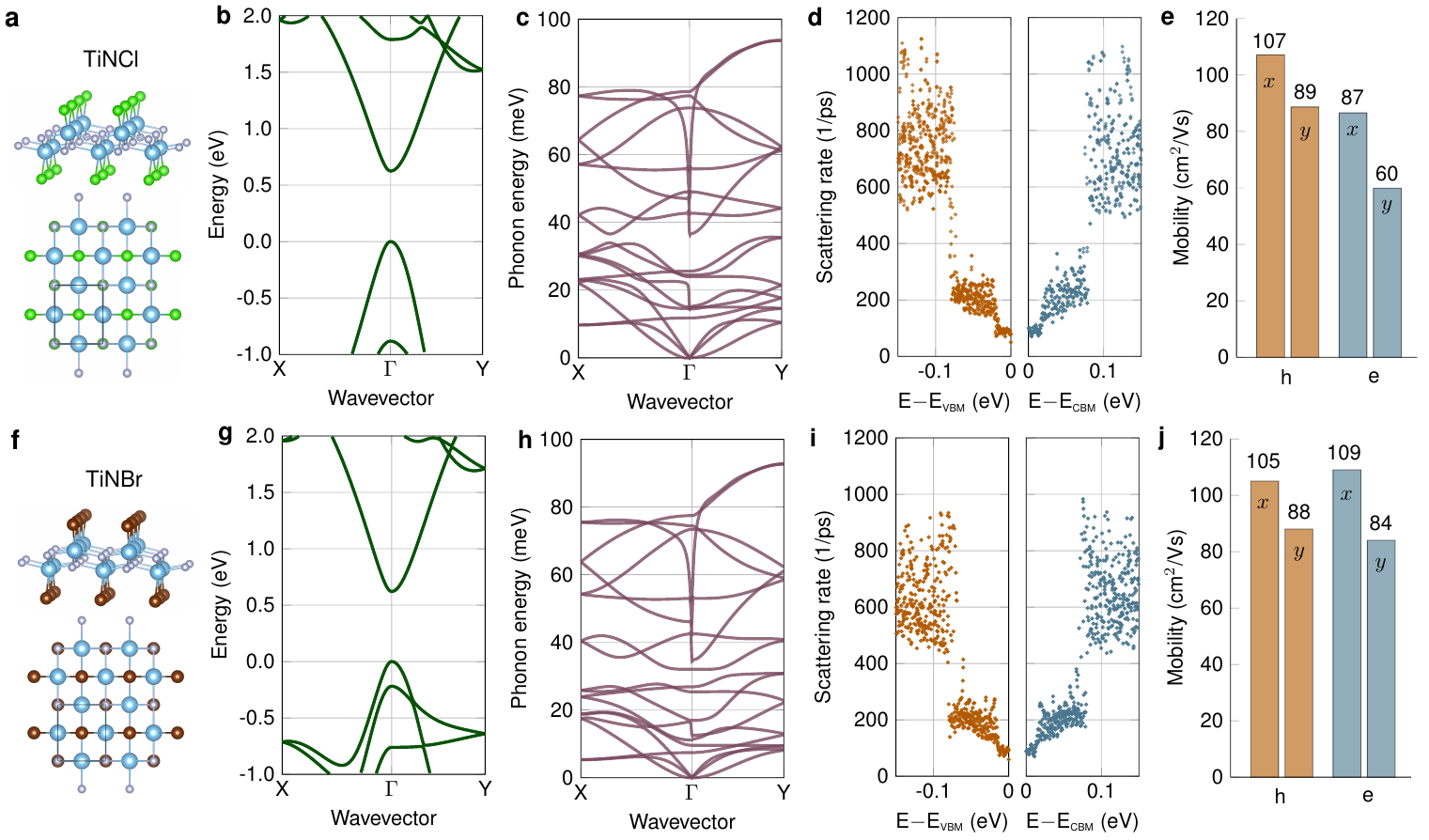}\vspace{-5pt}
\vspace{-5pt}
\caption{\label{TiNCl_TiNBr_mobility}\textbf{Carrier mobilities of \ce{TiNCl} and \ce{TiNBr} monolayers}. 
(a) Side and top views of monolayer \ce{TiNCl}. Ti is in blue, N is in white, and Cl is in green. (b) DFT electronic band structure. (c) Phonon dispersion relations. (d) Carrier-phonon scattering rates of holes (orange) and electrons (blue). (e) Phonon-limited carrier mobility of monolayer \ce{TiNCl} from the \textit{ai}BTE: holes (orange) and electrons (blue), evaluated at 300~K and for a carrier density of 10$^{10}$~cm$^{-2}$. The $x$ direction is aligned with the rows of Cl atoms. (f) Ball-stick models of \ce{TiNBr}, with Br in brown. (g)-(j) Same as (b)-(e) but for \ce{TiNBr}.\vspace{-10pt}} 
\end{center}
\end{figure*}

\begin{figure*}
\begin{center}
\includegraphics[width=0.8\linewidth]{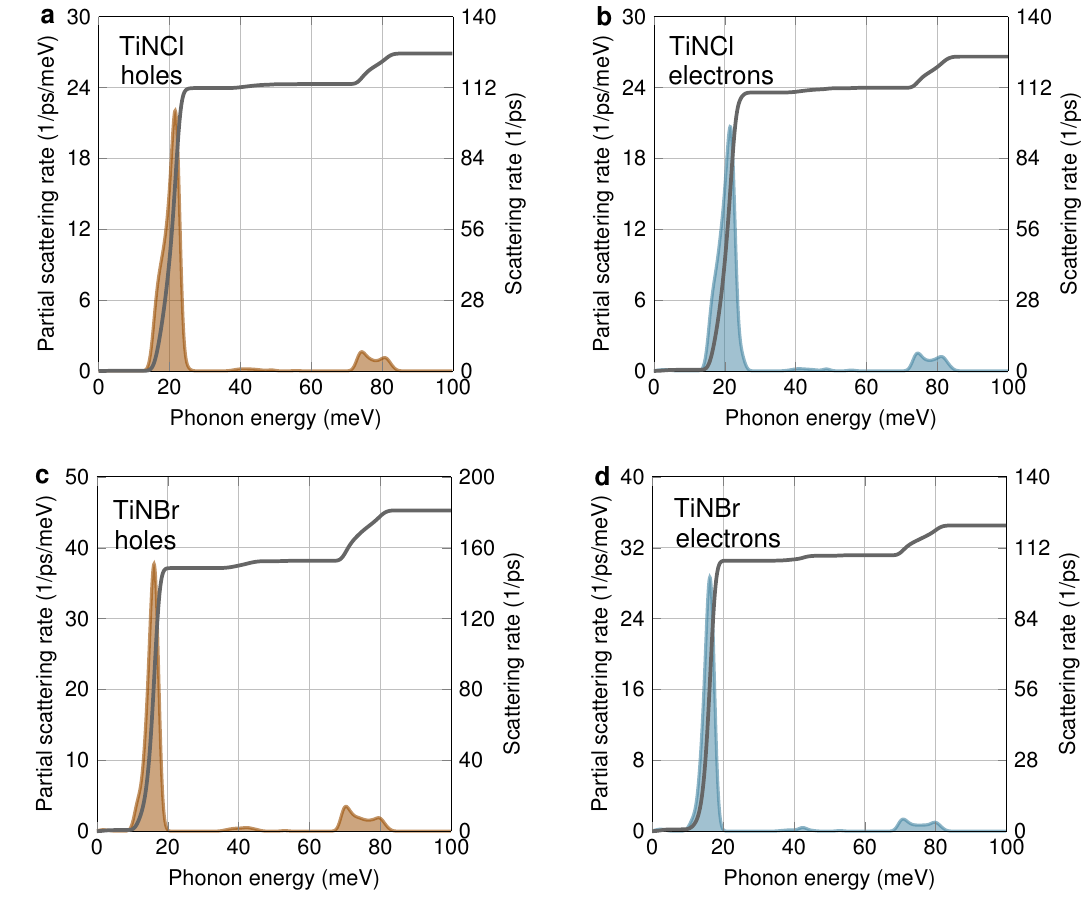}
\end{center}
\caption{\textbf{Analysis of scattering rates for TiNCl and TiNBr}. (a) Partial scattering rates of holes by phonons in TiNCl, resolved by phonon mode energy (orange line and filled area). The gray line is the cumulative scattering rate, i.e. the running integral of the blue curve, as reported on the right axis. (b) Partial scattering rates of electrons by phonons in TiNCl (blue line and filled area). The gray line is the cumulative rate (right axis). (c) and (d): Same format as for (a) and (b), but for holes and electrons in TiNBr, respectively. All rates are calculated as averages over carriers with energy $k_{\rm B} T$ from the band edges, with $T=300$\,K, for a carrier concentration of $10^{10}$\,cm$^{-2}$.}
\label{spectral_decomp_TNCB}
\end{figure*}

%
%
\textbf{Titanium nitride chloride and bromide}. TiNCl and TiNBr consist of TiN bilayers decorated two halogen layers, as shown in \figref{TiNCl_TiNBr_mobility}(a) and (f). The parent 3D compounds are well known for their superconducting properties~\cite{Yamanaka2010}, but the synthesis of individual monolayers has not been reported yet. These 2D materials have been investigated theoretically for a number of applications, spanning photovoltaics, thermoelectrics, and electronics~\cite{Liang2018, Wang2020, RostamiOsanloo2022}. One interesting feature of \ce{TiNCl} and \ce{TiNBr} is that they possess parabolic bands with a direct gap at $\Gamma$, unlike all the other compounds considered in the present work [\figref{TiNCl_TiNBr_mobility}(b) and (g)]. The electron-phonon coupling is very strong in both cases, and is dominated by LO stretching modes at 20~meV and 18~meV for TiNCl and TiNBr, respectively (Figure~\ref{spectral_decomp_TNCB}). As a result, the carrier scattering rates are very high, as seen in \figref{TiNCl_TiNBr_mobility}(d) and (i). The directionally-averaged mobilities at room temperature are all in the order of 100~\mob, namely 98~\mob\ and 74~\mob\ for holes and electrons in TiNCl, and 96~\mob\ and 97~\mob\ for holes and electrons in TiNBr. Our calculations are at odds with recent theoretical predictions placing the mobilities of these materials in the range 600-1500~\mob~\cite{Wang2020}. This discrepancy can be traced back to the use of an acoustic deformation potential scattering model in prior work, which misses the key contribution of optical phonons as seen in Figure~\ref{spectral_decomp_TNCB}.

\begin{figure*}
\begin{center}
\includegraphics[width=0.95\linewidth]{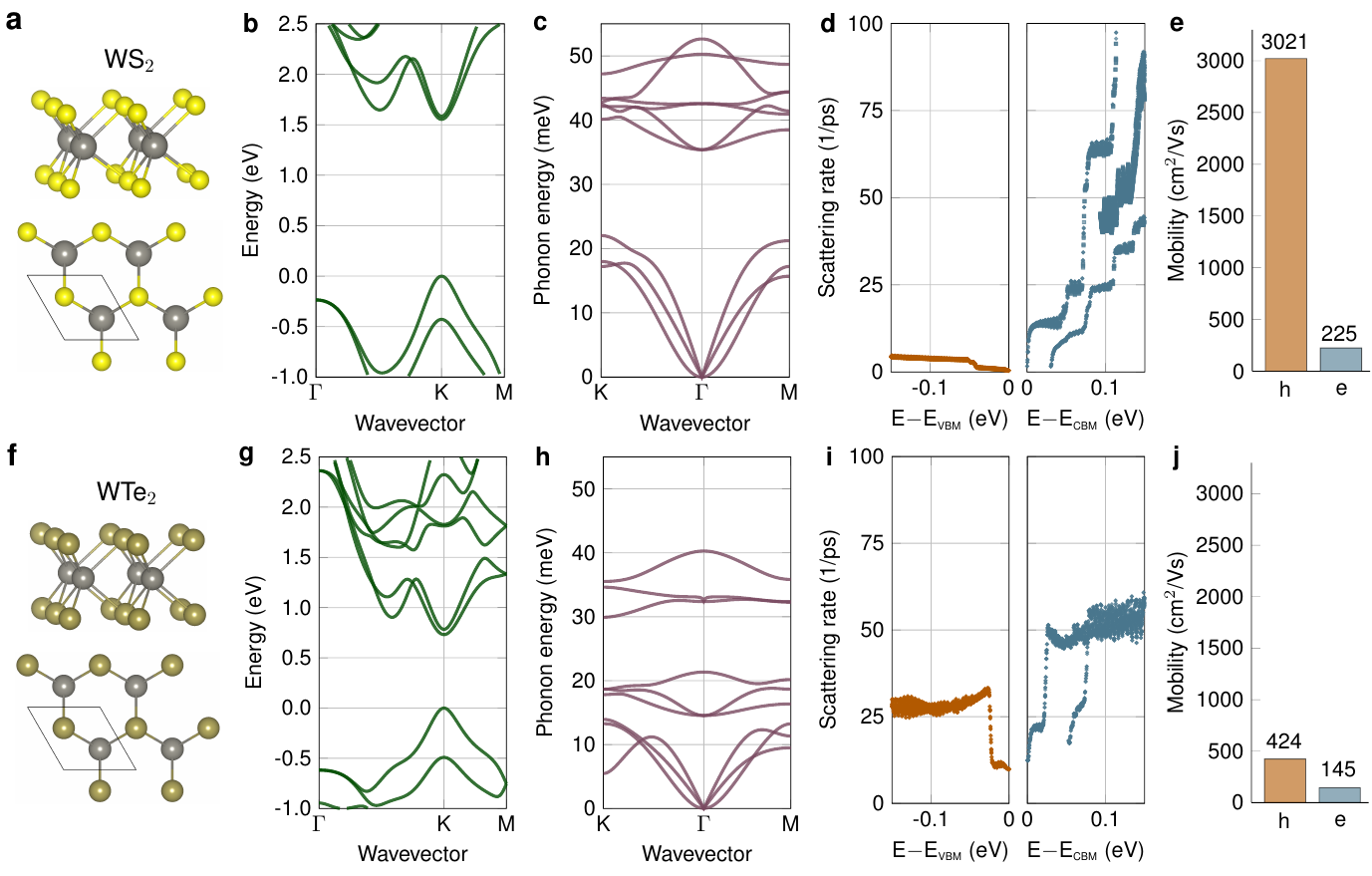}\vspace{-5pt}
\vspace{-5pt}
\end{center}
\caption{\label{WS2_WTe2_mobility}\textbf{Carrier mobilities of \ce{WS2} and \ce{WTe2} monolayers}. 
(a) Side and top views of monolayer \ce{WS2}. W is gray, S is yellow. (b) DFT band structure. (c) Phonon dispersions. (d) Carrier-phonon scattering rates of holes (orange) and electrons (blue). (e) Phonon-limited carrier mobility of monolayer \ce{WS2} from the \textit{ai}BTE: holes (orange) and electrons (blue), evaluated at 300~K and for 10$^{10}$~cm$^{-2}$ carriers. (f) Ball-stick models of \ce{WTe2}, with Te in olive. (g)-(j) Same as (b)-(e) but for \ce{WTe2}.\vspace{-10pt}} 
\end{figure*}

%
%
\textbf{Tungsten disulfide and ditelluride}. \ce{WS2} and \ce{WTe2} are common TMDs and are isostructural with the prototypical compound \ce{MoS2}, as shown in \figref{WS2_WTe2_mobility}(a) and (f), respectively. SOC does not affect the phonon dispersions of these compounds, but plays an important role in the electronic band structures, especially the valence bands. In fact, \figref{WS2_WTe2_mobility}~(b) and (g) show large splittings at the K-valleys, of 440~meV and 490~meV for \ce{WS2} and \ce{WTe2}, respectively. These values are an order of magnitude larger than the highest phonon energy, therefore the incorporation of SOC completely suppresses intravalley interband scattering at K. This effects results in very low hole scattering rates, as shown in \figref{WS2_WTe2_mobility}(d) and (i).
It is worth noting that the spin-splitting at the K-valleys in has been confirmed by angle-resolved photoemission spectroscopy (ARPES) measurements~\cite{Hinsche2017}. Figure~\ref{spectral_decomp_WS2.WTe2}(a) and (b) show that both holes and electrons in WS$_2$ are primarily scattered by acoustic phonons (holes: 65\%; electrons: 88\% ) since Fr\"ohlich coupling is very small in this material (see Table~\ref{tableS3}). Conversely, transport in WTe$_2$ is dominated by polar phonon scattering, as seen in Figure~\ref{spectral_decomp_WS2.WTe2}(c) and (d). Our calculated room-temperature hole mobilities are 3021~\mob\ for \ce{WS2}, and 424~\mob\ for \ce{WTe2}. 

The very high mobility of holes in \ce{WS2} will be analyzed in detail in the next section. In comparison, prior theoretical work reported much lower values, namely 988~\mob\ for \ce{WS2} in Ref.~\citenum{Hinsche2017} and 900~\mob\ and 200~\mob\ for \ce{WS2} and \ce{WSe2}, respectively in Ref.~\citenum{Zhang2023b}. These discrepancies can be ascribed to the fact that the authors of Refs.~\citenum{Zhang2023b} and~\citenum{Hinsche2017} used the relaxation time approximation and a high carrier concentration of 10$^{13}$~cm$^{-2}$. Using similar settings, we obtain a hole mobility of 1549~\mob\ for \ce{WS2}; the residual discrepancy is likely a numerical convergence issue.
Recently, the authors of Ref.~\citenum{Sohier2023} investigated the implications of SOC and valley degeneracy in the transport properties of 2D TMDs. In the case of \ce{WS2}, they reported an extremely high room-temperature hole mobility around 5000 cm$^2$/Vs, which is much higher than in our calculations. We suspect that this high mobility results from the choice made in Ref.~\citenum{Sohier2023} to compute electron-phonon matrix elements in the presence of free carriers already at the DFPT level. This choice tends to suppress electron-phonon couplings and favors very high mobilities. It is unclear to us whether this choice is the most realistic, since DFPT screening is static in nature (while carrier screening of electron-phonon coupling is a dynamic process) and since at such high carrier concentrations one might also need to consider scattering by carrier plasmons. SOC also plays a smaller but non-negligible role in electron transport: from Figure~\ref{bandstr_wo_wSOC}(k) and (l) we see that SOC induces K-valley splittings of the order of 50~meV for both compounds, which is comparable to the optical phonon energies and leads to a partial suppression of interband scattering. Our calculated electron mobilities for \ce{WS2} and \ce{WTe2} are 225~\mob\ and 145~\mob, respectively, and are in good agreements with the values reported in Ref.~\citenum{Zhang2023b} (200~\mob\ and 150~\mob, respectively).

\begin{figure*}
\begin{center}
\includegraphics[width=0.95\linewidth]{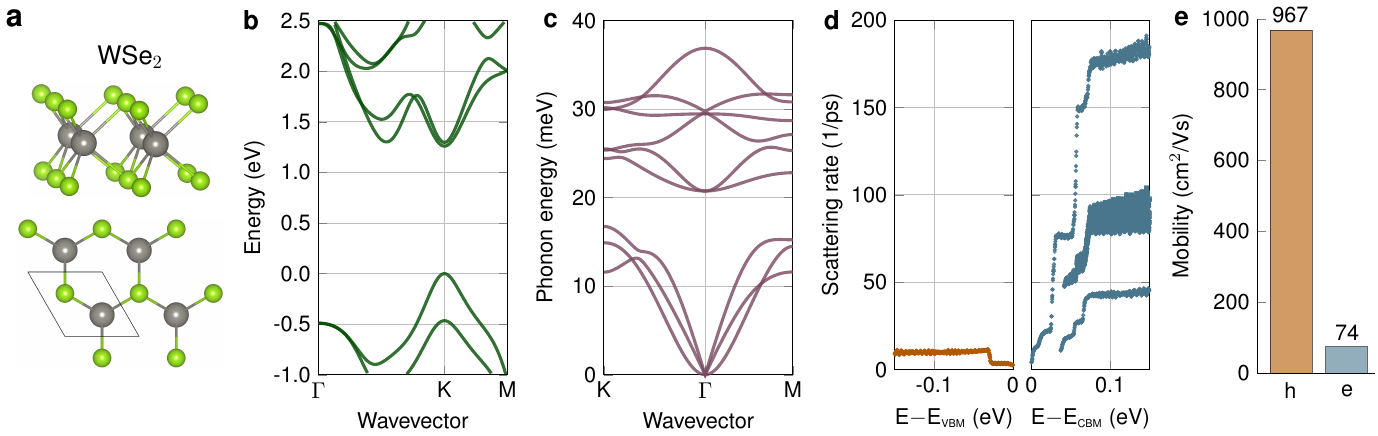}
\end{center}
\vspace{-10pt}
\caption{\label{WSe2_mobility}\textbf{Carrier mobilities of WSe$_2$ monolayer}. 
(a) Side and top views of monolayer WSe$_2$. W is gray, Se is green. (b) DFT electronic band structure. (c) Phonon dispersion relations. (d) Carrier-phonon scattering rates of holes (orange) and electrons (blue). (e) Phonon-limited carrier mobility of monolayer WSe$_2$ from the \textit{ai}BTE: holes (orange) and electrons (blue), evaluated at 300~K and for a carrier concentration of 10$^{10}$~cm$^{-2}$. The case of \ce{WSe2} is very similar to that of \ce{WS2} analyzed in the main text. In particular, also in this case the valence bands have their maximum at the K-valley, and K-$\Gamma$ intervalley scattering is forbidden by energy conservation. The coupling to polar phonons is weak, and the hole mobility is exceptionally high, as in \ce{WS2}. Previous theoretical work reported a hole mobility of 600\,\mob~\cite{Zhang2023b}, which is lower than our value because the mobility was computed within the (self-energy) relaxation time approximation (SERTA) as wells as at higher carrier concentration of $10^{13}$ cm$^{-2}$. Our calculations at this high carrier density yield mobilities of 477 and 827 \mob\ for SERTA and \textit{ai}BTE, respectively. Quadrupole corrections are expected to reduce the mobility, as in the case of \ce{WS2}. Measured hole mobilities are in the range 100-250\,\mob~\cite{Fang2012, Chen_2014, Allain_2014, Chuang2016}, and are likely dominated by extrinsic effects as already discussed for \ce{WS2}. Measured electron mobilities are in the range 10-200\,\mob~\cite{Chen_2014, Allain_2014,Liu_2013}, in line with our calculations. Astonishingly, in a very recent work~\cite{Joe2024}, high quality \ce{WSe2} samples were successfully synthesized, and an exceptional high hole mobility of 1217 \mob\ at temperature 200 K for a hole concentration of 8$\times 10^{12}$ cm$^{-2}$ was measured. To compare with this latest study, we performed one additional calculation using the same temperature and carrier concentration, and we obtained a mobility of 1375 cm$^2$/Vs, which is in good agreement with the experiments (we note that this calculation does not include GW corrections and quadrupoles). This latest experiment thus confirms the reliability of our independent predictions.
}
\end{figure*}

For completeness, in Figure~\ref{WSe2_mobility} we also analyze the carrier mobility of \ce{WSe2}, which belongs to the same group as \ce{WS2} and \ce{WTe2}, but does not appear in Set~D since its estimated electron mobility via Emin's formula is slightly below 100~\mob\ [Fig.~\ref{eq.emin}(b)]. Figures~\ref{WSe2_mobility}, \ref{bandstr_wo_wSOC_3}(c), \ref{spectral_decomp_WS2.WTe2}(e) and \ref{spectral_decomp_WS2.WTe2}(f) report our results for this additional compound. We find a very high hole mobility of 967~\mob\ at room temperature, making this system a very promising $p$-type channel material, comparable to h-Sb.

%
%
\vspace{20pt}
\noindent\textbf{In-depth analysis for \textit{p}-type WS$_2$}
\vspace{5pt}

Our systematic and unbiased screening singles out monolayer \ce{WS2} as a potential ultra-high mobility semiconductor with $\mu_{\rm h} = 1549$-3021~\mob\ at room temperature, depending on carrier concentration. If experimentally realized, this exceptional mobility could open the way to ground-breaking new applications in ultra-scaled electronics, therefore our data call for a more in-depth analysis and higher-level calculations.

\begin{figure*}
  \begin{center}
  \includegraphics[width=\textwidth]{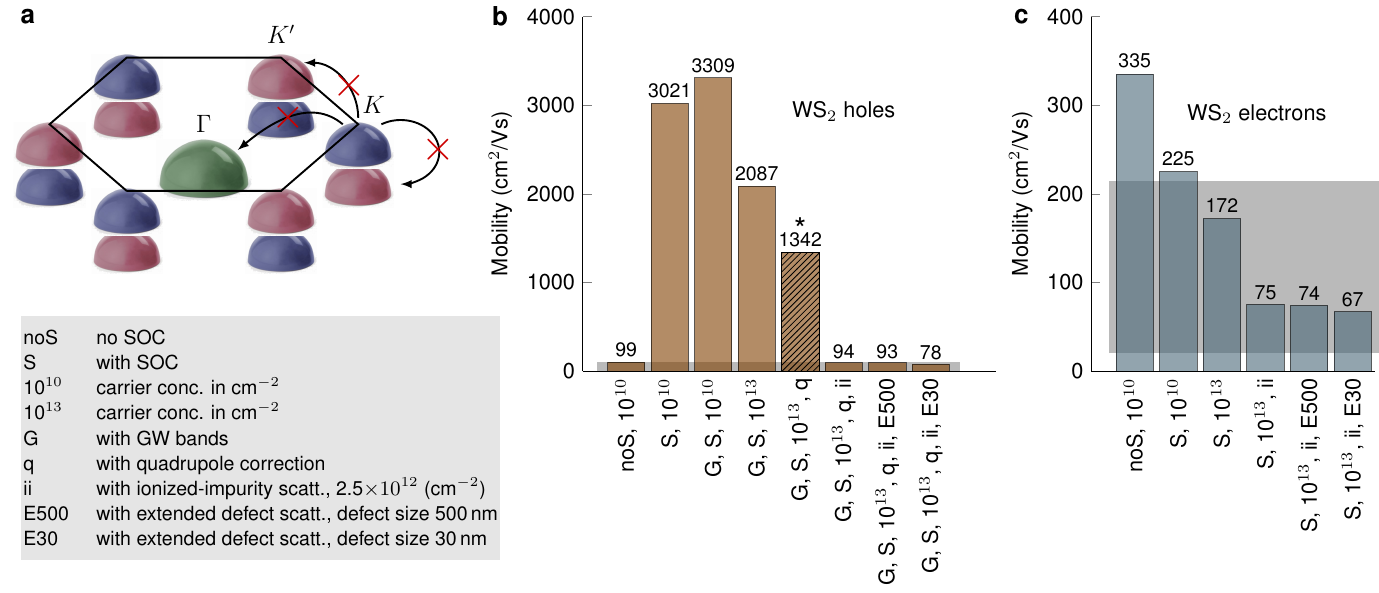}\vspace{-5pt}
  \caption{\textbf{In-depth analysis of the carrier mobility in \ce{WS2}}. (a) Schematic representation of the effect of SOC on the valence band valleys of monolayer \ce{WS2}. The hexagon denotes the first Brillouin zone, the blue and red paraboloids represent K/K$'$ valleys with opposite spin orientation, green denotes the $\Gamma$ valley. Crossed arrows indicate hole-phonon scattering processes that are forbidden by either energy or spin conservation. (b) Calculated hole mobility of monolayer \ce{WS2} under different scenarios and levels of theory. From left to right: DFT calculation without including SOC and low carrier concentration (10$^{10}$\,cm$^{-2}$); SOC included on top of the previous result; GW corrections included; As in the previous bar, but this time for a high density of carriers (10$^{13}$\,cm$^{-2}$); As previous bar, but with quadrupole corrections; Ionized impurity scattering added to the previous result; Scattering by extended defects (e.g, grain boundaries or ripples) added to the previous result, with large defect size of 500\,nm; As in the previous bar, but with small defect size of 30\,nm. The horizontal gray bar indicates the range of experimental data from Refs.~\citenum{Jo_2014, Braga_2012, Cao2017, Hou2020, Xie2022, Qi2022}. The striped bar with the star on top indicates our best estimate for the intrinsic hole mobility. (c) Same notation as in (b), but for the electron mobility in monolayer \ce{WS2}. The horizontal gray bar indicates the range of experimental data from Refs.~\citenum{Alharbi_2016, Reale_2017, Yun_2015, Iqbal2015, Cui2015, Iqbal_2016, Wang_2021}.\vspace{-10pt}}
  \label{ws2.summary}
  \end{center}
\end{figure*}

One might ask what is the origin of such a high hole mobility. In Fig.~\ref{ws2.summary}(b) we show that the calculated mobility \textit{without} including SOC is dramatically lower, 99~\mob\ at 300\,K and for a low carrier concentration of 10$^{10}$\,cm$^{-2}$. This result indicates that SOC plays an essential role in achieving high mobility. To understand this effect, in Fig.~\ref{ws2.summary}(a) we show a schematic of the valence band valleys of \ce{WS2} in the presence of SOC. Without SOC, the $\Gamma$-valley and the K-valleys are at nearly the same energy, and the K-valleys are doubly degenerate. Upon including SOC, the $\Gamma$-valley is pushed down by 235~meV below the top of the K-valleys, rendering hole-phonon scattering from K to $\Gamma$ forbidden. Furthermore, the degeneracy of the K-valley is lifted by a SOC splitting of 428~meV, making intravalley interband hole-phonon scattering forbidden as well. Lastly, the spin-momentum locking that is common to all TMDs induces opposed spin textures on the K and K$'$ valleys, therefore intervalley K-K$'$ hole-phonon scattering also becomes forbidden. These three SOC-related mechanisms contribute to propelling the hole mobility of \ce{WS2} to very high values. A similar mechanism is operational also in \ce{WSe2}, where SOC induces a K-valley splitting of 465~meV; however, in this case the higher ionic contribution to the dielectric constant (Table~\ref{tableS3}) makes the Fr\"ohlich coupling stronger than in \ce{WS2}, leading to a still-impressive but lower intrinsic mobility (see Figures~\ref{WSe2_mobility} and \ref{spectral_decomp_WS2.WTe2}).

There exists abundant experimental literature on the electron mobility of \ce{WS2}, with room-temperature values ranging between 20~\mob\ and 213~\mob~\cite{Alharbi_2016, Reale_2017, Yun_2015, Iqbal2015, Cui2015, Iqbal_2016, Wang_2021}. The record-high electron mobility of 214~\mob\ was obtained for exfoliated monolayer \ce{WS2} and Ohmic contacts to Al/Au electrodes~\cite{Iqbal2015}. This experimental value is remarkably close to our calculated mobility of 225~\mob, suggesting that the samples of Ref.~\citenum{Iqbal2015} have a relatively low defect density.

Unlike electron mobilities, experimental studies reported measurements of hole mobility in \ce{WS2} with low values ranging between 4~\mob\ and 105~\mob~\cite{Jo_2014, Braga_2012, Cao2017, Hou2020, Xie2022, Qi2022}. 
Possible sources of discrepancy between our very high predicted mobility and these experimental data are: (i) inaccuracy of DFT band structure, (ii) difference in carrier concentration, (iii) effect of quadrupolar couplings, (iv) scattering by defects, (v) scattering by extended defects such as grain boundaries or ripples, (vi) contact resistance in devices, and (vii) gate dielectric and substrate effects. In the following we analyze each of these effects in turn.

\textbf{Inaccuracy of DFT band structure}. As discussed above, the high hole mobility of \ce{WS2} is closely related to the energetics of the $\Gamma$ and $K$ valleys in this compound. Since these energies are all evaluated at the DFT level, it is important to check band structures with a higher level theory. To this end, we perform GW calculations and obtain the band structures shown in Figure~\ref{gw.bs.MoS2.WS2}(b). With the incorporation of GW quasiparticle corrections in the calculations, we obtain a room-temperature mobility of 3309~\mob, which is 10\% higher than our DFT value owing to a slightly lighter hole mass (0.36~$m_e$ with GW vs.\ 0.39~$m_e$ in DFT).

\textbf{Difference in carrier concentration}. Our calculations are performed using a carrier concentration of 10$^{10}$~cm$^{-2}$, while the experimental data reported in Refs.~\citenum{Jo_2014, Braga_2012} correspond to the much higher carrier density of 10$^{13}$~cm$^{-2}$. At this high density, the quasi-Fermi level lies 69~meV below VBM at K, therefore the carrier distribution function extends beyond the threshold for polar phonon emission around 50~meV [Fig.~\ref{WS2_WTe2_mobility}(d)], leading to a significant increase in the scattering rates. At this higher concentration, and including GW quasiparticle corrections, we calculate a hole mobility of 2087~\mob. This result represents a 37\% reduction over the mobility at low carrier concentration, see Fig.~\ref{ws2.summary}(b).

\begin{figure*}
\begin{center}
\includegraphics[width=0.6\textwidth]{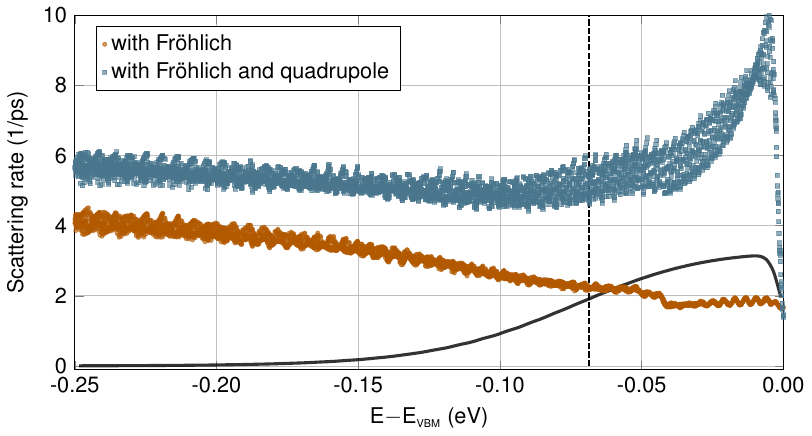}
\end{center}
\caption{\textbf{Hole scattering rates in WS$_2$ including quadrupole corrections}. Computed scattering rates of holes in monolayer WS$_2$, as a function of energy with respect to the valence band top. These calculations are performed for a carrier concentration of $10^{13}$\,cm$^{-2}$ and include GW corrections. Orange data point are the scattering rates obtained by including Fr\"ohlich long-range corrections, the blue data points include both Fr\"ohlich and quadrupole corrections. The solid line is the hole distribution function at 300\,K (arb. units), and the vertical line indicates the Fermi level.}
\label{quad.WS2}
\end{figure*}

\textbf{Effect of quadrupolar couplings}. Long-range electron-phonon couplings such as the Fr\"ohlich interaction introduce non-analyticities in the electron-phonon coupling matrix elements that necessitate a specialized treatment (see Methods). In most cases, the Fr\"ohlich electron-phonon interaction dominates over other non-analytic terms, but in materials with weak polar couplings the next-to-leading-order contribution can play a significant role~\cite{Park2020,Brunin2020}. This contribution results from dynamical quadrupoles and relates to piezoelectric effects~\cite{Royo2019}. \ce{WS2} stands out among the materials in Set~D for having a very small ionic contribution to the dielectric constant ($\epsilon_{\rm ion} = 0.05$, Table~\ref{tableS2}), which implies weak Fr\"ohlich couplings, as confirmed by the frequency-resolved scattering rates in Figure~\ref{spectral_decomp_WS2.WTe2}(a). In order to check for the effect of dynamical quadrupoles, we recompute the room-temperature hole mobility for a carrier concentration of 10$^{13}$\,cm$^{-2}$. Upon including quadrupole correction, we observe a 36\% reduction of the mobility, from 2087 to 1342~\mob\ with quadrupoles [see Fig.~\ref{ws2.summary}(b)]. This result can be rationalized by noting that piezoacoustic scattering increases the intra-valley scattering rates in the energy range where polar phonon scattering is very weak (see Figure~\ref{quad.WS2}).

We point out that these calculations do not take into account screening of piezoacoustic and polar optical phonon scattering by free carriers, therefore it is possible that these effects be somewhat overestimated.

\textbf{Scattering by defects}. Real \ce{WS2} samples contain a variety of intrinsic and extrinsic point defects such as S vacancies, W vacancies, W/S antisite defects, and substitutional O defects~\cite{Lin2017, Jeong2017, deGraaf2021, Wang2021, Schuler2019}. To estimate the impact of defects on the mobility, we incorporate ionized-impurity scattering as an upper bound to the defect-induced scattering rates. When we consider a typical defect concentration of 2.5$\times$10$^{12}$\,cm$^{-2}$~\cite{Wan_2022}, we obtain a defect-limited hole mobility of 101~\mob. By combining this value with the intrinsic, phonon-limited mobility of 1342~\mob\ via Matthiessen’s rule, we obtain 94~\mob. Therefore, the inclusion of defect scattering reduces the intrinsic mobility by a factor 14.3 [see Fig.~\ref{ws2.summary}(b)]. This finding indicates that reducing the defect density will be essential to enable high-mobility \textit{p}-type \ce{WS2}.
\begin{figure}
  \begin{center}
  \includegraphics[width=0.9\columnwidth]{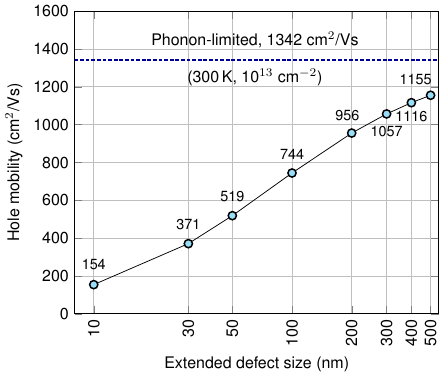}\vspace{-5pt}
  \caption{\textbf{Scattering by extended defects} in \ce{WS2}. Calculated hole mobility of \ce{WS2} as a function of extended defect size (discs). The calculations are performed at room temperature and for a carrier concentration of 10$^{13}$\,cm$^{-2}$. We include SOC, GW quasiparticle corrections, and electron-phonon scattering. The horizontal dashed line is the intrinsic, phonon-limited mobility without extended defects.}\vspace{-10pt}
  \label{WS2_mob.gb}
  \end{center}
\end{figure}

\textbf{Scattering by extended defects}. Another important source of scattering in TMDs is provided by one-dimensional and two-dimensional extended defects such as grain boundaries and ripples~\cite{Brivio_2011}. Since grain sizes in CVD-grown \ce{WS2} range from 30~nm to 500~nm~\cite{Groven2018,Naylor2023}, and ripples range from 100~nm to 600~nm~\cite{Zhang_2016, Lee_2020, Pang_2021}, in Fig.~\ref{WS2_mob.gb} we show calculations of the hole mobility for extended defect sizes in the range 10-500~nm. It is seen that this scattering channel reduces the hole mobility by almost an order of magnitude over the range considered. In Fig.~\ref{ws2.summary}(b) we show the hole mobilities obtained by considering phonons, defects, and extended defects altogether. We see that, for an average size of extended defects of 30~nm, the hole mobility drops from 94~\mob\ (without such defects) to 78~\mob, therefore the presence of small grains or ripples causes a twofold reduction of the mobility. This latter value is very close to the highest reported experimental value of 105~\mob~\cite{Braga_2012}. We point out that this experimental value is for an exfoliated \ce{WS2} sample with liquid electrolyte gate, therefore a precise comparison is not possible at present. However, it is clear from the present analysis that the incorporation of phonons, SOC, GW band structures, carrier concentration, quadrupoles, defects, and extended defects all contribute to bringing the theoretical value very close to experiments. 

A summary view of the relative importance of each of the above effects is provided in Fig.~\ref{ws2.summary}(b). In the same panel, we highlight with stripes our best theoretical estimate for the intrinsic (defect-free) hole mobility of suspended singe-crystal \ce{WS2}, i.e. 1342~\mob\ at 300~K, for a carrier density of 10$^{13}$\,cm$^{-2}$.

Unlike the hole mobility, in the case of the electron mobility electron-phonon scattering is strong enough to dominate over all other effects, as shown in Figure~\ref{ws2.summary}(c). In this case, our calculations fall within the range of experimentally measured mobilities.

\textbf{Contact resistance in devices}. A significant obstacle toward using TMDs in \textit{p}-type transistor channels is to realize Ohmic metal-semiconductor contacts. For example, in Refs.~\citenum{Jo_2014} the authors showed that transistors with Schottky-type \ce{WS2}/Au contacts carry a high contact resistance. When extracting the channel mobility from these device measurements, contact resistance often introduces an additional spurious contribution that tends to lower the apparent mobility~\cite{Mitta2020}. Realizing Ohmic \emph{p}-contacts with \ce{WS2} is particularly challenging owing to its high ionization energy of 5.74~eV~\cite{Keyshar2017,Kang_2013, Kim2021}, which exceeds the work function of most elemental metals~\cite{Yang2022}. 

The importance of contacts can be gauged by considering the related examples of few-layer \ce{MoS2}, for which the Ohmic contacts were realized using degenerately \textit{p}-doped 2D \ce{Mo_{0.995}Nb_{0.005}S2}~\cite{Chuang2016} or very high work function Pt~\cite{Liu2018}, and yielded a high hole mobility around 200~\mob. More recently, Ohmic contacts have been successfully realized for $p$-type monolayer \ce{WSe2} using Pt electrode combined with h-BN encapsulation~\cite{Liu2022, Joe2024} as well as with $\alpha$-\ce{RuCl3}~\cite{Pack2024}. These studies demonstrated record-high hole mobilities, ranging from 655 to 1260~\mob\ at room temperature. With these advances in synthesis and techniques being achieved, the realization of Ohmic contacts for $p$-type monolayer \ce{WS2}~\cite{Rai2018, Zheng2021} is expected to come in the near future.  

\textbf{Gate dielectric and substrate}. Our calculations describe a scenario where monolayer \ce{WS2} is suspended in vacuum. In practical field-effect devices the 2D layer sits on top of a dielectric substrate, and is insulated from the gate electrode by an oxide, for example \ce{HfO2}~\cite{Radisavljevic2011}. The dielectric environment that surrounds the 2D channel may have beneficial effects in terms of screening the scattering potential of ionized impurities. However, it was pointed out early on that such a dielectric medium can also degrade mobility by introducing remote phonon scattering~\cite{Hess1979,Fischetti2001}. This scattering mechanism originates from the long-range electrostatic field generated by plasmon-phonon polaritons; these hybrid excitations result from the hybridization of free-carrier plasmons in the channel and polaron optical phonons in the insulator. 

The coupling matrix elements for remote phonon scattering scales as $g_{\rm RP} = \hbar\omega [(\epsilon_\infty+\epsilon_\infty^{\rm ins})^{-1}-(\epsilon_\infty+\epsilon_0^{\rm ins})^{-1}] $, where $\epsilon_\infty$ is the high-frequency dielectric constant of the 2D channel and $\epsilon_\infty^{\rm ins}$, $\epsilon_0^{\rm ins}$ are the static and high-frequency dielectric constants of the insulator, respectively~\cite{Fischetti2001}. Taking \ce{HfO2} as a representative example ($\epsilon_\infty^{\rm ins}=5.03$, $\epsilon_0^{\rm ins}=22.0$, $\hbar\omega =12.4$\,meV, Ref.~\cite{Fischetti2022}), and using the parameters for \ce{WS2} in Table~\ref{tableS3}, we find $g_{\rm RP}=0.21$\,meV. This energy scale is significantly larger than the characteristic energy of the Fr\"ohlich coupling included in our calculations, $g_{\rm F}=\hbar\omega(1/\epsilon_\infty-1/\epsilon_0) \simeq 0.02$\,meV (using parameters from Table~\ref{tableS3}). 

The above order-of-magnitude comparison suggests that remote phonon scattering may play a significant role in the mobility of \ce{WS2}. This conclusion is supported by recent calculations of the hole mobility of \ce{WS2} embedded in a model dielectric environment describing a \ce{HfO2}/\ce{WS2}/\ce{SiO2} stack, which reported a nearly fivefold mobility reduction by remote phonon scattering~\cite{Fischetti2022}. 

\textbf{Strain}. Strain engineering is widely used in the semiconductor industry to modify the electronic properties of bulk materials, and offers a pathway to fine-tune transport in 2D semiconductors~\cite{Dai2019, Du2020, S.Yang2021, Qi2023}. In the present work, we do not consider the effect of strain since our calculations are performed for ideal, unstrained monolayer \ce{WS2}. However, it is conceivable that unintentional strain in monolayer \ce{WS2} samples, which may result from substrate lattice mismatch or wrinkles~\cite{Dai2019, S.Yang2021}, could reduce the experimentally measured hole mobility. The impact of strain will be systematically investigated in our future works.
%
%
\section{Discussion and Conclusions}\label{Conclusions}
\begin{figure*}
  \begin{center}
  \includegraphics[width=\textwidth]{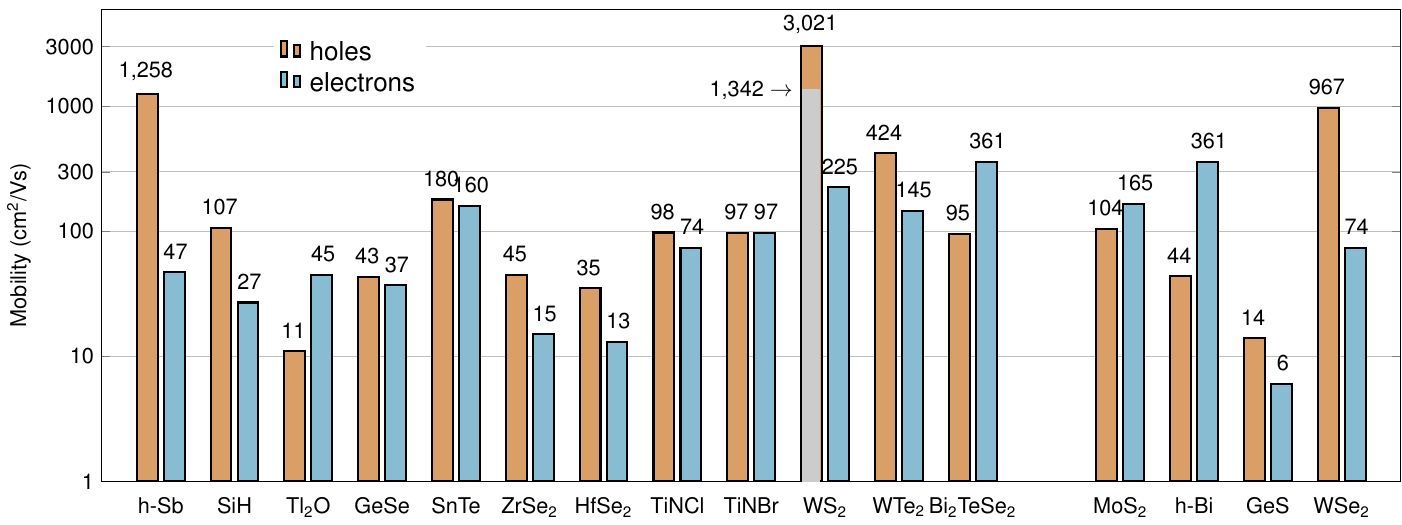}\vspace{-5pt}
  \caption{\textbf{Summary view of \textit{ab initio} mobilities}. Hole and electron mobilities of the 12 2D compounds in Set~D (left) plus the additional compounds MoS$_2$, h-Bi, GeS, WSe$_2$ (right). Calculations are performed using the \aibte, they include only electron-phonon scattering and 2D Fr\"ohlich corrections, and are performed at 300\,K for a low carrier concentration of 10$^{-10}$\,cm$^{-2}$. Note the logarithmic scale. The gray bar in the case of \ce{WS2} indicates that its mobility decreases to 1342~\mob\ upon taking into account GW corrections, high-carrier density, and quadrupole effects. Similar reductions of mobility are expected for \ce{WSe2} and {h-Sb}.\vspace{-10pt}}
  \label{top13}
  \end{center}
\end{figure*}

Starting from the 5,619 layered compounds collected in the MC2D database, we performed a first-principles HT screening of potential high-mobility 2D materials using the tiered approach illustrated in Fig.~\ref{HT_tier}. We filtered putative 2D semiconductors by combining \textit{ab initio} data reported in MC2D and a large number of new first-principles calculations. This analysis led to down-selection of 16 compounds (12 in Set~D plus 4 closely-related compounds) for which we proceeded to perform state-of-the-art transport calculations using the \textit{ai}BTE approach implemented in the \texttt{EPW} code. Our three main findings are as follows: 

(i) Even though we started from a very large library of putative as well as existing materials, only a handful of compounds exhibit very high mobilities. In particular, considering a temperature or 300\,K and a low carrier concentration of 10$^{10}$\,cm$^{-2}$, we found very high intrinsic hole mobilities in antimonene h-Sb (1258\,\mob), \ce{WSe2} (with 967\,\mob), and \ce{WS2} (3021\,\mob); in the case of electron mobilities, the top candidates are bismuthene h-Bi (361\,\mob) and skippenite \ce{Bi2TeSe2} (361\,\mob). These data are summarized in Fig.~\ref{top13}. That the highest hole mobility exceeds the best electron mobility by almost $10\times$ is unexpected, and is ground for optimism in view of realizing \textit{p}-channel 2D devices. 

(ii) The intrinsic theoretical hole mobility of \ce{WS2} is exceptionally high. To be confident about this surprising result, we have performed an extensive range of validation tests. For example, we showed that our calculations for \ce{MoS2} are very close to available experimental data; furthermore, our calculations for the elecron mobility of \ce{WS2} are also in the same range as experimental data. To rationalize the large discrepancy between the intrinsic hole mobility predicted here and measured data for \textit{p}-type \ce{WS2}, we extended our calculations to include the effects of spin-orbital coupings, quasiparticle band structures, carrier concentration, quadrupole corrections, impurity scattering, and scattering by extended defects. We find that each of these effects can significantly reduce the hole mobility, and that calculations including all these effects simultaneously are close to available experimental estimates. Our best theoretical estimate for the intrinsic hole mobility of \ce{WS2} including SOC, GW bands, and quadrupoles is 1342~\mob\ at 300~K and for 10$^{13}$\,cm$^{-2}$ carriers. This value should constitute the ultimate mobility limit for the ideal, suspended single-crystal and defect-free monolayer.

We also pointed out that the hole mobility is especially prone to underestimation due to the contact resistance and the scarcity of high-workfunction metals suitable for \ce{WS2}. Another important mechanism that is not included in the present study is the effect of gate dielectric and substrate, and in particular remote phonon scattering. Similarly, higher-order phonon scattering processes, such as electron-two-phonon scattering~\cite{Lee2020,Minnich2022}, could also play a role and deserve a detailed investigation.

Overall, the present analysis suggests that \textit{extrinsic} effects and \textit{interfaces} could play a major role in the observed hole mobility of \ce{WS2}, and that by controlling these effects it might be possible to unlock unprecedented performance. Therefore, the present work calls for renewed efforts in the experimental synthesis and processing of \ce{WS2} for \textit{p}-type transistor channels. 

\begin{figure}
  \begin{center}
  \includegraphics[width=0.9\columnwidth]{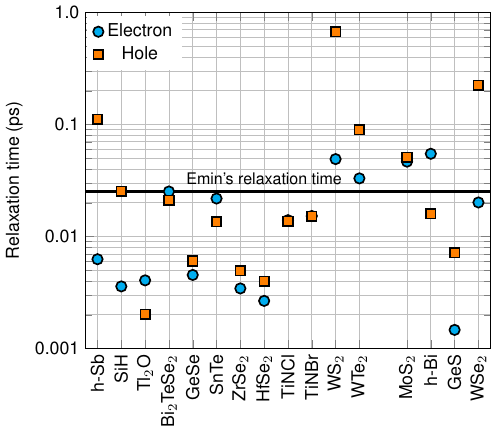}\vspace{-5pt}
  \caption{\textbf{Relaxation times for compounds in Set~D}. Computed average relaxation times of electrons and holes for the 12 compounds in Set~D. For completeness, we also show data for the 4 additional compounds included in Fig.~\ref{top13}. The average relaxation time $\tau$ is obtained from Drude's formula via the computed \aibte\ mobilities $\mu$ and conductivity effective masses $m^*$, $\tau = \mu m^* / e$ ($e$ is the electron charge). For comparison, we also show the simple estimate via Emin's formula (thick horizontal line).\vspace{-10pt}}
  \label{tau.extracted.aiBTE}
  \end{center}
\end{figure}

(iii) This study shows that reliable predictions of carrier mobilities in 2D materials ultimately require full-blown \textit{ab initio} calculations. In particular, we find that the scattering rates are extremely sensitive to materials structure and chemistry, and show significant variations between compounds of the same family and even between electron and holes of the same compound (see Fig.~\ref{tau.extracted.aiBTE}). Therefore, simplified approaches based on constant relaxation-time approximations and related simplifications may not be sufficient in quantitative studies. Furthermore, our in-depth analysis of \ce{WS2} reveals that focusing on intrinsic phonon-limited carrier mobilities may not be fully representative of the experimental reality, especially in the case of compounds with weak electron-phonon couplings where extrinsic effects might ultimately dominate. The lesson for future HT searches is that a blanket approach might be too coarse to identify promising materials, and deep-dives into the specifics of individual compounds and their interfaces with contacts and substrates will be necessary going forward.

In summary, our findings indicate that \ce{WS2} is inherently an ultra-high mobility semiconductor, but its true potential is hindered by \textit{defects} and \textit{interfaces}. Hence, our study suggests that the the key to high-mobility transistors based on 2D materials may not lie in the discovery of entirely new compounds, but rather in managing defect density and co-optimizing channel, contacts, and dielectrics. We propose that this approach may hold greater promise for achieving groundbreaking advances in ultra-scaled electronics based on 2D materials.
%
%
%
\section{Methods}\label{Methods} 

%
%
\noindent\textbf{Computational setup}
\vspace{5pt}

In this section we describe the computational setup employed for performing calculations of the electronic structure, lattice dynamics, electron-phonon couplings, and carrier transport coefficients in the present study.

We perform DFT~\cite{Hohenberg1964, Kohn1965} and density-functional perturbation theory (DFPT)~\cite{Baroni1987, Giannozzi1991, Gonze1997a, Gonze1997b, Baroni2001} calculations using planwaves basis sets and pseudopotentials, as implemented in the \texttt{Quantum ESPRESSO} suite~\cite{Giannozzi2017}. We describe exchange and correlation effects using the PBE functional~\cite{Perdew1996}. For HT calculations of band structures and effective masses without SOC, we employ pseudopotentials from the Standard Solid-State Pseudopotentials (SSSP) library~\cite{Prandini2018}, which afford a low planewaves kinetic energy cutoff and are advantageous for calculations at scale. For HT calculations of effective masses, phonon dispersion relations, electron-phonon couplings, and carrier transport coefficients including SOC, we employ optimized norm-conserving pseudopotentials (ONCV)~\cite{D.R.Hamann-PRB13} from the PseudoDojo library~\cite{Lejaeghere2016, vanSetten2018} to achieve higher accuracy at the price of higher computational cost. In all cases, we use the planewaves kinetic energy cutoffs recommended in these libraries, which have been determined via extensive testing~\cite{Prandini2018, vanSetten2018}. 
For structural optimization, we set the energy threshold to $10^{-5}$~Ry, and the force threshold to $10^{-4}$~Ry/bohr when performing HT calculations. These thresholds are reduced by one order of magnitude for calculations of phonons and transport coefficients. To eliminate spurious Coulomb interactions among periodic slab replicas along the $z$ direction, we apply 2D truncation in all DFT and DFPT calculations using the approach of Ref.~\citenum{Sohier2017}.

We calculate conductivity effective masses [see Eq.~\eqref{cond_effmass}] using the \texttt{BoltzTrap} code~\cite{Madsen2006}. The advantage of this approach is that BoltzTrap does not require Wannier-Fourier interpolation, therefore it can be used for HT calculations on large materials libraries~\cite{Ricci2017}. To validate our results, we collected effective masses in the literature for 16 compounds (see \tabref{mobility_paras}) and presented a comparison with our data in Table~\ref{effective_mass}. Our conductivity effective masses are generally consistent with previous works.
\begin{table}
\caption{\textbf{Brillouin zone sampling}. Brillouin zone grids employed for \aibte\ calculations of carrier mobility of compounds in Set~D, plus the four additional compounds \ce{MoS2}, h-Bi, GeS, and WSe$_2$. Unless otherwise specified, calculations are performed at room temperature (300~K), for a carrier concentration of $10^{10}$~cm$^{-2}$. $\{\mathbf{k}_{\rm c}\}$ and $\{\mathbf{q}_{\rm c}\}$ refer to the coarse Brillouin zone grids for electron and phonon wavevectors in DFT and DFPT calculations, respectively. $\{\mathbf{k}_{\rm f}\}$ and $\{\mathbf{q}_{\rm f}\}$ indicate to the electron and phonon fine grids employed with Wannier-Fourier interpolation in \texttt{EPW}, respectively. Fine grids are determined by converging mobility values to an accuracy of less than 1~\mob.}
\vspace{4pt}
\centering
\begin{tabular}{ l r r r r }
\toprule\\[-7pt]
Compound & $\{\mathbf{k}_{\rm c}\}$ & $\{\mathbf{q}_{\rm c}\}$ & Carrier type & $\{\mathbf{k}_{\rm f}\}=\{\mathbf{q}_{\rm f}\}$ \\[3pt]
\midrule\\[-7pt]
h-Sb           & $24^2$       & $12^2$       & electron & $480^2$        \\
               &              &              & hole     & $1560^2$       \\[4pt]
SiH            & $24^2$       & $12^2$       & electron & $240^2$        \\
               &              &              & hole     & $840^2$        \\[4pt]
\ce{Tl2O}      & $24^2$       & $12^2$       & electron & $600^2$        \\
               &              &              & hole     & $360^2$        \\[4pt]
\ce{Bi2TeSe2}  & $18^2$       & $9^2$        & electron & $1260^2$       \\
               &              &              & hole     & $450^2$        \\[4pt]
GeSe           & 24$\times$20 & 12$\times$10 & electron & 600$\times$500 \\
               &              &              & hole     & 600$\times$500 \\[4pt]
SnTe           & $20^2$       & $10^2$       & electron & $500^2$        \\
               &              &              & hole     & $700^2$        \\[4pt]
\ce{ZrSe2}     & $24^2$       & $12^2$       & electron & $480^2$        \\
               &              &              & hole     & $720^2$        \\[4pt]
\ce{HfSe2}     & $24^2$       & $12^2$       & electron & $360^2$        \\
               &              &              & hole     & $720^2$        \\[4pt]
TiNCl          & 24$\times$20 & 12$\times$10 & electron & 360$\times$300 \\
               &              &              & hole     & 480$\times$400 \\[4pt]
TiNBr          & 24$\times$20 & 12$\times$10 & electron & 480$\times$400 \\
               &              &              & hole     & 480$\times$400 \\[4pt]
\ce{WS2}       & $24^2$       & $12^2$       & electron & $720^2$        \\
               &              &              & hole     & $1320^2$       \\[4pt]
\ce{WTe2}      & $24^2$       & $12^2$       & electron & $600^2$        \\
               &              &              & hole     & $720^2$        \\[-1pt]
\midrule\\[-10pt]
\ce{MoS2}      & $24^2$       & $12^2$       & electron & $720^2$        \\
               &              &              & hole     & $480^2$        \\[4pt]
\ce{h-Bi}      & $12^2$       & $12^2$       & electron & $1200^2$       \\
               &              &              & hole     & $960^2$        \\[4pt]
\ce{GeS}       & 24$\times$20 & 12$\times$10 & electron & 720$\times$600 \\
               &              &              & hole     & 600$\times$500 \\[4pt]
\ce{WSe2}      & $12^2$       & $24^2$       & electron & $720^2$        \\
               &              &              & hole     & $840^2$        \\
\bottomrule
\end{tabular}
\label{mobility_paras}
\end{table}

We compute electrical transport properties by solving the \aibte~\cite{Li2015,Ponce2018} as implemented in the \texttt{EPW} code~\cite{Lee2023}. \texttt{EPW} employs Wannier-Fourier interpolation by calling the \texttt{Wannier90} code in library mode~\cite{Marzari1997, Souza2001, Mostofi2008, Marzari2012, Mostofi2014, Pizzi2020}. The calculations involve determining electron wavefunctions, vibrational eigenmodes, and their couplings on coarse Brillouin zone grids, and interpolating these quantities onto ultra-dense grids by means of maximally-localized Wannier functions~\cite{Giustino2007, Marzari2012}. The interpolation accuracy and reliablity is thoroughly checked in all cases~\cite{Ponce2021}. Owing to their Fermi-Dirac occupation factors and the energy selection rule for electron-phonon scattering [see Eqs.~\eqref{BTE_gen} and \eqref{scattering_rate} below], only electronic states in the vicinity of the band edges contribute to the carrier mobility. Accordingly, we consider states within a 300~meV energy window from either the conduction band minimum (CBM) or the valence band maximum (VBM) for electron and hole mobility calculations, respectively. We have performed convergence tests to verify that this choice yields accurate mobility values. To ensure convergence of calculated mobilities with respect to the Brillouin zone sampling, we systematically increase the number of points in the fine grid and repeat the calculations until the mobility changes by less than 1~\mob. The grid sizes obtained using this criterion and employed throughout this study are reported in \tabref{mobility_paras} for reproducibility purposes. Brillouin-zone integrals are evaluated using the adaptive smearing of Refs.~\citenum{Li2014,Ponce2021} which has been shown to provide the most reliable results. Long-range Fr\"ohlich electron-phonon couplings \cite{Kaasbjerg2012,Verdi2015,Sjakste2015,Sohier2016} are included by using the method of Ref.~\citenum{Sio2022}, which fully takes into account the dipolar fields generated by out-of-plane vibrations unlike earlier approaches. For a few test cases, we confirmed that the  alternative approach of Refs.~\citenum{Ponce2023a,Ponce2023b} yields very similar results. Long-range quadrupole corrections \cite{Park2020,Brunin2020} are not included except for \ce{WS2}; for this compound we included quadrupoles using the method of Ref.~\citenum{Ponce2023b}, with quadrupole tensors computed via the \texttt{Abinit} code~\cite{abinit2020}. All calculations of carrier mobility are performed by including spin-orbit coupling (SOC). 

Quasiparticle GW calculations for \ce{WS2} are performed using the \texttt{BerkeleyGW} code~\cite{Hybertsen1986, Deslippe2012}. We employ a dielectric matrix kinetic energy cutoff of 25~Ry, 390 empty bands for the summation over states, and the Hybertsen-Louie plasmon-pole model. GW band structures are used to obtain more accurate values for the effective masses, and to solve \aibte\ equations with the highest precision in the case of WS$_2$.

%
%
\vspace{20pt}
\noindent\textbf{\textit{Ab initio} Boltzmann transport equation}
\vspace{5pt}

In this section, we briefly summarize the methodology underlying the \aibte\ approach, as well as its simplified version provided by the relaxation time approximation for comparison with previous work. 
A detailed derivation of the \aibte\ formalism can be found in Refs.~\citenum{Ponce2018,Ponce2020}. The carrier mobility tensor in the weak-field limit is computed as:
\begin{equation}\label{mu_BTE}
\mu_{\alpha\beta} = \frac{1}{n_{\rm c}\,\Omega}\frac{1}{N_\bk}\sum_{n\bk}
v_{n\bk,\alpha} \partial_{E_\beta} f_{n\bk},
\end{equation}
where Greek indices indicate Cartesian directions, $n_{\rm c}$ is the carrier density, $\Omega$ is the volume of the crystal unit cell, and $N_\bk$ is the number of electron wavevectors $\bk$ in a uniform Brillouin zone grid; $v_{n\mathbf{k},\alpha}$ denotes the electron group velocity along the Cartesian direction $\alpha$ for the electronic state with band $n$ and wavevector $\bk$; $\partial_{E_\beta} f_{n\bk}$ is the derivative of the electron occupation function with respect to the $\beta$-component of the electric field, evaluated at vanishing field. In the case of calculations without SOC, an additional factor of 2 is needed in Eq.~\eqref{mu_BTE} to account for spin degeneracy.

The variation $\partial_{E_\beta} f_{n\bk}$ is obtained from the self-consistent solution of the equation:
\begin{eqnarray}\label{BTE_gen}
   -e v^{\beta}_{n\bk} \frac{\partial f^0_{n\bk}}{\partial \varepsilon_{n\bk}}
  &=& \frac{1}{N_\bq}\sum_{m\bq}\,
     \big( \ratemn \, \partial_{E_{\beta}}f_{m\bk+\bq} \nonumber \\
    &&\hspace{32pt} -\ratenm \, \partial_{E_{\beta}}f_{n\bk} \big),
\end{eqnarray}
where $e$ is the electron charge, $f^0_{n\bk}$ denotes the Fermi-Dirac occupation of the state $n\bk$, and $\ratenm$ denotes the transition rate from the state $n\bk$ to the state $m\bk+\bq$. The electron-phonon scattering rate is written as:
\begin{eqnarray}\label{scattering_rate}
  &&\ratenm^{\rm (ph)} = \sum_{\nu} \frac{2\pi}{\hbar} \left| g_{mn\nu}(\bk,\bq) \right|^2 \nonumber \\
  &&\hspace{10pt}\times \big[ (n_{\bq\nu}+1-f^0_{m\bk+\bq})\delta(\varepsilon_{n\bk}\!-\!\varepsilon_{m\bk+\bq}-\hbar\omega_{\bq\nu}) \nonumber \\
  &&\hspace{32pt}+ (n_{\bq\nu}+f^0_{m\bk+\bq})\delta(\varepsilon_{n\bk}\!-\!\varepsilon_{m\bk+\bq}+\hbar\omega_{\bq\nu}) \big],
\end{eqnarray}
where $\hbar$ is the reduced Planck constant, and $g_{mn\nu}(\bk,\bq)$ is the electron-phonon matrix element for the scattering between Kohn-Sham states $n\bk$ and $m\bk+\bq$ via the phonon of wavevector $\bq$ in branch $\nu$~\cite{Giustino2017}, $n_{\bq\nu}$ denotes the Bose-Einstein occupation of this mode, and $\omega_{\bq\nu}$ is the corresponding vibrational frequency; $\varepsilon_{n\bk}$ and $\varepsilon_{m\bk+\bq}$ denote DFT or GW eigenvalues.

The evaluation of the electron-phonon matrix elements $g_{mn\nu}(\bk,\bq)$ using Wannier-Fourier interpolation requires some caution in polar materials. In fact, in 3D materials this matrix element diverges as $1/|\bq|$ for $\bq \rightarrow 0$ in the case of longitudinal-optical (LO) phonons~\cite{Verdi2015,Sjakste2015}. This effect is a consequence of the long-range electric field established by a LO phonon at long wavelength. In 2D materials, there is no such singularity, by the matrix element exhibits a discontinuity in the long-wavelength limit~\cite{Sohier2016,Deng2021,Sio2022,Ponce2023b}. To handle these non-analyticities, the matrix element is separated into a short-range and a long-range component, $g=g_\mathcal{S} + g_\mathcal{L}$~\cite{Verdi2015}. The short-range component is dealt with via standard Wannier-Fourier interpolation~\cite{Giustino2007}, while the long-range component is evaluated using an exact formula that contains the LO phonon frequency, Born effective charges, and electronic dielectric permittivity tensor for the dipole term~\cite{Verdi2015,Sjakste2015}, and additionally the Born dynamical quadrupole tensors for the quadrupole component~\cite{Park2020,Brunin2020,Ponce2021}.

The evaluation of the long-range component of the matrix element, $g_\mathcal{L}$, in 2D polar materials is made challenging by the presence of a dielectric interface between the 2D material and vacuum (or another dielectric substrate). This problem has been addressed by resorting to various approximations, e.g. using an infinitesimally-thin 2D later~\cite{Peeters1986}, a square-well potential along the direction perpendicular to the slab~\cite{Sarma1985, Hahn2018}, and a uniform macroscopic polarization within the 2D slab~\cite{Sohier2016}. More recent work included the effect of out-of-plane dipoles~\cite{Deng2021, Royo2021, Ponce2023b}, and a unified approach to both 3D and 2D polaron electron-phonon interactions~\cite{Sio2022}. In this work, we evaluate $g_\mathcal{L}$ using Eq.~(34) of Ref.~\citenum{Sio2022}, as implemented in \texttt{EPW}~v5.8~\cite{Lee2023}.

As for Fr\"ohlich interactions, long-range quadrupole corrections also exhibit non-analiticities at long wavelenght, which can be captured in 2D materials using the method of Ref.~\citenum{Ponce2023b}. These effects are not expected to play a significant role for the low-mobility compounds of Set~D, but are important in the case of WS$_2$ where long-range Fr\"ohlich couplings are weak and low-energy acoustic modes dominate. The computed tensors of dynamical quadrupoles for \ce{WS2} are reported in Table~\ref{quad_tensor}. We expect corrections of similar magnitude for the related high-mobility compound WSe$_2$, which also exhibits weak Fr\"ohlich couplings. The present calculations of long-range electron-phonon couplings do not take into account free-carrier screening. Screening is expected to play a role for carrier densities of 10$^{13}$\,cm$^{-2}$, for which the Fermi level lies within the bands, and will likely weaken long-range electron-phonon scattering.

In order to check for scattering mechanisms beyond electron-phonon processes, we also perform calculations of scattering by ionized impurities in selected cases. Ionized impurities lead to strong carrier scattering due to the long-range nature of their associated Coulomb potential~\cite{Lu2022,Leveillee2023}. In Ref.~\citenum{Leveillee2023} the following ionized-impurity scattering rate was derived starting from the Kohn-Luttinger ensamble average procedure~\cite{Kohn1957}:
\begin{eqnarray}\label{eq:tauave3}
 && \ratenm^{\text{(ii)}} =
 n_{\rm imp}
  \frac{2\pi e^4 Z^2}{\hbar\,\epsilon_0^2\,\Omega}
 \nonumber \\ && \times \!\!\!\sum_{\bf G \ne -{\bf q}}
  \frac{|\langle u_{m\bk+\bq} | e^{i{\bf G}\cdot {\bf r}}|u_{n\bk} \rangle|^2}
     {|(\bq+\bG)\cdot\!\bm\epsilon^0\!\cdot(\bq+\bG)|^2}
  \,\delta(\varepsilon_{n\bk}-\varepsilon_{m\bk+\bq}),
\end{eqnarray}
where $n_{\rm imp}$ is the impurity concentration, $\epsilon_0$ and $\bm\epsilon^0$ are vacuum permittivity and the electronic dielectric permittivity tensor, respectively. $\bG$ denote reciprocal lattice vectors, and $u_{n\bk}$ is the lattice-periodic part of the Kohn-Sham state. Equation~\eqref{eq:tauave3} was derived for bulk crystals; a generalization of this equation for 2D materials should proceed along the same lines as for the Fr\"ohlich matrix element in Ref.~\citenum{Sio2022}, but is not available yet. Therefore, here we compute ionized-impurity scattering in the bulk parent compounds, and we estimate the effect of impurity scattering in 2D by using Matthiessen's rule rule~\cite{Leveillee2023}.

We take into account scattering by extended defects using the simplest possible model, whereby the relaxation time is given by the time needed for the carrier to traverse the entire defect length $L$. Given the carrier velocity ${\bf v}_{n\mathbf{k}}$, the scattering rate in this model is $\Gamma_{n\mathbf{k}}^{(\rm ed)}=|{\bf v}_{n\mathbf{k}}|/L$. This model is popular in the study of thermal transport~\cite{Mingo2005, Lindsay2009, Lindsay2010, Lee2023b}. To obtain  electron velocities, we employ Wannier-Fourier interpolation within the \texttt{EPW} code. The defect size $L$ is an external parameter of the model.

%
%
\vspace{20pt}
\noindent\textbf{Transport descriptors for HT screening}
\vspace{5pt}

Accurate mobility calculations using Eqs.~\eqref{BTE_gen} are too expensive for HT approaches. In order to perform a rapid screening large materials libraries, it is necessary to down-select the most promising materials via descriptors that are computationally affordable. In the simplest approximation, the mobility is expressed by Drude's law as $\mu = e \tau / m^*$, where $\tau$ and $m^*$ are an effective relaxation time and an average effective mass. The effective mass can in principle be obtained from the band curvature at the VBM or CBM edges; however, this textbook approach does not take into account possible band degeneracy and multi-valley band extrema. A more useful definition which automatically includes these effects is provided by the conductivity effective mass~\cite{Madsen2006,Hautier2014}:
\begin{equation}
    \left[m^*\right]^{-1}_{\alpha\beta} = 
    - \left(\sum_{n\bk}  f_{n\bk}^0\right)^{\!\!-1}\!\!\!\!
    \sum_{n\bk} v_{n\bk}^{\alpha} v_{n\bk}^{\beta} \,\frac{\partial f_{n\bk}^0}{ \partial \varepsilon_{n\bk} }.
\label{cond_effmass}
\end{equation}
This expression reduces to the textbook definition of effective mass in the case of a single parabolic band, and provide an averaged quantity in more complex scenarios. Formally, this definition can be obtained from Eqs.~\eqref{mu_BTE} and \eqref{BTE_gen} by neglecting the $\partial_{E_\beta}f_{m\bk+\bq}$ term in the latter equation, replacing $\Gamma_{n\bk\rightarrow m\bk+\bq}$ by a constant rate, and assuming Drude's formula to hold. We note that, by definition, the conductivity effective mass depends on temperature and carrier concentration.

For the average relaxation time $\tau$ to be used for HT screening, we test two possible approximations: (i)~We evaluate the relaxation time corresponding to polar phonon scattering in 2D, using a model electron-phonon matrix element derived in Ref.~\citenum{Sio2022}; (ii)~We use the much simpler Emin's formula~\cite{Emin2013}.

The scattering rate for polar phonon scattering in 2D can be derived from Fermi's Golden Rule by assuming a dispersionless polar phonon of frequency $\omega$ and an electron in a parabolic band of mass $m^*$. Using the matrix element in Eqs.~(62) and (63) of Ref.~\citenum{Sio2022}, the rate for carriers at the energy $\varepsilon$ above the CBM reads:
\begin{eqnarray}\label{eq.frohlichmodel}
   \tau^{-1}(\varepsilon) &=& 
   \frac{e^2 m^* \omega d}{8\pi\epsilon_0 \hbar^2}
    \epsilon^{\rm ion}
       \Big\{ \theta( \varepsilon\!-\!\hbar\omega) [1-f(\varepsilon\!-\!\hbar\omega)+n(\hbar\omega)]
        \nonumber \\ &\times& g(\varepsilon,\hbar\omega) 
     + [f(\varepsilon+\hbar\omega)+n(\hbar\omega)] g(\varepsilon,-\hbar\omega)
       \Big\},
\end{eqnarray}
where $d$ is the effective thickness of the 2D material, $\epsilon^{\rm ion}$ is the lattice contribution to the dielectric constant, and $f$, $n$, $\theta$ are Fermi-Dirac, Bose-Einstein, and Heaviside functions, respectively. The two terms in the square brackets in Eq.~\eqref{eq.frohlichmodel} correspond to phonon emission and absorption, respectively. The auxiliary function $g$ is:
\begin{equation}\label{eq.gxy}
    g(x,y) =
        \int_{ 2x -y  -2  \sqrt{x} \sqrt{x-y} }^{2x -y  +2  \sqrt{x} \sqrt{x-y}} 
               \frac{[4 xu- (y + u)^2]^{-1/2}}{(1+\sqrt{u/\eta})^2}du,
\end{equation}
where $\eta =\hbar^2 [4\epsilon^\infty/(2\epsilon^{\infty,2}-1)d]^2/2m^*$~\cite{Sio2022}. The quadrature in Eq.~\eqref{eq.gxy} is performed numerically. The average scattering rate needed in Drude's formula is obtained from $\tau(\varepsilon)$ in Eq.~\eqref{eq.frohlichmodel} via:
  \begin{equation}\label{eq.taufr}
  \tau = \int_0^\infty \!\! dx \,x\, e^{-x} \,\tau(x\kt).
  \end{equation}
Mobility estimates using this model only require the knowledge of the effective mass, slab thickness, dielectric constants, and phonon frequency. 

Emin's formula for the relaxation time is considerably simpler, as it does not take into account any materials parameters. It is given by~\cite{Emin2013}:
 \begin{equation}\label{eq.emin}
   \tau = \hbar/\kt .
 \end{equation}
The advantages of this expression are that (i) it establishes a simple conversion formula between effective mass and mobility, $\mu = e \hbar / m^* \kt$, (ii) it exhibits a reasonable dependence on temperature, and (iii) it does not require any materials data other than band structures.
In this manuscript we use both Eqs.~\eqref{eq.frohlichmodel} and \eqref{eq.emin} for the initial HT screening of 2D materials.

\section{Acknowledgments}
This research is supported by SUPREME, one of seven centers in JUMP 2.0, a Semiconductor Research Corporation (SRC) program sponsored by DARPA (HT search for 2D materials); and by the Computational Materials Science program of U.S. Department of Energy, Office of Science, Basic Energy Sciences under Award DE-SC0020129 (2D Fr\"ohlich implementation and \texttt{EPW} software development). Computational resources were provided by the National Energy Research Scientific Computing Center (a DOE Office of Science User Facility supported under Contract No. DE-AC02-05CH11231), the Argonne Leadership Computing Facility (a DOE Office of Science User Facility supported under Contract DE-AC02-06CH11357), and the Texas Advanced Computing Center (TACC) at The University of Texas at Austin. We acknowledge Eric Pop and Tomas Palacios for fruitful discussions and for feedback on experimental mobility data.

\section{Author Contributions}
V.-A.H. performed calculations and data analysis. F.G. designed and supervised the project. Both authors contributed to the writing of the manuscript.
\section{Competing Interests}
The authors declare no competing interests.
\section{Data Availability}
For reproducibility of transport calculations, we provide all necessary input files stored in the archive of Materials Cloud [\href{https://doi.org/10.24435/materialscloud:aw-d3}{Materials Cloud Archive 2024.154 (2024)}].
\section{Code availability}
The codes used in this work, namely EPW, Quantum ESPRESSO, Boltztrap, Wannier90, BerkeleyGW, and ABINIT, are all open-source software and are freely available on their corresponding websites.

\section{Appendices}\label{appendices}
This section provides all supplementary information for the results presented above. 

\begin{table*}
\caption{\textbf{Band gap and effective masses of materials in Set~A}. Calculated band gap and conductivity effective masses along the $x$ and $y$ directions, as well as their average, for all materials in Set~A (166 compounds). Effective masses are evaluated at room temperature, for a carrier concentration of $10^{10}$~cm$^{-2}$. Spin-orbit coupling (SOC) is not included in these calculations. Magnetic compounds appear at the bottom of the table.}
\label{tableI}
\vspace{4pt}
\centering
\begin{longtable}{@{\extracolsep{\fill}} l l r r r r r r r}
\hline\\[-8.5pt] \hline\\[-5pt]
\multicolumn{1}{l}{Unit cell formula}    & \multicolumn{1}{l}{Reduced formula}  & \multicolumn{1}{c}{Gap (eV)} & \multicolumn{3}{c}{Electron mass ($m_e$)}        & \multicolumn{3}{c}{Hole mass ($m_e$)} \\[2pt]
&&&$x$&$y$&ave.&$x$&$y$&ave.\\[2pt]
\cline{4-6} \cline{7-9}\\[-7pt]
\ce{Ag2Br2}   &  \ce{AgBr}   &  1.302  &  0.296  &  0.305  &  0.300  &  0.430  &  0.676  &  0.553\\
\ce{Ag2I2}   &  \ce{AgI}   &  2.176  &  0.461  &  0.461  &  0.461  &  1.040  &  1.040 &  1.040\\
\ce{AgNO2}   &  \ce{AgNO2}   &  1.663  &  0.826  &  2.483  &  1.655  &  0.586  &  2.253  &  1.419\\
\ce{AgO4Cl}   &  \ce{AgO4Cl}   &  2.966  &  0.562  &  0.562  &  0.562  &  1.178  &  1.178  &  1.178\\
\ce{Al2O2Cl2}   &  \ce{AlOCl}   &  5.759  &  0.393  &  0.538  &  0.466  &  0.944  &  3.360  &  2.152\\
\ce{As4}   &  \ce{As}   &  0.755  &  0.270  &  1.217  &  0.744  &  0.118  &  0.371  &  0.244\\
\ce{Au2Br2}   &  \ce{AuBr}   &  2.031  &  1.784  &  2.324  &  2.054  &  0.474  &  0.736  &  0.605\\
\ce{Au2I2}   &  \ce{AuI}   &  1.979  &  0.788  &  4.118  &  2.453  &  0.317  &  0.852  &  0.585\\
\ce{Au2Se2}   &  \ce{AuSe}   &  1.172  &  0.367  &  1.921  &  1.144  &  0.943  &  1.838  &  1.391\\
\ce{BN}   &  \ce{BN}   &  4.624  &  0.956  &  0.956  &  0.956  &  0.665  &  0.665  &  0.665\\
\ce{Ba2H2I2}   &  \ce{BaHI}   &  3.339  &  0.449  &  0.449  &  0.449  &  2.282  &  2.282  &  2.282\\
\ce{Bi2}   &  \ce{Bi}   &  0.561  &  0.055  &  0.055  &  0.055  &  0.379  &  0.379  &  0.379\\
\ce{Bi2O2Br2}   &  \ce{BiOBr}   &  2.451  &  0.242  &  0.242  &  0.242  &  1.342  &  1.342  &  1.342\\
\ce{Bi2O2Cl2}   &  \ce{BiOCl}   &  2.757  &  0.285  &  0.285  &  0.285  &  0.998  &  0.998  &  0.998\\
\ce{Bi2O2I2}   &  \ce{BiOI}   &  1.503  &  0.194  &  0.194  &  0.194  &  0.510  &  0.510  &  0.510\\
\ce{Bi2Se3}   &  \ce{Bi2Se3}   &  0.937  &  0.128  &  0.128  &  0.128  &  0.994  &  0.994  &  0.994\\
\ce{Bi2Te2S}   &  \ce{Bi2Te2S}   &  1.007  &  0.118  &  0.118  &  0.118  &  0.359  &  0.359  &  0.359\\
\ce{Bi2Te2S}   &  \ce{Bi2Te2S}   &  1.170  &  0.185  &  0.185  &  0.185  &  0.707  &  0.707  &  0.707\\
\ce{Bi2Te2Se}   &  \ce{Bi2Te2Se}   &  0.933  &  0.114  &  0.114  &  0.114  &  0.383  &  0.383  &  0.383\\
\ce{Bi2Te3}   &  \ce{Bi2Te3}   &  1.025  &  0.113  &  0.113  &  0.113  &  1.040  &  1.040  &  1.040\\
\ce{Bi2TeSe2}   &  \ce{Bi2TeSe2}   &  1.150  &  0.109  &  0.109  &  0.109  &  1.490  &  1.490  &  1.490\\
\ce{BiTeCl}   &  \ce{BiTeCl}   &  1.830  &  0.169  &  0.169  &  0.169  &  1.748  &  1.748  &  1.748\\
\ce{BiTeI}   &  \ce{BiTeI}   &  1.575  &  0.219  &  0.219  &  0.219  &  0.856  &  0.856  &  0.856\\
\ce{CaH2O2}   &  \ce{CaH2O2}   &  3.703  &  0.764  &  0.764  &  0.764  &  2.335  &  2.335  &  2.335\\
\ce{Ca2H2Br2}   &  \ce{CaHBr}   &  4.220  &  0.628  &  0.628  &  0.628  &  3.649  &  3.649  &  3.649\\
\ce{Ca2H2I2}   &  \ce{CaHI}   &  3.691  &  0.716  &  0.716  &  0.716  &  0.588  &  0.588  &  0.588\\
\ce{CaI2}   &  \ce{CaI2}   &  3.816  &  0.658  &  0.658  &  0.658  &  1.013  &  1.013  &  1.013\\
\ce{CdH2O2}   &  \ce{CdH2O2}   &  2.250  &  0.444  &  0.444  &  0.444  &  1.853  &  1.853  &  1.853\\
\ce{CdBr2}   &  \ce{CdBr2}   &  3.214  &  0.398  &  0.398  &  0.398  &  1.825  &  1.825  &  1.825\\
\ce{CdCl2}   &  \ce{CdCl2}   &  3.876  &  0.496  &  0.496  &  0.496  &  2.574  &  2.574  &  2.574\\
\ce{CdI2}   &  \ce{CdI2}   &  2.383  &  0.504  &  0.504  &  0.504  &  1.293  &  1.293  &  1.293\\
\ce{Cu2Br2}   &  \ce{CuBr}   &  1.642  &  0.434  &  0.434  &  0.434  &  1.95  &  1.950  &  1.950\\
\ce{Cu2I2}   &  \ce{CuI}   &  2.105  &  0.354  &  0.354  &  0.354  &  1.773  &  1.773  &  1.773\\
\ce{Cu2I2}   &  \ce{CuI}   &  2.195  &  0.364  &  0.389  &  0.377  &  0.908  &  1.233  &  1.071\\
\ce{Cu4Te2}   &  \ce{Cu2Te}   &  0.251  &  0.386  &  0.672  &  0.529  &  0.361  &  0.448  &  0.404\\
\ce{Ga2Ge2Te2}   &  \ce{GaGeTe}   &  0.752  &  0.065  &  0.065  &  0.065  &  1.602  &  1.602  &  1.602\\
\ce{Ga2S2}   &  \ce{GaS}   &  2.212  &  0.223  &  0.223  &  0.223  &  3.503  &  3.503  &  3.503\\
\ce{Ga2S2}   &  \ce{GaS}   &  2.341  &  0.236  &  0.236  &  0.236  &  3.372  &  3.372  &  3.372\\
\ce{Ga2Se2}   &  \ce{GaSe}   &  1.785  &  0.168  &  0.168  &  0.168  &  2.867  &  2.867  &  2.867\\
\ce{Ga2Te2}   &  \ce{GaTe}   &  1.408  &  0.294  &  0.294  &  0.294  &  2.111  &  2.111  &  2.111\\
\ce{Ga2Te2Cl2}   &  \ce{GaTeCl}   &  2.259  &  0.241  &  0.764  &  0.502  &  0.182  &  1.060  &  0.621\\
\ce{Ge2S2}   &  \ce{GeS}   &  1.660  &  0.226  &  0.464  &  0.345  &  0.267  &  0.594  &  0.430\\
\ce{GeI2}   &  \ce{GeI2}   &  2.045  &  0.715  &  0.715  &  0.715  &  0.677  &  0.677  &  0.677\\
\ce{GeI2}   &  \ce{GeI2}   &  2.150  &  0.259  &  0.259  &  0.259  &  0.929  &  0.929  &  0.929\\
\ce{Hf2N2Br2}   &  \ce{HfNBr}   &  2.171  &  0.321  &  0.368  &  0.345  &  0.351  &  0.535  &  0.443\\
\ce{Hf2N2Br2}   &  \ce{HfNBr}   &  2.049  &  0.648  &  0.648  &  0.648  &  1.779  &  1.779  &  1.779\\
\ce{Hf2N2Cl2}   &  \ce{HfNCl}   &  2.386  &  0.635  &  0.635  &  0.635  &  3.200  &  3.200  &  3.200\\
\ce{Hf2N2I2}   &  \ce{HfNI}   &  0.981  &  0.698  &  0.698  &  0.698  &  0.592  &  0.592  &  0.592\\
\ce{HfS2}   &  \ce{HfS2}   &  1.265  &  0.440  &  0.440  &  0.440  &  0.368  &  0.368  &  0.368\\
\ce{HfSe2}   &  \ce{HfSe2}   &  0.630  &  0.353  &  0.353  &  0.353  &  0.271  &  0.271  &  0.271\\
\ce{HgI2}   &  \ce{HgI2}   &  1.804  &  0.518  &  0.518  &  0.518  &  1.153  &  1.153  &  1.153\\
\ce{In2O2Br2}   &  \ce{InOBr}   &  2.219  &  0.338  &  0.421  &  0.380  &  0.484  &  2.244  &  1.364\\
\ce{In2Se2}   &  \ce{InSe}   &  1.480  &  0.199  &  0.199  &  0.199  &  3.746  &  3.746  &  3.746\\
\ce{In2Se2}   &  \ce{InSe}   &  1.687  &  0.236  &  3.190  &  1.713  &  0.369  &  4.043  &  2.206\\
\ce{In2Se3}   &  \ce{In2Se3}   &  0.984  &  0.526  &  0.539  &  0.532  &  0.627  &  1.347  &  0.987\\
\ce{K2Ag2Se2}   &  \ce{KAgSe}   &  0.575  &  0.243  &  0.243  &  0.243  &  0.647  &  0.647  &  0.647\\
\ce{K2Tl2O2}   &  \ce{KTlO}   &  1.537  &  0.511  &  0.644  &  0.578  &  2.695  &  12.025  &  7.360\\[5pt]
\hline\\[-8.5pt] \hline\\[-7pt]
\end{longtable}
\end{table*}
\begin{table*}
\centering
\begin{longtable}{@{\extracolsep{\fill}} l l r r r r r r r}
\hline\\[-8.5pt] \hline\\[-7pt]
\multicolumn{1}{l}{Unit cell formula}    & \multicolumn{1}{l}{Reduced formula}  & \multicolumn{1}{c}{Gap (eV)} & \multicolumn{3}{c}{Electron mass ($m_e$)}        & \multicolumn{3}{c}{Hole mass ($m_e$)} \\[2pt]
&&&$x$&$y$&ave.&$x$&$y$&ave.\\[2pt]
\cline{4-6} \cline{7-9}\\[-5pt]
\ce{La2GeI2}   &  \ce{La2GeI2}   &  0.360  &  0.282  &  0.282  &  0.282  &  0.828  &  0.828  &  0.828\\
\ce{La2O2Br2}   &  \ce{LaOBr}   &  4.022  &  0.842  &  0.842  &  0.842  &  1.814  &  1.814  &  1.814\\
\ce{La2O2I2}   &  \ce{LaOI}   &  3.399  &  0.757  &  0.757  &  0.757  &  0.725  &  0.725  &  0.725\\
\ce{Li2H2O2}   &  \ce{LiHO}   &  3.932  &  0.881  &  0.881  &  0.881  &  3.339  &  3.339  &  3.339\\
\ce{LiAlTe2}   &  \ce{LiAlTe2}   &  0.918  &  0.158  &  0.158  &  0.158  &  0.869  &  0.869  &  0.869\\
\ce{LiAuI4}   &  \ce{LiAuI4}   &  0.937  &  0.864  &  19.359  &  10.112  &  1.448  &  2.424  &  1.936\\
\ce{LiBH4}   &  \ce{LiBH4}   &  6.344  &  0.925  &  1.003  &  0.964  &  1.456  &  2.379  &  1.917\\
\ce{Lu2CCl2}   &  \ce{Lu2CCl2}   &  0.947  &  0.334  &  0.334  &  0.334  &  1.731  &  1.731  &  1.731\\
\ce{Lu2O2Br2}   &  \ce{LuOBr}   &  4.364  &  0.360  &  0.360  &  0.360  &  2.326  &  2.326  &  2.326\\
\ce{Lu2O2I2}   &  \ce{LuOI}   &  3.255  &  1.387  &  1.387  &  1.387  &  0.480  &  0.480  &  0.480\\
\ce{MgO2H2}   &  \ce{MgH2O2}   &  3.330  &  0.745  &  0.745  &  0.745  &  3.281  &  3.281  &  3.281\\
\ce{MgBr2}   &  \ce{MgBr2}   &  4.780  &  0.333  &  0.333  &  0.333  &  1.962  &  1.962  &  1.962\\
\ce{MgCl2}   &  \ce{MgCl2}   &  6.017  &  0.417  &  0.417  &  0.417  &  2.575  &  2.575  &  2.575\\
\ce{MgI2}   &  \ce{MgI2}   &  3.579  &  0.422  &  0.422  &  0.422  &  1.471  &  1.471  &  1.471\\
\ce{MoS2}   &  \ce{MoS2}   &  1.631  &  0.486  &  0.486  &  0.486  &  1.980  &  1.980  &  1.980\\
\ce{MoSe2}   &  \ce{MoSe2}   &  1.489  &  0.608  &  0.608  &  0.608  &  0.763  &  0.763  &  0.763\\
\ce{MoTe2}   &  \ce{MoTe2}   &  1.081  &  0.660  &  0.660  &  0.660  &  0.833  &  0.833  &  0.833\\
\ce{Na2H2O2}   &  \ce{NaHO}   &  2.776  &  0.707  &  0.709  &  0.708  &  1.640  &  3.918  &  2.779\\
\ce{NaCN}   &  \ce{NaCN}   &  4.870  &  0.704  &  0.876  &  0.790  &  1.160  &  5.417  &  3.288\\
\ce{NiO2}   &  \ce{NiO2}   &  1.297  &  1.191  &  1.191  &  1.191  &  4.109  &  4.109  &  4.109\\
\ce{P2}   &  \ce{P}   &  1.907  &  0.245  &  0.245  &  0.245  &  1.452  &  1.452  &  1.452\\
\ce{P4}   &  \ce{P}   &  0.892  &  0.161  &  1.277  &  0.719  &  0.154  &  3.287  &  1.720\\
\ce{Pb2F2Br2}   &  \ce{PbFBr}   &  3.038  &  0.439  &  0.439  &  0.439  &  1.205  &  1.205  &  1.205\\
\ce{Pb2F2Cl2}   &  \ce{PbFCl}   &  3.535  &  0.566  &  0.566  &  0.566  &  1.299  &  1.299  &  1.299\\
\ce{Pb2F2I2}   &  \ce{PbFI}   &  2.312  &  0.309  &  0.309  &  0.309  &  0.550  &  0.550  &  0.550\\
\ce{Pb2O2}   &  \ce{PbO}   &  2.441  &  0.476  &  0.476  &  0.476  &  6.001  &  6.001  &  6.001\\
\ce{PbF4}   &  \ce{PbF4}   &  2.420  &  1.049  &  1.049  &  1.049  &  13.271  &  13.271  &  13.271\\
\ce{PbI2}   &  \ce{PbI2}   &  2.542  &  0.299  &  0.299  &  0.299  &  1.473  &  1.473  &  1.473\\
\ce{PbTe}   &  \ce{PbTe}   &  1.617  &  0.307  &  0.307  &  0.307  &  0.532  &  0.532  &  0.532\\
\ce{Pd2Cl4}   &  \ce{PdCl2}   &  0.841  &  1.350  &  1.782  &  1.566  &  1.039  &  2.051  &  1.545\\
\ce{Pd2S4}   &  \ce{PdS2}   &  1.163  &  0.558  &  1.486  &  1.022  &  1.843  &  5.879  &  3.861\\
\ce{PtO2}   &  \ce{PtO2}   &  1.691  &  1.849  &  1.849  &  1.849  &  2.804  &  2.804  &  2.804\\
\ce{PtS2}   &  \ce{PtS2}   &  1.791  &  0.420  &  0.420  &  0.420  &  1.311  &  1.311  &  1.311\\
\ce{PtSe2}   &  \ce{PtSe2}   &  1.350  &  0.326  &  0.326  &  0.326  &  2.978  &  2.978  &  2.978\\
\ce{Rb2Cl2}   &  \ce{RbCl}   &  4.611  &  0.700  &  0.700  &  0.700  &  4.982  &  4.982  &  4.982\\
\ce{Rh2Te2Cl2}   &  \ce{RhTeCl}   &  1.111  &  0.381  &  0.770  &  0.576  &  2.684  &  4.414  &  3.549\\
\ce{Sb2}   &  \ce{Sb}   &  1.232  &  0.230  &  0.230  &  0.230  &  0.347  &  0.347  &  0.347\\
\ce{Sb2Te2I2}   &  \ce{SbTeI}   &  0.879  &  0.226  &  0.250  &  0.238  &  0.578  &  2.262  &  1.420\\
\ce{Sb2Te2Se}   &  \ce{Sb2Te2Se}   &  0.658  &  0.123  &  0.123  &  0.123  &  0.374  &  0.374  &  0.374\\
\ce{Sb2Te3}   &  \ce{Sb2Te3}   &  0.676  &  0.132  &  0.132  &  0.132  &  0.544  &  0.544  &  0.544\\
\ce{Sb2TeSe2}   &  \ce{Sb2TeSe2}   &  0.744  &  0.105  &  0.105  &  0.105  &  0.815  &  0.815  &  0.815\\
\ce{Sb2TeSe2}   &  \ce{Sb2TeSe2}   &  0.711  &  0.157  &  0.157  &  0.157  &  0.582  &  0.582  &  0.582\\
\ce{Sc2CCl2}   &  \ce{Sc2CCl2}   &  0.876  &  0.436  &  0.436  &  0.436  &  1.279  &  1.279  &  1.279\\
\ce{Sc2O2Br2}   &  \ce{ScOBr}   &  3.221  &  0.814  &  2.101  &  1.458  &  0.767  &  2.908  &  1.837\\
\ce{Si2H2}   &  \ce{SiH}   &  2.190  &  0.235  &  0.235  &  0.235  &  0.392  &  0.392  &  0.392\\
\ce{Sn2O2}   &  \ce{SnO}   &  2.110  &  0.560  &  0.692  &  0.626  &  0.399  &  3.981  &  2.190\\
\ce{Sn2O2}   &  \ce{SnO}   &  3.005  &  0.884  &  0.884  &  0.884  &  6.902  &  6.902  &  6.902\\
\ce{SnF4}   &  \ce{SnF4}   &  3.921  &  0.798  &  0.798  &  0.798  &  14.07  &  14.07  &  14.07\\
\ce{SnS2}   &  \ce{SnS2}   &  1.506  &  0.400  &  0.400  &  0.400  &  0.593  &  0.593  &  0.593\\
\ce{SnSe2}   &  \ce{SnSe2}   &  0.756  &  0.344  &  0.344  &  0.344  &  0.489  &  0.489  &  0.489\\
\ce{Sr2Br2F2}   &  \ce{SrBrF}   &  5.283  &  0.410  &  0.410  &  0.410  &  1.812  &  1.812  &  1.812\\
\ce{Sr2H2Br2}   &  \ce{SrHBr}   &  4.354  &  0.530  &  0.530  &  0.530  &  5.633  &  5.633  &  5.633\\
\ce{Sr2H2I2}   &  \ce{SrHI}   &  3.908  &  0.456  &  0.456  &  0.456  &  2.585  &  2.585  &  2.585\\
\ce{Sr2I4}   &  \ce{SrI2}   &  3.813  &  0.630  &  1.007  &  0.819  &  0.285  &  11.245  &  5.765\\
\ce{Ti2N2Br2}   &  \ce{TiNBr}   &  0.616  &  0.193  &  0.370  &  0.281  &  0.186  &  0.362  &  0.274\\
\ce{Ti2N2Cl2}   &  \ce{TiNCl}   &  0.619  &  0.218  &  0.452  &  0.335  &  0.193  &  0.300  &  0.246\\
\ce{TiS2}   &  \ce{TiS2}   &  0.045  &  0.519  &  0.519  &  0.519  &  0.157  &  0.157  &  0.157\\
\ce{Tl2F2}   &  \ce{TlF}   &  3.612  &  0.387  &  0.387  &  0.387  &  1.582  &  1.582  &  1.582\\
\ce{Tl2O}   &  \ce{Tl2O}   &  1.044  &  0.161  &  0.161  &  0.161  &  0.343  &  0.343  &  0.343\\
\ce{Tl2S}   &  \ce{Tl2S}   &  1.386  &  0.305  &  0.305  &  0.305  &  2.491  &  2.491  &  2.491\\[5pt]
\hline\\[-8.5pt] \hline\\[-7pt]
\end{longtable}
\end{table*}
%
\begin{table*}
\centering
\begin{longtable}{@{\extracolsep{\fill}} l l r r r r r r r}
\hline\\[-8.5pt] \hline\\[-7pt]
\multicolumn{1}{l}{Unit cell formula}    & \multicolumn{1}{l}{Reduced formula}  & \multicolumn{1}{c}{Gap (eV)} & \multicolumn{3}{c}{Electron mass ($m_e$)}        & \multicolumn{3}{c}{Hole mass ($m_e$)} \\[2pt]
&&&$x$&$y$&ave.&$x$&$y$&ave.\\[2pt]
\cline{4-6} \cline{7-9}\\[-5pt]
\ce{WS2}   &  \ce{WS2}   &  1.848  &  0.329  &  0.329  &  0.329  &  0.640  &  0.640  &  0.640\\
\ce{WSe2}   &  \ce{WSe2}   &  1.600  &  0.394  &  0.394  &  0.394  &  0.513  &  0.513  &  0.513\\
\ce{WTe2}   &  \ce{WTe2}   &  1.081  &  0.349  &  0.349  &  0.349  &  0.495  &  0.495  &  0.495\\
\ce{Y2O2Cl2}   &  \ce{YOCl}   &  4.381  &  0.889  &  0.889  &  0.889  &  3.403  &  3.403  &  3.403\\
\ce{YbI2}   &  \ce{YbI2}   &  0.716  &  0.455  &  0.455  &  0.455  &  33.205  &  33.205  &  33.205\\
\ce{ZnBr2}   &  \ce{ZnBr2}   &  3.429  &  0.549  &  0.549  &  0.549  &  1.222  &  1.222  &  1.222\\
\ce{ZnCl2}   &  \ce{ZnCl2}   &  4.252  &  0.510  &  0.510  &  0.510  &  1.807  &  1.807  &  1.807\\
\ce{ZnCl2}   &  \ce{ZnCl2}   &  4.460  &  0.455  &  0.455  &  0.455  &  1.711  &  1.711  &  1.711\\
\ce{ZnI2}   &  \ce{ZnI2}   &  1.992  &  0.407  &  0.407  &  0.407  &  0.689  &  0.689  &  0.689\\
\ce{Zr2I4}   &  \ce{ZrI2}   &  0.413  &  0.383  &  2.224  &  1.304  &  1.097  &  1.189  &  1.143\\
\ce{Zr2N2Br2}   &  \ce{ZrNBr}   &  1.822  &  0.337  &  0.34  &  0.339  &  0.374  &  0.868  &  0.621\\
\ce{Zr2N2Br2}   &  \ce{ZrNBr}   &  1.667  &  0.579  &  0.579  &  0.579  &  2.155  &  2.155  &  2.155\\
\ce{Zr2N2Cl2}   &  \ce{ZrNCl}   &  1.925  &  0.579  &  0.579  &  0.579  &  2.199  &  2.199  &  2.199\\
\ce{Zr2N2I2}   &  \ce{ZrNI}   &  1.297  &  0.294  &  0.302  &  0.298  &  0.196  &  1.838  &  1.017\\
\ce{Zr2N2I2}   &  \ce{ZrNI}   &  0.734  &  0.606  &  0.606  &  0.606  &  0.621  &  0.621  &  0.621\\
\ce{ZrCl2}   &  \ce{ZrCl2}   &  0.996  &  1.274  &  1.274  &  1.274  &  0.481  &  0.481  &  0.481\\
\ce{ZrNBr}   &  \ce{ZrNBr}   &  2.818  &  0.543  &  0.543  &  0.543  &  0.695  &  0.695  &  0.695\\
\ce{ZrNCl}   &  \ce{ZrNCl}   &  3.258  &  0.555  &  0.555  &  0.555  &  2.194  &  2.194  &  2.194\\
\ce{ZrS2}   &  \ce{ZrS2}   &  1.149  &  0.520  &  0.520  &  0.520  &  0.351  &  0.351  &  0.351\\
\ce{ZrSe2}   &  \ce{ZrSe2}   &  0.498  &  0.399  &  0.399  &  0.399  &  0.251  &  0.251  &  0.251\\
\ce{Ge2Se2}   &  \ce{GeSe}   &  1.054  &  0.193  &  0.305  &  0.249  &  0.170  &  0.300  &  0.235\\
\ce{KTlCl4}   &  \ce{KTlCl4}   &  2.770  &  2.654  &  2.654  &  2.654  &  26.038  &  26.038  &  26.038\\
\ce{Sn2Te2}   &  \ce{SnTe}   &  0.683  &  0.226  &  0.241  &  0.233  &  0.084  &  0.148  &  0.116\\
\ce{FeCl2}   &  \ce{FeCl2}   &  0.853  &  1.660  &  1.660  &  1.660  &  5.391  &  5.391  &  5.391\\[2pt]
\hline\\[-8.5pt]
magnetic compounds & & & & & & & & \\[3pt]
\ce{CdOCl}   &  \ce{CdClO}   &  0.183  &  1.987  &  1.987  &  1.987  &  2.246  &  2.246  &  2.246\\
\ce{CoBr2}   &  \ce{CoBr2}   &  0.200  &  2.732  &  2.732  &  2.732  &  5.358  &  5.358  &  5.358\\
\ce{CoCl2}   &  \ce{CoCl2}   &  0.176  &  2.565  &  2.565  &  2.565  &  4.633  &  4.633  &  4.633\\
\ce{CuCl2}   &  \ce{CuCl2}   &  0.233  &  0.631  &  284.576  &  142.603  &  0.414  &  263.509  &  131.962\\
\ce{Er2S2Cl2}   &  \ce{ErSCl}   &  0.197  &  4.382  &  19.902  &  12.142  &  0.303  &  1.948  &  1.125\\
\ce{Ho2S2I2}   &  \ce{HoSI}   &  0.519  &  9.297  &  19.138  &  14.218  &  0.367  &  1.093  &  0.730\\
\ce{NiBr2}   &  \ce{NiBr2}   &  0.761  &  1.456  &  1.456  &  1.456  &  1.516  &  1.516  &  1.516\\
\ce{NiCl2}   &  \ce{NiCl2}   &  1.098  &  2.036  &  2.036  &  2.036  &  1.732  &  1.732  &  1.732\\
\ce{NiI2}   &  \ce{NiI2}   &  0.324  &  1.036  &  1.036  &  1.036  &  1.144  &  1.144  &  1.144\\
\ce{Cr2O2Br2}   &  \ce{CrOBr}   &  0.488  &  0.908  &  1.198  &  1.053  &  0.367  &  3.686  &  2.026\\
\ce{Cr2O2Cl2}   &  \ce{CrOCl}   &  0.640  &  0.794  &  1.824  &  1.309  &  0.393  &  3.238  &  1.816\\
\ce{Cr2S2Br2}   &  \ce{CrSBr}   &  0.442  &  0.225  &  6.754  &  3.489  &  0.298  &  3.044  &  1.671\\
\ce{LaBr2}   &  \ce{LaBr2}   &  0.628  &  2.770  &  2.770  &  2.770  &  0.916  &  0.916  &  0.916\\
\ce{La2Br2}   &  \ce{LaBr}   &  0.583  &  1.953 &  1.953 &  1.953  &  2.651  &  2.651  &  2.651\\
\ce{CrBr2}   &  \ce{CrBr2}   &  0.822  &  0.879  &  127.796  &  64.338  &  2.127  &  28.718  &  15.423\\
\ce{CrI2}   &  \ce{CrI2}   &  0.704  &  0.600  &  30.442  &  15.521  &  4.038  &  23.220  &  13.629\\
\ce{MnBr2}   &  \ce{MnBr2}   &  1.886  &  0.813 &  603.380  &  302.097  &  1.947  &  3.422 & 2.685\\
\ce{MnCl2}   &  \ce{MnCl2}   &  2.094  & 0.975 & 213.706 & 107.340 & 3.260 & 4.893 & 4.077\\
\ce{MnI2}   &  \ce{MnI2}   &  1.434  &  0.603  &  81.015  &  40.809  &  0.460  &  4.173  &  2.317\\
\ce{MnH2O2}   &  \ce{MnH2O2}   &  0.528  &  0.665  &  0.676  &  0.670  &  0.901  &  24.358  &  12.629\\
\ce{VBr2}   &  \ce{VBr2}   &  1.234  &  1.215  &  1.953  &  1.584  &  3.710  &  9.013  &  6.362\\
\ce{VCl2}   &  \ce{VCl2}   &  1.364  &  1.501  &  3.079  &  2.290  &  3.361  &  17.908  &  10.635\\
\ce{VI2}   &  \ce{VI2}   &  1.155  & 1.001 & 1.249 & 1.125 & 3.442 & 3.537 & 3.490\\
\ce{VOBr2}   &  \ce{VOBr2}   &  0.876  &  1.030  &  3.608  &  2.319  &  1.662  &  22.135  &  11.899\\
\ce{VOCl2}   &  \ce{VCl2O}   &  0.911  &  1.075  &  3.532  &  2.303  &  1.610  &  22.886  &  12.248\\[5pt]
\hline\\[-8.5pt] \hline
\end{longtable}
\end{table*}
\setcounter{table}{2}
%
\begin{table*}
\caption{\textbf{Band gap and effective masses of materials in Set~B}. Calculated band gap and conductivity effective masses along the $x$ and $y$ directions, as well as their average, for all materials in Set~B (95 compounds). Effective masses are evaluated at room temperature, for a carrier concentration of $10^{10}$~cm$^{-2}$. Materials with both hole and electron effective masses $m_{\rm e}^*< m_e$ and $m_{\rm h}^*<m_e$, which constitute Set~C, are highlighted in bold font. These calculations include SOC.} \label{tableS2}
\vspace{4pt}
\centering
\begin{longtable}{@{\extracolsep{\fill}} l l r r r r r r r}
\hline\\[-8.5pt] \hline\\[-5pt]
\multicolumn{1}{l}{Unit cell formula}    & \multicolumn{1}{l}{Reduced formula}  & \multicolumn{1}{c}{Gap (eV)} & \multicolumn{3}{c}{Electron mass ($m_e$)}        & \multicolumn{3}{c}{Hole mass ($m_e$)} \\[2pt]
&&&$x$&$y$&ave.&$x$&$y$&ave.\\[2pt]
\cline{4-6} \cline{7-9}\\[-7pt]
\ce{Ag2Br2}   &  \ce{AgBr}   &  1.291  &  0.300  &  0.312  &  \textbf{0.306}  &  0.464  &  0.717  &  \textbf{0.591}\\
\ce{As4}   &  \ce{As}   &  0.772  &  0.261  &  1.188  &  \textbf{0.724}  &  0.121  &  0.374  &  \textbf{0.247}\\
\ce{Bi2}   &  \ce{Bi}   &  0.465  &  0.267  &  0.267  &  \textbf{0.267}  &  0.639  &  0.639  &  \textbf{0.639}\\
\ce{Bi2O2I2}   &  \ce{BiOI}   &  1.150  &  0.218  &  0.218  &  \textbf{0.218}  &  0.619  &  0.619  &  \textbf{0.619}\\
\ce{Bi2Te2S}   &  \ce{Bi2Te2S}   &  0.274  &  0.105  &  0.105  &  \textbf{0.105}  &  0.518  &  0.518  &  \textbf{0.518}\\
\ce{Bi2Te2S}   &  \ce{Bi2Te2S}   &  0.257  &  0.142  &  0.142  &  \textbf{0.142}  &  0.757  &  0.757  &  \textbf{0.757}\\
\ce{Bi2Te2Se}   &  \ce{Bi2Te2Se}   &  0.230  &  0.105  &  0.105  &  \textbf{0.105}  &  0.673  &  0.673  &  \textbf{0.673}\\
\ce{BiTeI}   &  \ce{BiTeI}   &  0.615  &  0.418  &  0.418  &  0.418  &  1.039  &  1.039  &  1.039\\
\ce{Ga2Te2Cl2}   &  \ce{GaTeCl}   &  2.158  &  0.259  &  1.372  &  \textbf{0.816}  &  0.356  &  1.040  &  \textbf{0.698}\\
\ce{Ge2S2}   &  \ce{GeS}   &  1.753  &  0.251  &  0.612  &  \textbf{0.431}  &  0.378  &  1.364  &  \textbf{0.871}\\
\ce{Ge2Se2}   &  \ce{GeSe}   &  1.142  &  0.165  &  0.272  &  \textbf{0.219}  &  0.187  &  0.303  &  \textbf{0.245}\\
\ce{GeI2}   &  \ce{GeI2}   &  1.843  &  0.753  &  0.753  &  \textbf{0.753}  &  0.563  &  0.563  &  \textbf{0.563}\\
\ce{Hf2N2Br2}   &  \ce{HfNBr}   &  2.178  &  0.324  &  0.376  &  \textbf{0.350}  &  0.355  &  0.542  &  \textbf{0.449}\\
\ce{Hf2N2I2}   &  \ce{HfNI}   &  0.593  &  0.694  &  0.694  &  \textbf{0.694}  &  0.373  &  0.373  &  \textbf{0.373}\\
\ce{HfS2}   &  \ce{HfS2}   &  1.267  &  0.456  &  0.456  &  \textbf{0.456}  &  0.327  &  0.327  &  \textbf{0.327}\\
\ce{HfSe2}   &  \ce{HfSe2}   &  0.483  &  0.361  &  0.361  &  \textbf{0.361}  &  0.201  &  0.201  &  \textbf{0.201}\\
\ce{K2Ag2Se2}   &  \ce{KAgSe}   &  0.492  &  0.238  &  0.238  &  \textbf{0.238}  &  0.548  &  0.548  &  \textbf{0.548}\\
\ce{LiAlTe2}   &  \ce{LiAlTe2}   &  0.642  &  0.189  &  0.189  &  \textbf{0.189}  &  0.561  &  0.561  &  \textbf{0.561}\\
\ce{MoS2}   &  \ce{MoS2}   &  1.617  &  0.497  &  0.497  &  \textbf{0.497}  &  0.865  &  0.865  &  \textbf{0.865}\\
\ce{MoSe2}   &  \ce{MoSe2}   &  1.335  &  0.583  &  0.583  &  \textbf{0.583}  &  0.675  &  0.675  &  \textbf{0.675}\\
\ce{MoTe2}   &  \ce{MoTe2}   &  0.950  &  0.610  &  0.610  &  \textbf{0.610}  &  0.756  &  0.756  &  \textbf{0.756}\\
\ce{Pb2F2I2}   &  \ce{PbFI}   &  1.922  &  0.371  &  0.371  &  \textbf{0.371}  &  0.955  &  0.955  &  \textbf{0.955}\\
\ce{PbTe}   &  \ce{PbTe}   &  1.070  &  0.172  &  0.172  &  \textbf{0.172}  &  0.493  &  0.493  &  \textbf{0.493}\\
\ce{Sb2}   &  \ce{Sb}   &  1.022  &  0.235  &  0.235  &  \textbf{0.235}  &  0.156  &  0.156  &  \textbf{0.156}\\
\ce{Sb2Te2Se}   &  \ce{Sb2Te2Se}   &  0.445  &  0.114  &  0.114  &  \textbf{0.114}  &  0.475  &  0.475  &  \textbf{0.475}\\
\ce{Sb2Te3}   &  \ce{Sb2Te3}   &  0.405  &  0.151  &  0.151  & \textbf{0.151}  &  0.666  &  0.666  &  \textbf{0.666}\\
\ce{Sb2TeSe2}   &  \ce{Sb2TeSe2}   &  0.596  &  0.153  &  0.153  &  \textbf{0.153}  &  0.954  &  0.954  &  \textbf{0.954}\\
\ce{Sb2TeSe2}   &  \ce{Sb2TeSe2}   &  0.525  &  0.320  &  0.320  &  \textbf{0.320}  &  0.651  &  0.651  &  \textbf{0.651}\\
\ce{Si2H2}   &  \ce{SiH}   &  2.170  &  0.234  &  0.234  &  \textbf{0.234}  &  0.418  &  0.418  &  \textbf{0.418}\\
\ce{Sn2Te2}   &  \ce{SnTe}   &  0.657  &  0.224  &  0.259  &  \textbf{0.241}  &  0.110  &  0.156  &  \textbf{0.133}\\
\ce{SnS2}   &  \ce{SnS2}   &  1.558  &  0.403  &  0.403  &  \textbf{0.403}  &  0.582  &  0.582  &  \textbf{0.582}\\
\ce{SnSe2}   &  \ce{SnSe2}   &  0.752  &  0.354  &  0.354  &  \textbf{0.354}  &  0.633  &  0.633  &  \textbf{0.633}\\
\ce{Ti2N2Br2}   &  \ce{TiNBr}   &  0.621  &  0.191  &  0.366  &  \textbf{0.278}  &  0.187  &  0.368  &  \textbf{0.277}\\
\ce{Ti2N2Cl2}   &  \ce{TiNCl}   &  0.625  &  0.219  &  0.451  &  \textbf{0.335}  &  0.192  &  0.300  &  \textbf{0.246}\\
\ce{Tl2O}   &  \ce{Tl2O}   &  0.934  &  0.159  &  0.159  &  \textbf{0.159}  &  0.325  &  0.325  &  \textbf{0.325}\\
\ce{WS2}   &  \ce{WS2}   &  1.584  &  0.384  &  0.384  &  \textbf{0.384}  &  0.391  &  0.391  &  \textbf{0.391}\\
\ce{WSe2}   &  \ce{WSe2}   &  1.260  &  0.479  &  0.479  &  \textbf{0.479}  &  0.407  &  0.407  &  \textbf{0.407}\\
\ce{WTe2}   &  \ce{WTe2}   &  0.765  &  0.401  &  0.401  &  \textbf{0.401}  &  0.373  &  0.373  &  \textbf{0.373}\\
\ce{ZnI2}   &  \ce{ZnI2}   &  1.793  &  0.424  &  0.424  &  \textbf{0.424}  &  0.574  &  0.574  &  \textbf{0.574}\\
\ce{Zr2N2Br2}   &  \ce{ZrNBr}   &  1.815  &  0.333  &  0.338  &  \textbf{0.336}  &  0.377  &  0.921  &  \textbf{0.649}\\
\ce{Zr2N2I2}   &  \ce{ZrNI}   &  0.365  &  0.604  &  0.604  &  \textbf{0.604}  &  0.386  &  0.386  &  \textbf{0.386}\\
\ce{ZrNBr}   &  \ce{ZrNBr}   &  2.600  &  0.541  &  0.541  &  \textbf{0.541}  &  0.484  &  0.484  &  \textbf{0.484}\\
\ce{ZrS2}   &  \ce{ZrS2}   &  1.154  &  0.537  &  0.537  &  \textbf{0.537}  &  0.346  &  0.346  &  \textbf{0.346}\\
\ce{ZrSe2}   &  \ce{ZrSe2}   &  0.353  &  0.404  &  0.404  &  \textbf{0.404}  &  0.194  &  0.194  &  \textbf{0.194}\\
\ce{Ag2I2}   &  \ce{AgI}   &  1.911  &  0.464  &  0.464  &  \textbf{0.464}  &  0.899  &  0.899  &  \textbf{0.899}\\
\ce{AgO4Cl}   &  \ce{AgO4Cl}   &  2.902  &  0.568  &  0.568  &  0.568  &  1.236  &  1.236  &  1.236\\
\ce{Au2Br2}   &  \ce{AuBr}   &  2.035  &  1.981  &  2.618  &  2.300  &  0.396  &  1.096  &  0.746\\
\ce{Au2I2}   &  \ce{AuI}   &  1.848  &  0.928  &  5.088  &  3.008  &  0.271  &  1.599  &  0.935\\
\ce{Bi2O2Br2}   &  \ce{BiOBr}   &  2.021  &  0.267  &  0.267  &  0.267  &  1.470  &  1.470  &  1.470\\
\ce{Bi2O2Cl2}   &  \ce{BiOCl}   &  2.394  &  0.322  &  0.322  &  0.322  &  1.086  &  1.086  &  1.086\\
\ce{Bi2Se3}   &  \ce{Bi2Se3}   &  0.385  &  0.162  &  0.162  &  0.162  &  1.183  &  1.183  &  1.183\\
\ce{Bi2Te3}   &  \ce{Bi2Te3}   &  0.190  &  0.114  &  0.114  &  0.114  &  1.372  &  1.372  &  1.372\\
\ce{Bi2TeSe2}   &  \ce{Bi2TeSe2}   &  0.221  &  0.123  &  0.123  &  \textbf{0.123}  &  0.390  &  0.390  &  \textbf{0.390}\\
\ce{BiTeCl}   &  \ce{BiTeCl}   &  0.829  &  0.257  &  0.257  &  \textbf{0.257}  &  0.992  &  0.992  &  \textbf{0.992}\\
\ce{CdH2O2}   &  \ce{CdH2O2}   &  2.291  &  0.449  &  0.449  &  0.449  &  1.909  &  1.909  &  1.909\\
\ce{CdI2}   &  \ce{CdI2}   &  2.188  &  0.536  &  0.536  &  \textbf{0.536}  &  0.585  &  0.585  &  \textbf{0.585}\\
\hline\\[-8.5pt] \hline\\[-7pt]
\end{longtable}
\end{table*}
\newpage
\begin{table*}
\centering
\begin{longtable}{@{\extracolsep{\fill}} l l r r r r r r r}
\hline\\[-8.5pt] \hline\\[-7pt]
\multicolumn{1}{l}{Unit cell formula}    & \multicolumn{1}{l}{Reduced formula}  & \multicolumn{1}{c}{Gap (eV)} & \multicolumn{3}{c}{Electron mass ($m_e$)}        & \multicolumn{3}{c}{Hole mass ($m_e$)} \\[2pt]
&&&$x$&$y$&ave.&$x$&$y$&ave.\\[2pt]
\cline{4-6} \cline{7-9}\\[-5pt]
\ce{Cu2Br2}   &  \ce{CuBr}   &  1.473  &  0.437  &  0.437  &  0.437  &  2.494  &  2.494  &  2.494\\
\ce{Cu2I2}   &  \ce{CuI}   &  1.806  &  0.363  &  0.363  &  \textbf{0.363}  &  0.606  &  0.606  &  \textbf{0.606}\\
\ce{Cu2I2}   &  \ce{CuI}   &  1.945  &  0.366  &  0.401  &  0.384  &  0.887  &  1.269  &  1.078\\
\ce{Ga2Ge2Te2}   &  \ce{GaGeTe}   &  0.632  &  0.064  &  0.064  &  0.064  &  1.954  &  1.954  &  1.954\\
\ce{Ga2S2}   &  \ce{GaS}   &  2.215  &  0.222  &  0.222  &  0.222  &  3.519  &  3.519  &  3.519\\
\ce{Ga2S2}   &  \ce{GaS}   &  2.352  &  0.240  &  0.240  &  0.240  &  3.410  &  3.410  &  3.410\\
\ce{Ga2Se2}   &  \ce{GaSe}   &  1.764  &  0.164  &  0.164  &  0.164  &  3.136  &  3.136  &  3.136\\
\ce{Ga2Te2}   &  \ce{GaTe}   &  1.310  &  0.368  &  0.368  &  0.368  &  2.414  &  2.414  &  2.414\\
\ce{GeI2}   &  \ce{GeI2}   &  1.994  &  0.265  &  0.265  &  0.265  &  1.124  &  1.124  &  1.124\\
\ce{Hf2N2Br2}   &  \ce{HfNBr}   &  2.006  &  0.646  &  0.646  &  0.646  &  1.785  &  1.785  &  1.785\\
\ce{Hf2N2Cl2}   &  \ce{HfNCl}   &  2.372  &  0.639  &  0.639  &  0.639  &  3.241  &  3.241  &  3.241\\
\ce{HgI2}   &  \ce{HgI2}   &  1.551  &  0.668  &  0.668  &  0.668  &  1.226  &  1.226  &  1.226\\
\ce{In2O2Br2}   &  \ce{InOBr}   &  2.252  &  0.332  &  0.413  &  0.372  &  1.176  &  1.549  &  1.362\\
\ce{In2Se2}   &  \ce{InSe}   &  1.361  &  0.194  &  0.194  &  0.194  &  4.109  &  4.109  &  4.109\\
\ce{In2Se3}   &  \ce{In2Se3}   &  1.078  &  0.464  &  0.559  &  0.512  &  0.718  &  1.452  &  1.085\\
\ce{K2Tl2O2}   &  \ce{KTlO}   &  1.538  &  0.528  &  0.647  &  0.588  &  2.54  &  11.733  &  7.136\\
\ce{Lu2CCl2}   &  \ce{Lu2CCl2}   &  0.952  &  0.335  &  0.335  &  0.335  &  1.808  &  1.808  &  1.808\\
\ce{Na2H2O2}   &  \ce{NaHO}   &  2.799  &  0.706  &  0.706  &  0.706  &  2.397  &  2.605  &  2.501\\
\ce{P2}   &  \ce{P}   &  1.921  &  0.245  &  0.245  &  0.245  &  1.436  &  1.436  &  1.436\\
\ce{P4}   &  \ce{P}   &  0.900  &  0.161  &  1.277  &  0.719  &  0.153  &  3.346  &  1.750\\
\ce{Pb2O2}   &  \ce{PbO}   &  2.418  &  0.599  &  0.599  &  0.599  &  5.867  &  5.867  &  5.867\\
\ce{PbI2}   &  \ce{PbI2}   &  1.772  &  0.364  &  0.364  &  0.364  &  1.934  &  1.934  &  1.934\\
\ce{PtS2}   &  \ce{PtS2}   &  1.729  &  0.448  &  0.448  &  0.448  &  2.145  &  2.145  &  2.145\\
\ce{PtSe2}   &  \ce{PtSe2}   &  1.200  &  0.335  &  0.335  &  0.335  &  3.519  &  3.519  &  3.519\\
\ce{Rh2Te2Cl2}   &  \ce{RhTeCl}   &  1.062  &  0.322  &  0.856  &  0.589  &  3.568  &  4.587  &  4.077\\
\ce{Sb2Te2I2}   &  \ce{SbTeI}   &  0.822  &  0.223  &  0.349  &  0.286  &  0.907  &  2.261  &  1.584\\
\ce{Sc2CCl2}   &  \ce{Sc2CCl2}   &  0.889  &  0.445  &  0.445  &  0.445  &  1.365  &  1.365  &  1.365\\
\ce{Sn2O2}   &  \ce{SnO}   &  2.091  &  0.563  &  0.695  &  0.629  &  0.401  &  4.081  &  2.241\\
\ce{Tl2S}   &  \ce{Tl2S}   &  1.255  &  0.300  &  0.300  &  0.300  &  2.958  &  2.958  &  2.958\\
\ce{YbI2}   &  \ce{YbI2}   &  0.945  &  0.482  &  0.482  &  0.482  &  74.399  &  74.399  &  74.399\\
\ce{Zr2N2Br2}   &  \ce{ZrNBr}   &  1.631  &  0.577  &  0.577  &  0.577  &  2.115  &  2.115  &  2.115\\
\ce{Zr2N2Cl2}   &  \ce{ZrNCl}   &  1.922  &  0.577  &  0.577  &  0.577  &  2.258  &  2.258  &  2.258\\
\ce{Zr2N2I2}   &  \ce{ZrNI}   &  1.170  &  0.294  &  0.297  &  \textbf{0.295}  &  0.247  &  0.883  & \textbf{0.565}\\
\ce{ZrCl2}   &  \ce{ZrCl2}   &  0.981  &  1.326  &  1.326  &  1.326  &  0.475  &  0.475  &  0.475\\
\ce{Ho2S2I2}   &  \ce{HoSI}   &  0.595  &  14.507  &  47.072  &  30.79  &  0.345  &  1.211  &  0.778\\
\ce{LaBr2}   &  \ce{LaBr2}   &  0.709  &  2.817  &  2.817  &  2.817  &  0.816  &  0.816  &  0.816\\
\ce{Cu4Te2}   &  \ce{Cu2Te}   &  0.070  &  0.390  &  0.689  &  \textbf{0.539}  &  0.181  &  0.217  &  \textbf{0.199}\\
\ce{La2GeI2}   &  \ce{La2GeI2}   &  0.372  &  0.318  &  0.318  &  \textbf{0.318}  &  0.895  &  0.895  &  \textbf{0.895}\\
\ce{MnO2H2}   &  \ce{MnO2H2}   &  0.598  &  0.672  &  0.679  &  0.676  &  0.921  &  29.593  & 15.257\\
\hline\\[-8.5pt] \hline\\[-7pt]
\end{longtable}
\end{table*}
\setcounter{table}{3}
\begin{table*}
\caption{\textbf{Materials parameters and mobility estimates of compounds in Set~C}. Calculated band gap, directionally-averaged electron and hole effective mass, high-frequency dielectric constant, ionic contribution to the static dielectric constant, effective thickness, LO phonon energy, electron and hole mobilities from the Fr\"ohlich model, and electron and hole mobilities from Emin's formula, for 50 compounds in Set~C. Fr\"ohlich data for Bi and Sb are not given since these compounds are non-polar.}\label{tableS3}
\vspace{4pt}
\centering
\begin{longtable}{@{\extracolsep{\fill}} l r r r r r r r r r r r}
\hline\\[-8.5pt] \hline\\[-7pt]
  &  & &  &  & &  & & \multicolumn{2}{c}{Fr\"ohlich} & \multicolumn{2}{c}{Emin} \\[4pt]
\cline{9-10} \cline{11-12}\\[-5pt]
Formula  & Gap & $m_{\rm e}^*$ & $m_{\rm h}^*$  & $\epsilon^{\infty}$ & $\epsilon^{\rm ion}$ & $d$ & $\hbar\omega_{\rm LO}$ & $\mu_{\rm e}$ & $\mu_{\rm h}$
     & $\mu_{\rm e}$ & $\mu_{\rm h}$ \\[4pt]
  & (eV) & ($m_e$) & ($m_e$) & & & (\AA) & (meV) & (\mob) & (\mob)
     & (\mob) & (\mob)\\[4pt]
\ce{AgBr} & 1.291 & 0.306 & 0.590 & 7.609 & 6.785 & 4.761 & 23.644 & 15 & 6 & 146 & 76\\
\ce{As} & 0.772 & 0.725 & 0.247 & 22.612 & 1.994 & 4.282 & 31.699 & 80 & 321 & 62 & 181\\
\ce{BiOI} & 1.150 & 0.218 & 0.619 & 11.320 & 55.980 & 6.686 & 57.759 & 21 & 5 & 205 & 72\\
\ce{Bi2Te2S} & 0.274 & 0.105 & 0.518 & 29.863 & 29.939 & 7.274 & 26.194 & 93 & 13 & 428 & 86\\
\ce{Bi2Te2S} & 0.257 & 0.142 & 0.757 & 36.947 & 53.205 & 7.177 & 27.110 & 50 & 7 & 326 & 59\\
\ce{Bi2Te2Se} & 0.230 & 0.105 & 0.673 & 33.310 & 35.051 & 7.380 & 19.032 & 69 & 7 & 425 & 67\\
\ce{GaTeCl} & 2.158 & 0.816 & 0.698 & 8.857 & 8.976 & 5.301 & 46.239 & 9 & 11 & 55 & 64\\
\ce{GeS} & 1.753 & 0.431 & 0.871 & 13.984 & 11.960 & 4.308 & 40.954 & 21 & 8 & 104 & 51\\
\ce{GeSe} & 1.142 & 0.219 & 0.245 & 26.375 & 33.181 & 4.327 & 28.080 & 24 & 21 & 205 & 183\\
\ce{GeI2} & 1.843 & 0.753 & 0.563 & 10.248 & 8.542 & 5.388 & 23.300 & 4 & 7 & 59 & 80\\
\ce{HfNBr} & 2.178 & 0.350 & 0.449 & 8.010 & 15.859 & 6.079 & 84.390 & 77 & 55 & 128 & 100\\
\ce{HfNI} & 0.593 & 0.694 & 0.373 & 9.923 & 8.135 & 7.399 & 79.368 & 76 & 163 & 65 & 120\\
\ce{HfS2} & 1.267 & 0.456 & 0.327 & 13.815 & 28.641 & 4.206 & 39.613 & 7 & 12 & 98 & 137\\
\ce{HfSe2} & 0.483 & 0.361 & 0.201 & 19.659 & 44.826 & 4.476 & 27.486 & 6 & 14 & 124 & 223\\
\ce{KAgSe} & 0.492 & 0.238 & 0.548 & 13.970 & 4.448 & 6.187 & 29.696 & 83 & 27 & 188 & 82\\
\ce{LiAlTe2} & 0.642 & 0.189 & 0.561 & 8.510 & 3.560 & 5.564 & 56.422 & 256 & 56 & 237 & 80\\
\ce{MoSe2} & 1.335 & 0.583 & 0.675 & 22.431 & 0.684 & 4.716 & 42.541 & 539 & 451 & 77 & 66\\
\ce{MoTe2} & 0.950 & 0.610 & 0.756 & 26.359 & 2.122 & 5.170 & 34.908 & 153 & 118 & 73 & 59\\
\ce{PbFI} & 1.922 & 0.371 & 0.955 & 8.327 & 18.946 & 6.565 & 39.525 & 9 & 2 & 121 & 47\\
\ce{PbTe} & 1.070 & 0.172 & 0.493 & 14.863 & 3.069 & 4.054 & 18.790 & 136 & 29 & 260 & 91\\
\ce{SiH} & 2.170 & 0.234 & 0.418 & 11.115 & 1.080 & 3.476 & 66.000 & 1205 & 529 & 191 & 107\\
\ce{Sb2Te2Se} & 0.445 & 0.114 & 0.475 & 43.464 & 26.825 & 7.236 & 22.484 & 136 & 25 & 393 & 94\\
\ce{Sb2Te3} & 0.405 & 0.151 & 0.666 & 51.560 & 40.172 & 7.417 & 20.896 & 79 & 14 & 297 & 67\\
\ce{Sb2TeSe2} & 0.596 & 0.153 & 0.954 & 48.983 & 45.793 & 7.026 & 23.209 & 68 & 8 & 140 & 69\\
\ce{Sb2TeSe2} & 0.525 & 0.320 & 0.651 & 42.507 & 28.667 & 7.062 & 23.723 & 37 & 16 & 293 & 47\\
\ce{SnTe} & 0.657 & 0.241 & 0.133 & 61.844 & 381.357 & 4.768 & 18.696 & 4 & 9 & 186 & 338\\
\ce{SnS2} & 1.558 & 0.403 & 0.582 & 10.896 & 14.201 & 4.382 & 42.890 & 15 & 9 & 111 & 77\\
\ce{SnSe2} & 0.752 & 0.354 & 0.633 & 15.112 & 16.171 & 4.638 & 30.543 & 14 & 6 & 127 & 71\\
\ce{TiNBr} & 0.621 & 0.278 & 0.278 & 17.685 & 46.382 & 5.656 & 80.609 & 91 & 91 & 161 & 161\\
\ce{TiNCl} & 0.625 & 0.335 & 0.246 & 16.768 & 43.996 & 5.186 & 82.285 & 72 & 106 & 134 & 182\\
\ce{Tl2O} & 0.934 & 0.159 & 0.325 & 17.626 & 11.128 & 4.927 & 46.690 & 156 & 60 & 282 & 138\\
\ce{WS2} & 1.584 & 0.384 & 0.391 & 18.911 & 0.050 & 4.409 & 52.772 & 15459 & 15105 & 117 & 114\\
\ce{WSe2} & 1.260 & 0.479 & 0.407 & 20.391 & 0.297 & 4.742 & 37.082 & 1053 & 1296 & 94 & 110\\
\ce{WTe2} & 0.765 & 0.401 & 0.373 & 24.651 & 1.229 & 5.188 & 29.513 & 307 & 337 & 112 & 120\\
\ce{ZnI2} & 1.793 & 0.424 & 0.574 & 7.967 & 72.471 & 5.024 & 21.085 & 1 & 1 & 106 & 78\\
\ce{ZrNBr} & 1.815 & 0.336 & 0.649 & 8.829 & 18.239 & 6.072 & 79.697 & 66 & 28 & 133 & 69\\
\ce{ZrNI} & 0.365 & 0.604 & 0.386 & 10.981 & 9.493 & 7.353 & 78.697 & 87 & 151 & 74 & 116\\
\ce{ZrNBr} & 2.600 & 0.541 & 0.484 & 9.892 & 13.330 & 4.054 & 64.226 & 25 & 29 & 83 & 93\\
\ce{ZrS2} & 1.154 & 0.537 & 0.346 & 15.475 & 33.183 & 4.188 & 41.551 & 6 & 12 & 83 & 130\\
\ce{ZrSe2} & 0.353 & 0.404 & 0.194 & 23.451 & 56.247 & 4.461 & 31.045 & 6 & 16 & 111 & 230\\
\ce{AgI} & 1.911 & 0.464 & 0.899 & 5.928 & 1.590 & 5.326 & 21.515 & 25 & 9 & 96 & 50\\
\ce{Bi2TeSe2} & 0.221 & 0.123 & 0.390 & 33.530 & 55.875 & 7.160 & 20.311 & 37 & 9 & 364 & 115\\
\ce{BiTeCl} & 0.829 & 0.257 & 0.992 & 14.275 & 10.948 & 4.581 & 23.039 & 23 & 4 & 174 & 45\\
\ce{CdI2} & 2.188 & 0.536 & 0.585 & 7.012 & 7.493 & 5.187 & 17.714 & 5 & 4 & 83 & 77\\
\ce{CuI} & 1.806 & 0.363 & 0.606 & 7.410 & 4.097 & 5.484 & 21.103 & 18 & 8 & 124 & 74\\
\ce{MoS2} & 1.617 & 0.497 & 0.865 & 20.910 & 0.206 & 4.377 & 57.039 & 3843 & 1965 & 90 & 52\\
\ce{ZrNI} & 1.170 & 0.296 & 0.565 & 9.755 & 20.333 & 6.701 & 79.859 & 84 & 37 & 151 & 79\\
\ce{La2GeI2} & 0.372 & 0.318 & 0.895 & 56.275 & 170.245 & 6.964 & 22.884 & 9 & 3 & 141 & 50\\
\ce{Bi} & 0.465 & 0.267 & 0.639 & - & 0.000 & - & - & - & - & 168 & 70\\
\ce{Sb} & 1.022 & 0.235 & 0.156 & - & 0.000 & - & - & - & - & 191 & 287\\
\hline\\[-8.5pt] \hline\\[-7pt]
\end{longtable}
\end{table*}
\setcounter{table}{4}
\begin{table*}[!ht]
\caption{\textbf{Comparison of calculated effective masses with literature values}. We compare our calculated conductivity effective masses for all compounds in Table~I of the main text with previous work. As there are several ways to define and compute effective masses, in each case we indicate the method employed in previous work via superscripts, as follows: $\star$: Momentum average, $\dagger$: Parabolic fitting, $\perp$: Density of states effective mass, $\parallel$: Band structure curvature, $\square$: Experiment, $\times$: Unclear. All values in units of the free electron mass mass.}
\begin{center}
\renewcommand{\arraystretch}{1.3}
\begin{tabular*}{0.98\textwidth}{@{\extracolsep{\fill}} l l l l l l }
\hline
\hline
Compound & SOC & \multicolumn{2}{l}{Electron $m_{\rm e}^{\ast}$} & \multicolumn{2}{l}{Hole $m_{\rm h}^{\ast}$} \\[3pt]
         \cline{3-6}
         &     & This         &  Previous studies         & This           & Previous studies    \\[3pt]
         &     & work         &                           & work           &                     \\[3pt]
\hline\\[-7pt]
h-Sb     & No     & 0.230   & -    
                  & 0.347   & -        \\
         & Yes    & 0.235   & 0.219$^\star$~\cite{Cheng2019}, 0.215$^\dagger$, 0.14$^\perp$~\cite{Gjerding2021}  
                  & 0.156   & 0.12$^\star$~\cite{Cheng2019}, 0.13$^\dagger$, 0.13$^\perp$~\cite{Gjerding2021}, 0.129$^\star$~\cite{Zhang2023b} \\
SiH      & No     & 0.235   & 1.815$^\parallel$~\cite{Khatami2019}    
                  & 0.392   & 0.348$^\parallel$~\cite{Khatami2019}  \\
         & Yes    & 0.234   & 1.83$^\dagger$, 0.65$^\perp$~\cite{Gjerding2021}  
                  & 0.418   & 0.26$^\dagger$, 0.26$^\perp$~\cite{Gjerding2021} \\
\ce{Tl2O}& No     & 0.161   & -     
                  & 0.343   & - \\
         & Yes    & 0.159   & 0.253$^\parallel$~\cite{Ma2017}    
                  & 0.325   & 2.229$^\parallel$~\cite{Ma2017} \\
\ce{Bi2TeSe2}& No   & 0.109  & -       
                    & 1.490  & -   \\
             & Yes  & 0.123  & 0.081$^\parallel$~\cite{Lu2019}, 0.04$^\dagger$, 0.04$^\perp$~\cite{Gjerding2021}  
                    & 0.390  & 0.104$^\parallel$~\cite{Lu2019}, 0.05$^\dagger$, 0.05$^\perp$~\cite{Gjerding2021} \\
GeSe     & No     & 0.249    & 0.265$^\parallel$~\cite{Xu2017b}, 0.2$^\star$~\cite{Yang2020b}, 0.25$^\parallel$~\cite{Xiao2023}
                  & 0.235    & 0.29$^\parallel$~\cite{Xu2017b}, 0.19$^\star$~\cite{Yang2020b}, 0.222$^\parallel$~\cite{Xiao2023} \\
         & Yes    & 0.219    & 0.21$^\parallel$~\cite{Xu2017}, 0.18$^\dagger$, 0.17$^\perp$~\cite{Gjerding2021}, 0.19$^\parallel$~\cite{Xiao2023}
                  & 0.245    & 0.17$^\dagger$, 0.17$^\perp$~\cite{Gjerding2021}, 0.245$^\parallel$~\cite{Xu2017}, 0.19$^\parallel$~\cite{Xiao2023} \\
SnTe     & No     & 0.233    & 0.09$^\parallel$~\cite{Xu2017b}, 0.11$^\parallel$~\cite{Guo2019}, 0.094$^\parallel$~\cite{Xiao2023} 
                  & 0.116    & 0.07$^\parallel$~\cite{Xu2017b}, 0.114$^\parallel$~\cite{Guo2019}, 0.088$^\parallel$~\cite{Xiao2023} \\
         & Yes    & 0.241    & 0.08$^\dagger$, 0.07$^\perp$~\cite{Gjerding2021}, 0.094$^\parallel$~\cite{Xiao2023} 
                  & 0.133    & 0.065$^\dagger$, 0.06$^\perp$~\cite{Gjerding2021}, 0.087$^\parallel$~\cite{Xiao2023} \\
\ce{ZrSe2}& No    & 0.399    & 1.15$^\parallel$~\cite{Bao2024-lm}    
                  & 0.251    & 0.52$^\parallel$~\cite{Bao2024-lm} \\
          & Yes   & 0.404    & 1.04$^\dagger$, 0.61$^\perp$~\cite{Gjerding2021}  
                  & 0.194    & 0.181$^\times$~\cite{Cheng2018}, 0.2$^\dagger$, 0.2$^\perp$~\cite{Gjerding2021}  \\
\ce{HfSe2}& No    & 0.353    & 1.055$^\parallel$~\cite{Bao2024-lm}   
                  & 0.271    & 0.45$^\parallel$~\cite{Bao2024-lm}  \\
          & Yes   & 0.361    & 1.18$^\dagger$, 0.61$^\perp$~\cite{Gjerding2021}  
                  & 0.201    & 0.181$^\times$~\cite{Cheng2018}, 0.2$^\dagger$, 0.2$^\perp$~\cite{Gjerding2021}  \\
TiNCl     & No    & 0.335    & -       
                  & 0.246    & -    \\
          & Yes   & 0.335    & 0.28$^\dagger$, 0.25$^\perp$~\cite{Gjerding2021}    
                  & 0.246    & 0.22$^\dagger$, 0.2$^\perp$~\cite{Gjerding2021} \\
TiNBr     & No    & 0.281    & -       
                  & 0.274    & -         \\
          & Yes   & 0.278    & 0.27$^\dagger$, 0.25$^\perp$~\cite{Gjerding2021}     
                  & 0.277    & 0.24$^\dagger$, 0.22$^\perp$~\cite{Gjerding2021} \\
\ce{WS2}  & No    & 0.329    & 0.33$^\times$~\cite{Jiang2015}, 0.314$^\times$~\cite{Sohier2018}    
                  & 0.640    & 0.43$^\times$~\cite{Jiang2015} \\
          & Yes   & 0.384    & 0.38$^\dagger$, 0.37$^\perp$~\cite{Gjerding2021}  
                  & 0.391    & 0.334$^\times$~\cite{Cheng2018}, 0.66$^\dagger$, 0.66$^\perp$~\cite{Gjerding2021}   \\
\ce{WTe2} & No    & 0.349    & 0.33$^\times$~\cite{Jiang2015}      
                  & 0.495    & 0.43$^\times$~\cite{Jiang2015}        \\
          & Yes   & 0.401    & 0.45$^\dagger$, 0.45$^\perp$~\cite{Gjerding2021}   
                  & 0.373    & 0.29$^\dagger$, 0.29$^\perp$~\cite{Gjerding2021}   \\
\hline\\[-10pt]
\ce{MoS2} & No    & 0.486    & 0.51$^\times$~\cite{Jiang2015}, 0.417$^\times$~\cite{Sohier2018}  
                  & 1.980    & 0.6$^\times$~\cite{Jiang2015}    \\
          & Yes   & 0.497    & 0.43$^\dagger$, 0.43$^\perp$~\cite{Gjerding2021}, 0.42$^\star$~\cite{Ponce2023b}, 0.428$^\star$~\cite{Zhang2023b} 
                  & 0.865    & 0.56$^\times$~\cite{Cheng2018}, 0.53$^\dagger$, 0.53$^\perp$~\cite{Gjerding2021}, 0.52$^\star$~\cite{Ponce2023b}  \\
\ce{h-Bi} & No    & 0.055    & -       
                  & 0.379    & -        \\
          & Yes   & 0.267    & 0.250$^\star$~\cite{Cheng2019}     
                  & 0.639    & 0.549$^\star$~\cite{Cheng2019}        \\
\ce{GeS}  & No    & 0.345    & 0.345$^\parallel$~\cite{Xu2017b}, 0.39$^\star$~\cite{Yang2020b}, 0.33$^\parallel$~\cite{Xiao2023}   
                  & 0.430    & 0.56$^\parallel$~\cite{Xu2017b}, 0.46$^\star$~\cite{Yang2020b}, 0.456$^\parallel$~\cite{Xiao2023}  \\
          & Yes   & 0.431    & 0.36$^\dagger$, 0.33$^\perp$~\cite{Gjerding2021}, 0.326$^\parallel$~\cite{Xiao2023}  
                  & 0.871    & 0.975$^\dagger$, 0.51$^\perp$~\cite{Gjerding2021}, 0.449$^\parallel$~\cite{Xiao2023}   \\
\ce{WSe2} & No    & 0.394    & 0.33$^\times$~\cite{Sohier2018}    
                  & 0.513    & -        \\
          & Yes   & 0.479    & 0.36$^\dagger$, 0.35$^\perp$~\cite{Gjerding2021}  
                  & 0.407    & 0.45$^\square$~\cite{Fallahazad2016}, 0.394$^\times$~\cite{Cheng2018}, 0.4$^\dagger$, 0.4$^\perp$~\cite{Gjerding2021}, 0.35$^\square$~\cite{Joe2024}  \\
\hline
\hline
\end{tabular*}
\label{effective_mass}
\end{center}
\end{table*}
%
%
%
\begin{table*}[ht!]
\caption{\textbf{Dynamical quadrupole tensors of monolayer WS$_2$}. Calculated dynamical quadrupole tensors for monolayer WS$_2$, in units of $e\cdot {\rm bohr}$. Calculations performed using the DFPT framework of Ref.~\citenum{Royo2019}, as implemented in the \texttt{Abinit} code~\cite{Gonze2020, Romero2020}.} \label{quad_tensor}
\begin{center}
\begin{tabular*} {0.9\textwidth} {@{\extracolsep{\fill}} l c r r r r r r}
\hline\\[-10.5pt] \hline\\[-9pt]
Atom &  Cart. Dir.   &   $Q_{xx}$   &    $Q_{yy}$  &     $Q_{zz}$   &     $Q_{yz}$  &    $Q_{xz}$   &    $Q_{xy}$ \\[3pt]
W    &  $x$    &   0.00000    &    0.00000   &     0.00000    &    0.00000    &    0.00000    &    3.21037  \\
     &  $y$    &   3.21037    &   -3.21037   &     0.00000    &    0.00000    &    0.00000    &    0.00000  \\
     &  $z$    &   0.00000    &    0.00000   &     0.00000    &    0.00000    &    0.00000    &    0.00000  \\
S-1  &  $x$    &   0.00000    &    0.00000   &     0.00000    &    0.00000    &   -1.16786    &    0.92645  \\
     &  $y$    &   0.92645    &   -0.92645   &     0.00000    &   -1.16786    &    0.00000    &    0.00000  \\
     &  $z$    &  -7.45908    &   -7.45908   &    -2.03160    &    0.00000    &    0.00000    &    0.00000  \\
S-2  &  $x$    &   0.00000    &    0.00000   &     0.00000    &    0.00000    &    1.16786    &    0.92645  \\
     &  $y$    &   0.92645    &   -0.92645   &     0.00000    &    1.16786    &    0.00000    &    0.00000  \\
     &  $z$    &   7.45908    &    7.45908   &     2.03160    &    0.00000    &    0.00000    &    0.00000  \\[2pt]
\hline\\[-10.5pt] \hline\\[-7pt]
\end{tabular*}
\end{center}
\end{table*}
%
%
%
%
%
%
%
%
%
%
\begin{figure*}
\begin{center}
\includegraphics[width=0.8\linewidth]{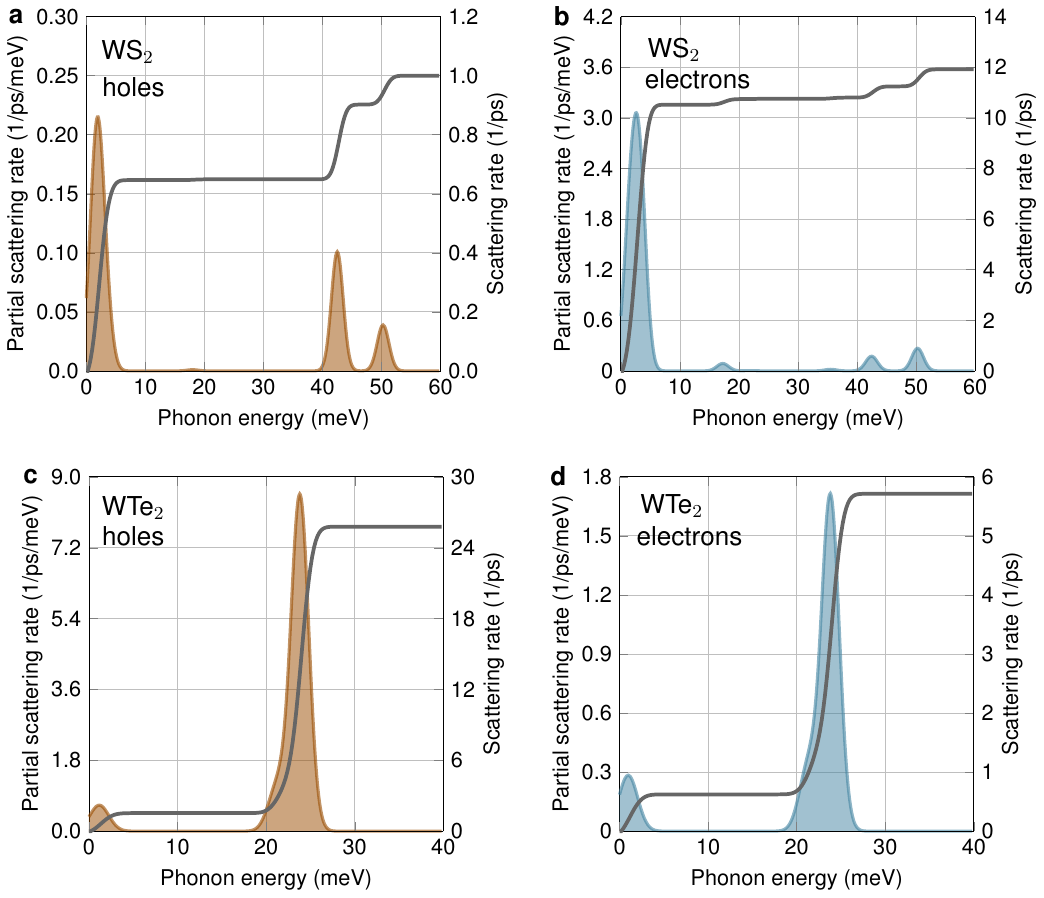}
\end{center}
\caption{\textbf{Analysis of scattering rates for \ce{WS2} and \ce{WTe2}}. (a) Partial scattering rates of holes by phonons in \ce{WS2}, resolved by phonon mode energy (orange line and filled area). The gray line is the cumulative scattering rate, i.e. the running integral of the blue curve, as reported on the right axis. (b) Partial scattering rates of electrons by phonons in \ce{WS2} (blue line and filled area). The gray line is the cumulative rate (right axis). (c) and (d): Same format as for (a) and (b), but for holes and electrons in \ce{WTe2}, respectively. All rates are calculated as averages over carriers with energy $k_{\rm B} T$ from the band edges, with $T=300$\,K, for a carrier concentration of $10^{10}$\,cm$^{-2}$.}
\label{spectral_decomp_WS2.WTe2}
\end{figure*}
%
%
%
%
%
\begin{figure*}
\begin{center}
\includegraphics[width=0.8\linewidth]{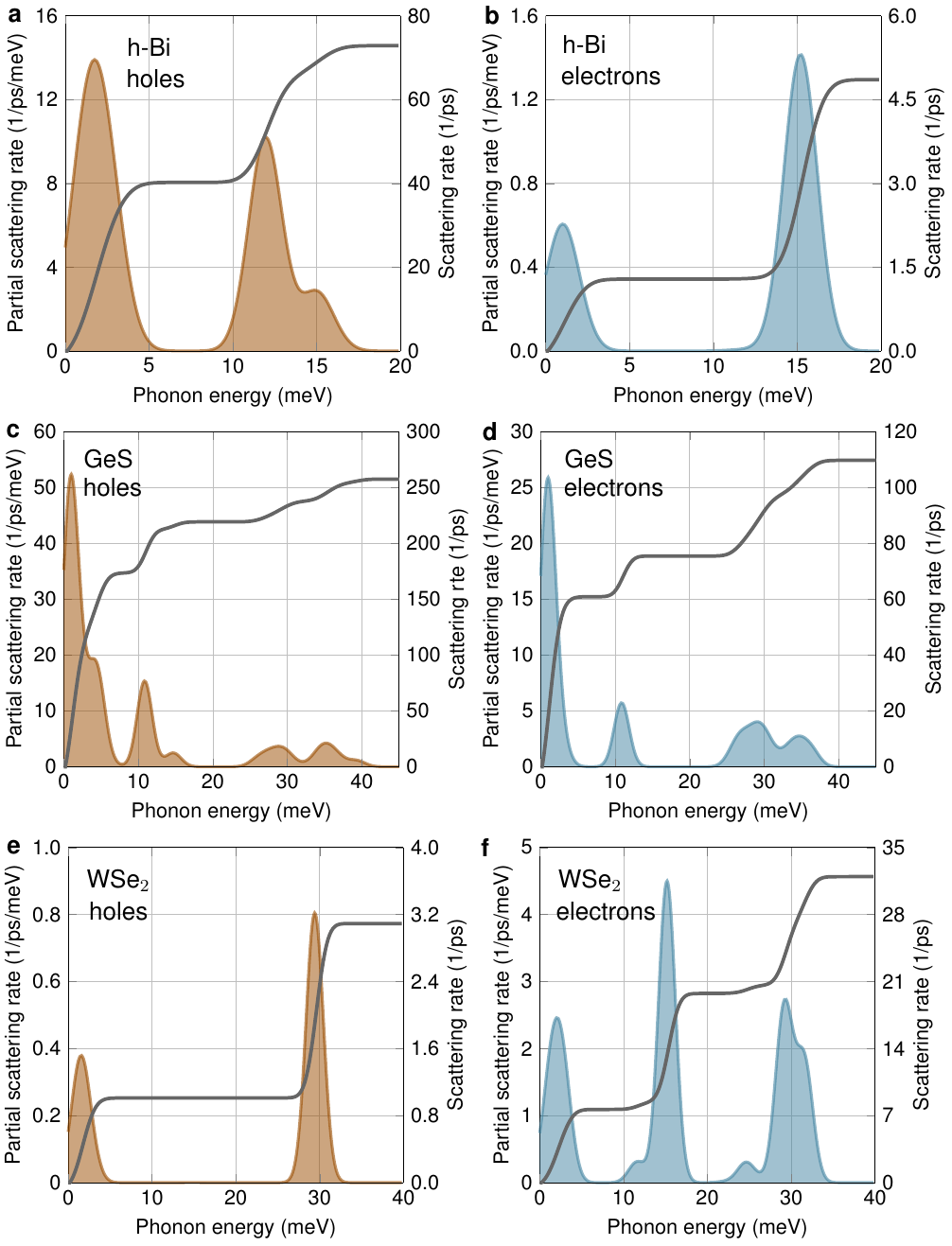}
\end{center}
\caption{\textbf{Analysis of scattering rates for h-Bi, GeS, and \ce{WSe2}}. (a) Partial scattering rates of holes by phonons in h-Bi, resolved by phonon mode energy (orange line and filled area). The gray line is the cumulative scattering rate, i.e. the running integral of the blue curve, as reported on the right axis. (b) Partial scattering rates of electrons by phonons in h-Bi (blue line and filled area). The gray line is the cumulative rate (right axis). (c) and (d): Same format as for (a) and (b), but for holes and electrons in GeS, respectively. (e) and (f): Same format as for (a) and (b), but for holes and electrons in \ce{WSe2}, respectively. All rates are calculated as averages over carriers with energy $k_{\rm B} T$ from the band edges, with $T=300$\,K, for a carrier concentration of $10^{10}$\,cm$^{-2}$.}
\label{spectral_decomp_WS2.WTe2}
\end{figure*}
%
\begin{figure*}
\begin{center}
\includegraphics[width=\linewidth]{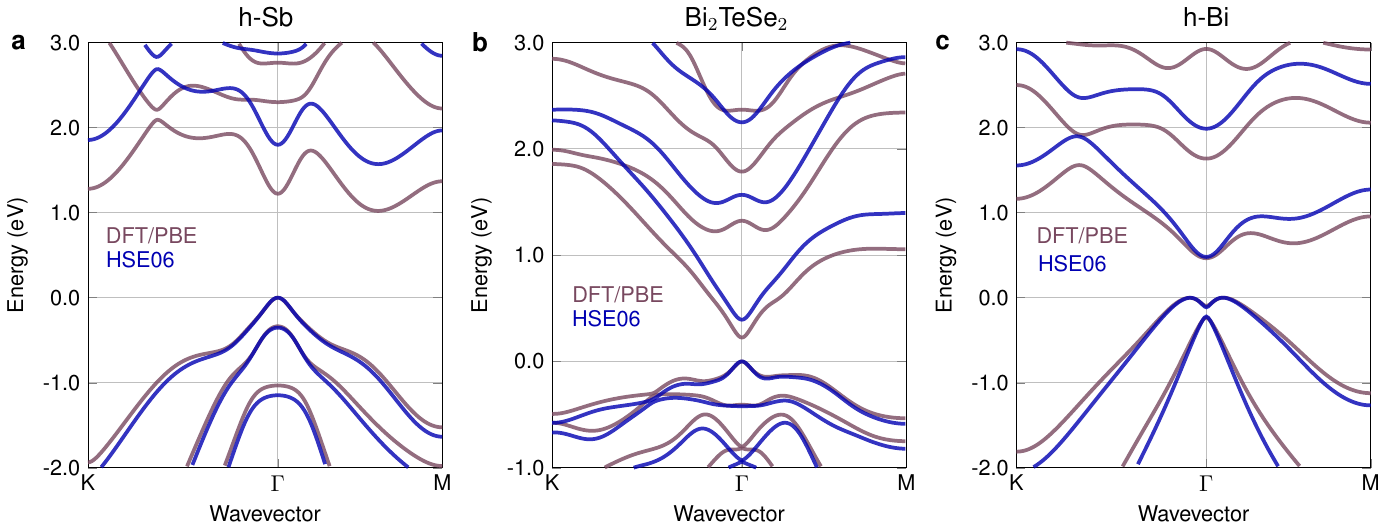}
\end{center}
\caption{\textbf{Band structure calculations using hybrid functionals}. (a) Comparison between the band structures of h-Sb calculated using DFT/PBE (red) and the hybrid HSE06 functional (blue)~\cite{Heyd2006}. (b) Comparison between DFT/PBE and HSE06 band structures of \ce{Bi2TeSe2}. (c) Comparison between DFT/PBE and HSE06~\cite{Heyd2003, Krukau2006} band structures of h-Bi. SOC is included in all calculations.}
\label{hse.bs.Sb-Bi-BTS}
\end{figure*}
%
%
\clearpage
\newpage
\bibliographystyle{apsrev4-1}
\bibliography{biblio}
\end{document}